\documentclass{pas}

\usepackage{multirow}
\usepackage{journals}
\usepackage{array}

\newcommand{\be}{\begin{enumerate}}
\newcommand{\ee}{\end{enumerate}}

\def\degrees{\hbox{$^\circ$}}
\def\arcsec{\hbox{$^{\prime\prime}$}}

\usepackage{tcolorbox}
\newtcolorbox[blend into=tables]{mytable}[2][]{colback=purple!5, ,colframe=purple!40!black, float=htb,  title={#2}, every float=\centering, #1}
\providecommand{\degr}{\hbox{$^{\circ}$}}

\providecommand{\arcsec}{\hbox{$^{\prime\prime}$}}
\providecommand{\mjybeam}{mJy\,beam$^{-1}$}
\providecommand{\ujy}{$\mu$Jy}
\providecommand{\ujybeam}{\ujy\,beam$^{-1}$}
\providecommand{\degsq}{deg$^2$}
\newcommand{\red}[1]{\textcolor{red}{#1}}

\begin{document}

\lefttitle{A New Look At The Dynamic Radio Sky}
\righttitle{Murphy \& Kaplan}

\jnlPage{1}{59}
\jnlDoiYr{2025}
\doival{10.1017/pasa.xxxx.xx}

\articletitt{Research Paper}

\title{The Dawes Review 13: A New Look at The Dynamic Radio Sky}

\author{\sn{Tara} \gn{Murphy}$^{1,3}$ and \sn{David L.} \gn{Kaplan}$^{2,3}$}

\affil{$^1$Sydney Institute for Astronomy, School of Physics, The University of Sydney, New South Wales 2006, Australia}

\affil{$^2$Department of Physics, University of Wisconsin-Milwaukee, P.O. Box 413, Milwaukee, WI 53201, USA}

\affil{$^3$ARC Centre of Excellence for Gravitational Wave Discovery (OzGrav), Hawthorn, VIC 3122, Australia}

\corresp{Tara Murphy, Email: tara.murphy@sydney.edu.au and David Kaplan, Email: kaplan@uwm.edu}

\citeauth{Murphy T and Kaplan D.L., A New Look at The Dynamic Radio Sky {\it Publications of the Astronomical Society of Australia} {\bf 00}, 1--12. https://doi.org/10.1017/pasa.xxxx.xx}

\history{(Received xx xx xxxx; revised xx xx xxxx; accepted xx xx xxxx)}

\begin{abstract}
Astronomical objects that change rapidly give us insight into extreme environments, allowing us to identify new phenomena, test fundamental physics, and probe the Universe on all scales. Transient and variable radio sources range from the cosmological, such as gamma-ray bursts, to much more local events, such as massive flares from stars in our Galactic neighbourhood. The capability to observe the sky repeatedly, over many frequencies and timescales, has allowed us to explore and understand dynamic phenomena in a way that has not been previously possible. 
In the past decade, there have been great strides forward as we prepared for the revolution in time domain radio astronomy that is being enabled by the SKA Observatory telescopes, the SKAO pathfinders and precursors, and other `next generation' radio telescopes. Hence it is timely to review the current status of the field, and summarise the developments that have happened to get to our current point. This review focuses on image domain (or `slow') transients, on timescales of seconds to years. We discuss the physical mechanisms that cause radio variability, and the classes of radio transients that result. We then outline what an ideal image domain radio transients survey would look like, and summarise the history of the field, from targeted observations to surveys with existing radio telescopes. We discuss methods and approaches for transient discovery and classification, and identify some of the challenges in scaling up current methods for future telescopes. 
Finally, we present our current understanding of the dynamic radio sky, in terms of source populations and transient rates, and look at what we can expect from surveys on future radio telescopes. 
\end{abstract}

\begin{keywords}
radio astronomy, radio transient sources, sky surveys, transient detection
\end{keywords}

\maketitle

\section{Introduction}
Transient radio emission is a signature of some of the most energetic and interesting events in our Universe: from stellar explosions \citep{weiler_radio_2002} and compact object mergers \citep{hallinan_radio_2017} to tidal disruption and accretion around supermassive black holes \citep{zauderer_birth_2011}. These dynamic events happen in extreme physical conditions and hence let us test fundamental physics such as mechanisms for magnetic field generation, particle acceleration, the strong-field gravity regime \citep{kramer_strong-field_2021} and the cosmological star formation history \citep{tanvir_-ray_2009,chandra_discovery_2010}. They also act as probes of the interplanetary \citep{jokipii_turbulence_1973}, interstellar \citep{rickett_radio_1990} and intergalactic media \citep{zhou_fast_2014}, giving us insight into the structure and composition of the Universe on all scales. 

Transient events have played an important role in astronomy since its early days. For example, there are numerous records of `guest stars' by Chinese astronomers in the first century of the Common Era \citep{stephenson_historical_2002}, some of which were later established to be historical supernovae. In Europe, Tycho Brahe's detailed observations of SN 1572 were an important step forward in challenging the Aristotelian/Ptolemaic paradigm of the `unchanging heavens' \citep{hinse_how_2023}. In the present era, targeted monitoring and surveys of optical transients have led to major discoveries: perhaps most notably the discovery of the accelerating expansion of the Universe through observations of distant supernovae \citep{riess_observational_1998, perlmutter_measurements_1999}, which won the 2011 Nobel Prize for Physics. Most recently, the detection of a gravitational wave transient, a binary black hole merger \citep{abbott_gw_2016}, led to the 2017 Nobel Prize for Physics and opened up the field of multi-messenger transients \citep{abbott_multi-messenger_2017}.

In contrast to optical and X-ray wavebands, the dynamic radio sky had been relatively unexplored (beyond targeted observations). The notable exception to this was pulsars, the discovery \citep{hewish_observation_1968} and study \citep{hulse_discovery_1975} of which have led to two Nobel Prizes, in 1974 and 1993 respectively. Looking beyond pulsars, the general lack of large-scale time domain studies at radio frequencies (until recent times --- see Section~\ref{s_history}) has been largely due to observational limitations: widefield, sensitive surveys with multiple epochs were difficult and time consuming to conduct. 

We now know the radio sky is variable on all timescales, from nanoseconds \citep{hankins_nanosecond_2003} through to months and years \citep{chandra_radio-selected_2012,zauderer_radio_2013}. The past decade has seen many advances in our understanding of radio transients, from the detection of radio emission from a binary neutron star merger \citep{hallinan_radio_2017}, to the real-time detection of an extreme scattering event \citep{bannister_real-time_2016} and the discovery of a new class of long period transients \citep{hurley-walker_radio_2022}. This review is being written as we prepare for the wave of discoveries that will come in the era of the SKA Observatory \citep[SKAO;][]{dewdney_ska1_2022}, which is expected to begin science verification observations in 2027 and 2029 for the SKA-Low and SKA-Mid telescopes, respectively\footnote{\url{https://www.skao.int/en/science-users/timeline-science}}. 

\subsection{The context for this review}
\begin{figure*}
    \centering
     \includegraphics[width=1.0\textwidth]{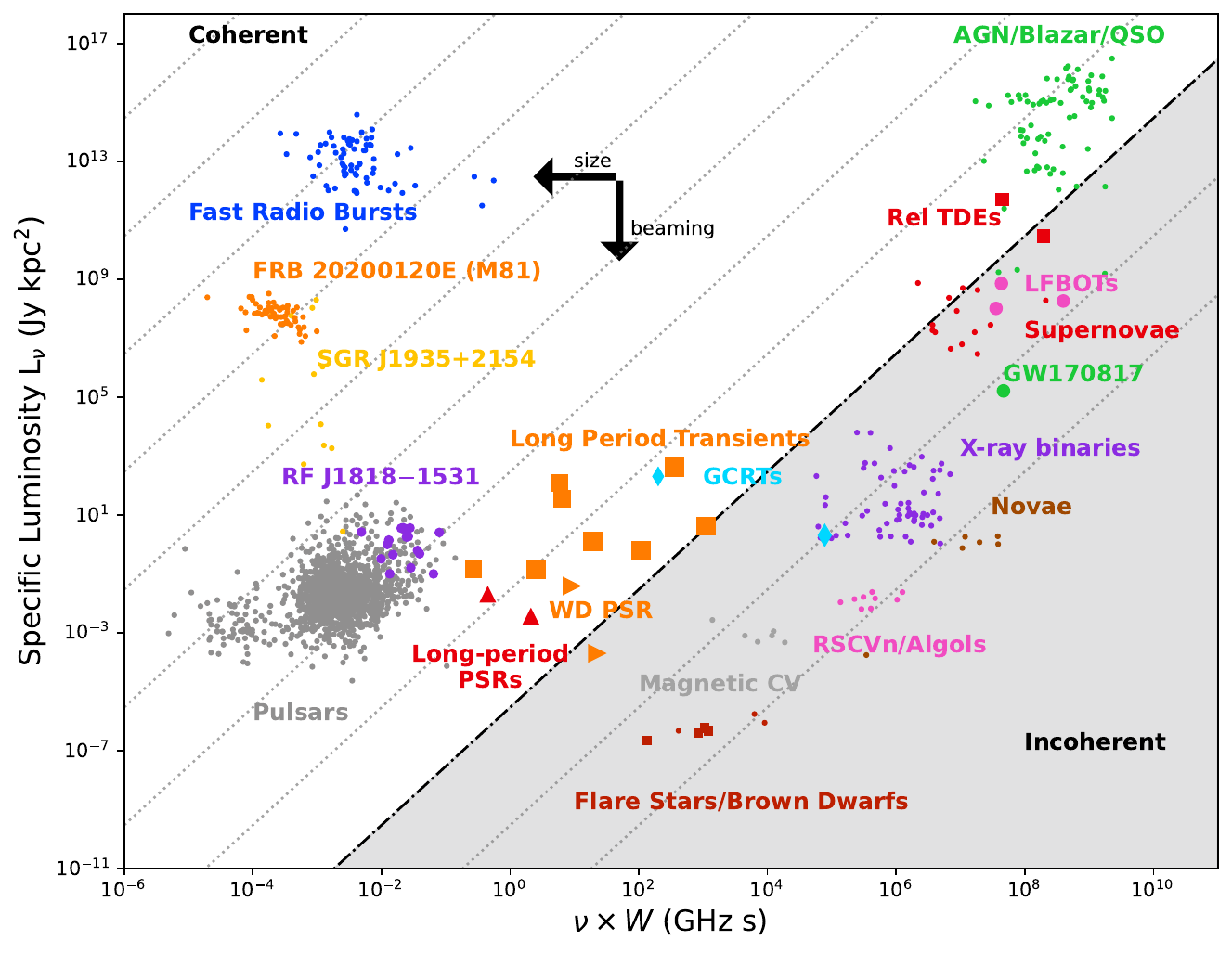}
    \caption{Transient phase space showing radio luminosity versus the product of timescale and observing frequency for different transient source classes, following \citet{cordes_dynamic_2004}.  Note that the luminosity assumes sources are beamed into only $1$\,sr and no relativistic beaming, which may or may not be appropriate for individual objects, while the timescales are just the observed variability timescales and ignore more constraining limits such as the finite sizes of e.g. stellar sources.  For sources with relativistic beaming the true brightness temperature could be significantly lower (e.g. \citealt{readhead_equipartition_1994}), while for some stellar sources the true brightness temperature could be significantly higher, as suggested by the arrows.  The diagonal lines show contours of brightness temperature, with coherent emitters having $T_B>10^{12}\,$K.  Adapted from \citet{pietka_variability_2015} and \citet{nimmo_burst_2022}, with select sources added:
    the binary neutron star merger GW170817 \citep{mooley_strong_2018}; LFBOTs \citep{ho_at2018cow_2019,coppejans_mildly_2020,ho_koala_2020}; relativistic TDEs \citep{mimica_radio_2015,andreoni_very_2022}; flare stars/brown dwarfs \citep{hallinan_periodic_2007,rose_periodic_2023,route_second_2016,zic_askap_2019}; long-period radio pulsars \citep{caleb_discovery_2022,wang_discovery_2025}; Galactic Centre radio transients \citep{hyman_powerful_2005,wang_discovery_2021}; white dwarf pulsars \citep{pelisoli_53-min-period_2023,de_ruiter_sporadic_2025}
    and long-period transients \citep{wang_detection_2025,lee_interpulse_2025,wang_discovery_2021,caleb_emission-state-switching_2024,hurley-walker_29_2024,hurley-walker_long-period_2023,hurley-walker_radio_2022,dong_chime_2025}.  In particular we highlight the range of sources from Table~\ref{t_lpts} that are filling out the centre of this space, straddling the coherent/incoherent divide: long-period radio pulsars are upward-pointing triangles, GCRTs are diamonds, pulsing white dwarf binaries are right-pointing triangles, and long period transients (LPTs) are squares.}
    \label{f_phase}
\end{figure*}

There have been a number of reviews of radio transient surveys, mostly written in anticipation of the advent of the `next-generation’ radio telescopes that would enable large-scale time-domain imaging surveys on the path to the SKAO. The first of these is \citet{cordes_dynamic_2004} who defined a phase space for the exploration of radio transients (we show an updated version of it in Figure~\ref{f_phase}), and proposed a metric for the optimisation of radio transients surveys (we discuss and expand on this in Section~\ref{s_motivation}), as well as discussing some of the key science cases for the SKAO. When \citet{fender_radio_2011} reviewed the field, they considered the time domain radio sky was still relatively unexplored. They highlighted the great potential of the major surveys about to commence on MeerKAT \citep{jonas_meerkat_2016}, the Australian SKA Pathfinder \citep[ASKAP;][]{hotan_australian_2021}, and other SKAO pathfinder and precursor telescopes, with a particular emphasis on the Low-Frequency Array \citep[LOFAR;][]{van_haarlem_lofar_2013}. However, their predictions were based primarily on the transient rates reported by \citet{bower_submillijansky_2007}, most of which were later shown to be spurious \citep[][see Section~\ref{s_archival}]{frail_revised_2012}.

The next review paper was the product of the `{\it Advancing Astrophysics with the Square Kilometre Array}' meeting in Italy in 2014 \citep{braun_advancing_2015}. \citet{fender_transient_2015} discuss the advantages of radio observations for understanding transient objects and present updated limits on radio transient rates from the untargeted transient surveys available at the time; these rates and limits were crude, owing to the lack of robust large-scale samples and reliable real-time identification of transients. Although not formally a review paper, \citet{mooley_caltech-nrao_2016} presents the most comprehensive recent compilation of radio transient surveys and detection rates. The online table\footnote{\url{http://www.tauceti.caltech.edu/kunal/radio-transient-surveys}} has been updated to 2025, and hence serves as a very useful summary of work in this area. We draw on this in our review of the history of radio transient surveys presented in Section~\ref{s_history}. 

Just over twenty years on from \citet{cordes_dynamic_2004} there are several major radio transient surveys well underway, with the capability to detect significant numbers of transient sources across the full range of their stated scientific goals. The SKAO will be operational within $\sim5-8$ years, and planning is underway for other future instruments including the next-generation Very Large Array \citep[ngVLA;][]{murphy_science_2018} and Deep Synoptic Array  \citep[DSA-2000;][]{hallinan_dsa-2000_2019}. Hence, it is timely to do a comprehensive review of this area. We hope it will be useful as a milestone in capturing the state of the field as we enter an era of transformational surveys.

The structure of the paper is as follows. In the rest of this section we outline the scope of the review and define some key terminology. In Section~\ref{s_causes} we discuss the key emission mechanisms that cause radio variability and in Section~\ref{s_classes} we describe each of the main classes of radio transients. In Section~\ref{s_surveys} we discuss the motivation for untargeted surveys, and present a history of radio transient surveys. In Section~\ref{s_detection} we discuss the main detection methods and approaches for radio transients. In Section~\ref{s_characterising} we discuss the metrics used to measure variability, and how we characterise the time variable sky.  In Section~\ref{s_future} we discuss SKAO-era telescopes and how we can incorporate what we have learned from previous work. Finally, in Section~\ref{s_conclusions} we present the future outlook for the field of radio transients.

\subsection{The scope of this review}\label{s_scope}
In this review we will focus on results from interferometers in a synthesis imaging mode (i.e. not single dish or coherent beamforming, although we do discuss those separately), often referred to as `image domain' or `slow' transients{\footnote{The term `slow transients' is widely, but not universally, used in the research community. It is defined in Section~\ref{s_terminology}.} or similar (also see Section~\ref{s_terminology}). Early observations of slow transients using single-dish telescopes \citep[e.g.][]{spangler_short-duration_1974}, made detections that were difficult to verify and distinguish from interference \citep{davis_interferometric_1978}.  Most recent studies use interferometric imaging at frequencies $<30\,$GHz, although  very large single-dish telescopes can offer advantages \citep[e.g.][]{route_5_2013,route_rise_2019}.  At higher frequencies bolometric imaging arrays are the best option for millimetre surveys \citep{whitehorn_millimeter_2016,naess_atacama_2021}.

For both scientific and technical reasons, this review focuses on astronomical phenomena that show variability at radio wavelengths on timescales of seconds to years. We will discuss the physical processes that cause this variability and summarise the main classes of objects that are detected as variable radio sources. We will also cover radio transients surveys, detection methods and population rates. 

To define a clear scope for the paper, we will \textit{not} discuss the following topics:
\begin{itemize}
    \item {\bf Solar, heliospheric, and ionospheric science:}
    The Sun is the brightest radio source in the sky, and was first detected in the early days of radio astronomy \citep{hey_solar_1946}. The proximity of the Sun to Earth means we are able to study its radio variability in great detail. Solar radio astronomy typically uses different observational approaches and techniques from the radio transients community. In fact, almost all continuum imaging observations (whether targeted, or untargeted surveys) deliberately avoid the Sun due to the impact on data quality. The main exceptions to this are recent programs that use interplanetary scintillation (IPS) caused by the solar wind (discovered by \citealt{clarke_thesis_1964}, and reported by \citealt{hewish_interplanetary_1964}), for example \citet{morgan_interplanetary_2018}, to both study the solar wind and to select the most compact extragalactic sources \citep[e.g.][]{chhetri_interplanetary_2018}. 
    Likewise, ionospheric variability is both interesting in its own right \citep[e.g.][]{loi_real-time_2015} and due to its impact on studying radio variability at low frequencies \citep[e.g.][]{loi_quantifying_2015}. However, for both scientific and methodological reasons we will not include solar, heliospheric and ionospheric radio astronomy in this review. For a review of this research area in the context of the SKAO, see \citet{nindos_solar_2019}. 

    \item {\bf Solar System objects:} the planets in our own Solar System are detectable at radio wavelengths. The five magnetised planets (Earth, Jupiter, Saturn, Uranus and Neptune) all show auroral radio emission \citep{zarka_auroral_1998} and emission from their radiation belts \citep{mauk_electron_2010}. The non-magnetised planets show radio emission from thermal heating of the planet by incident solar radiation \citep{kellermann_thermal_1966}. Many radio studies of solar system planets have been done with both ground-based telescopes and spacecraft \citep[e.g.\ the Voyager 1 and 2 missions;][]{warwick_voyager_1979}. 
    Because of their rapid movement across the sky, it is common to detect Solar System planets (in particular Jupiter) in widefield radio imaging surveys \citep[e.g.\ Table 2;][]{lenc_all-sky_2018}. Lessons learned from our solar system apply directly to exoplanetary systems (e.g., \citealt{kao_resolved_2023} reported the first detection of radiation belts around an ultracool dwarf star). However, planets (and the Moon) are typically excluded in the process of searching for Galactic and extragalactic transients and we will not cover them in this review.
    
    \item {\bf Asteroids and meteors:} When an asteroid (or material from an asteroid) enters the Earth’s atmosphere it vaporises, and can be seen as a streak of light across the sky. These large meteors are often called `fireballs', and there are regular monitoring programs to detect them, such as the NASA All-sky Fireball Network\footnote{\url{https://fireballs.ndc.nasa.gov}}, due to their potential impact on orbiting spacecraft \citep[e.g.][]{cooke_status_2012}. Very low frequency (10--88 MHz) observations with the Long Wavelength Array \citep{taylor_first_2012} established a correlation between low frequency transients and fireball trails, suggesting that the trails radiate at low radio frequencies \citep{obenberger_detection_2014,varghese_spatially_2024}. In addition, meteor trails are known to reflect low frequency radio transmissions \citep[e.g.][]{helmboldt_all-sky_2014}. This makes asteroids a foreground source of radio transient bursts that need to be considered at low frequencies. However, since our focus is on Galactic and extragalactic transients we will not discuss them further in this review. 

    \item {\bf Searches for extraterrestrial intelligence:} SETI has been an active area of research in radio astronomy since the early 1960s. Searching for extraterrestrial communications or other technosignatures requires many of the same approaches as searching for astronomical transients, and the history of SETI is interwoven with the history of radio transients. In some cases SETI projects have piggybacked on astronomy observations, or searched existing archival data. In other cases they have involved dedicated programs. Substantial funding for radio telescopes and instrumentation has also come from SETI-motivated philanthropy, for example the Allen Telescope Array \citep[ATA;][]{welch_allen_2009} and Breakthrough Listen \citep{worden_breakthrough_2017}. However, since the focus of this review is variable astronomical phenomena, we will not discuss SETI further. For reviews of SETI research, see \citet{tarter_search_2001} and \citet{ekers_seti_2002}. For a more recent discussion of approaches to detecting technosignatures, see \citet{wright_case_2022}. 

    \item {\bf Radio frequency interference:} RFI is an ever-present challenge for radio transient detection. Avoiding RFI was perhaps the most important factor in determining the scientific quality of sites for the SKAO telescopes \citep{schilizzi_square_2024}. Often appearing as extreme bursts of emission localised in frequency or time, RFI can mimic the astronomical signals we are trying to detect. This has resulted in false detections, such as the so-called `perytons,' which were later identified as RFI from microwaves ovens \citep{petroff_identifying_2015}. At longer timescales, RFI generally has less impact on transient detection as it is averaged out in the process of synthesis imaging. However, it can still have a significant impact on image quality. Reflection of low frequency radio signals off satellites in low Earth orbits can create transient objects in images, and is an increasing problem \citep{prabu_development_2020}. A very recent discovery is that of nanosecond electrostatic discharges from satellites \citep{james_nanosecond-duration_2025}. In this review we will only touch on RFI in the context of its impact on transient detection approaches, in Section~\ref{s_detection}. For an overview of RFI mitigation methods see \citet{fridman_rfi_2001} and \citet{offringa_post-correlation_2010}. \citet{lourenco_rfi_2024} and \citet{sihlangu_rfi_2022} provide recent analyses of the RFI environment on the ASKAP and MeerKAT sites, respectively. 

    \item {\bf Fast radio bursts:} FRBs are astronomical radio flashes with durations of milliseconds \citep{lorimer_bright_2007}. Their physical origin is as-yet unknown, although there are likely to be multiple classes of objects that cause FRBs \citep{pleunis_fast_2021}. While they are extremely interesting, and may be related to other types of radio transients \citep[e.g.,][]{bochenek_fast_2020,law_fast_2022}, fast radio bursts themselves are beyond the scope of this review.  This is partly for technical reasons  (FRBs are most often discovered with different instrumentation and survey methods than the slower transients below, for instance using single dish telescopes or interferometric voltage beams) and partly for practical reasons (the huge amount of 
    new science on FRBs and rapid developments necessitate their own reviews). For these reasons we point the reader to recent reviews such as \citet{cordes_fast_2019}, \citet{petroff_fast_2022}, or \citet{zhang_physics_2023}. 
\end{itemize}

A final note: in the sections discussing survey strategy and the history of radio transients surveys, our focus will be on widefield imaging surveys. Hence we will not discuss, for example, large targeted monitoring programs such as those for interstellar scintillation \citep{lovell_first_2003}, follow-up of sources detected in other wavebands, for example gamma-ray bursts \citep[GRBs;][]{frail_complete_2003} and large triggered follow-up programs \citep[e.g.][]{staley_ami_2013}. When discussing what we know about different variable source classes, we will of course incorporate all sources of information.

\subsection{Some notes on terminology}\label{s_terminology}
As the field of radio transients has evolved over the past few decades, there have been several important terms that are either (a) specific to this research area; (b) have been used in different ways in the community; or (c) have seen their usage evolve over this period. To avoid confusion we discuss and define them here. 
\begin{itemize}
    \item {\bf ‘Fast’ and ‘slow’ transients:} these terms are generally used to mean variability on timescales faster than and slower than the telescope correlator integration time. Fast transient searches need to use data at a pre-correlator stage, and for most radio telescopes this is of the order of seconds. For example, the ASKAP integration time is 10~seconds \citep{hotan_australian_2021} and MeerKAT is 2--8 seconds \citep{jonas_meerkat_2016}. For the VLA the default integration times for each configuration are between 2--5 seconds\footnote{\url{https://science.nrao.edu/facilities/vla/docs/manuals/oss/performance/tim-re}}. Due to the completely different data processing and analysis approaches (for example dedispersion is usually required in fast transient searches) it makes sense to consider these two domains separately. As a result, most of the major survey projects on SKAO pathfinder and precursor telescopes have split along these lines. For example, on ASKAP, the Commensal Real-time ASKAP Fast Transients Survey \citep[CRAFT;][]{macquart_commensal_2010} focuses on fast transients (also see \citealt{wang_craft_2025}) and the Variables and Slow Transients \citep[VAST;][]{murphy_vast_2013} survey focuses on slow transients. Likewise on MeerKAT, the Transients and Pulsars with MeerKAT \citep[TRAPUM;][]{stappers_update_2016} project focused on fast transients and the ThunderKAT project \citep{fender_thunderkat_2016} focused on slow transients. Conveniently, the behaviour of most astrophysical phenomena also fall into one of these two regimes (see Figure~\ref{f_phase}), although some such as pulsars can show variability on both millisecond timescales and much slower timescales of minutes to hours, and new discoveries (e.g. Section~\ref{s_lpts}) are further blurring the lines. This review will focus on slow transients.

    \item {\bf Transients and variables:} as the era of large-scale imaging surveys started, there was considerable debate about how the terms `transient' versus `variable' should be used. When considering the underlying astronomical objects themselves, it is possible to make a distinction between persistent objects that show variable behaviour (e.g. flaring stars) and explosive events that appear, then fade and ultimately disappear (e.g. gamma-ray bursts). However, observationally this distinction is less clear, since the time morphology of a particular dynamic event is highly dependent on the sensitivity limit and time sampling of the observations. For example, radio bursts from a star that has quiescent emission below the sensitivity limit of the image will appear as transient events. Hence in this review, as in much of the literature, we use the terms transient and highly variable, relatively interchangeably \citep[see, for example, comments in Section~5 of][]{mooley_caltech-nrao_2016}. 

    \item {\bf Commensal and piggyback observations:} The term `piggyback' refers to the situation in which observations that are taken for a particular scientific purpose can be used for a different purpose, without affecting the original observing schedule and specifications. In radio astronomy, this strategy has been extensively used by SETI; for example the SERENDIP project in which data collected by all science projects on Arecibo was analysed independently for narrowband radio signals \citep{werthimer_berkeley_2001}. The word `commensal' has its origins in biology. The Oxford Concise Medical Dictionary defines it as: `{\it an organism that lives in close association with another of a different species without either harming or benefiting it}'. It was introduced into the radio astronomy literature in the context of the plans for the Allen Telescope Array, to imply a more active form of piggybacking in which both parties had some influence over the observational strategy \citep{deboer_allen_2006}. The usage of commensal became more widespread in the literature after 2010, in the context of a number of new radio surveys being designed \citep[e.g.][]{macquart_commensal_2010, wayth_v-fastr_2011}. In the current literature commensal and piggyback observations are often used interchangeably. Note that readers outside this area of research can be confused by this technical meaning. 

    \item {\bf `Blind' and untargeted surveys:} The terms blind and untargeted are used interchangeably in the literature, to refer to a survey that has been designed without targeting specific objects, and with as few prior assumptions as possible. We prefer the term {\it untargeted survey} as it is more accurate and more inclusive. 

\end{itemize}

\section{Radio emission mechanisms and causes  of variability}\label{s_causes}

Astronomical objects that emit radio waves (from nearby planets and stars, to distant galaxies and active galactic nuclei; e.g. \citealt{kellermann_galactic_1988})  have a wide variety of emission types. In this section we focus on those types that produce transient and variable radio emission, either intrinsically, or as a result of propagation through an inhomogeneous ionised medium or other external causes.  In general the emission will come from plasma where the particles (primarily electrons) are in local thermodynamic equilibrium \citep{condon_essential_2016}, leading to \textit{thermal} emission, or are not in equilibrium, leading to  \textit{non-thermal} emission (note, though, that some authors restrict thermal emission to only blackbody or thermal bremsstrahlung radiation).  In most cases the sources of radio emission considered here arise from non-thermal plasmas.  

For all of these objects we can consider the \textit{brightness temperature} of the emission \citep{rybicki_radiative_1985,condon_essential_2016}, which is the temperature of a hypothetical optically-thick thermal plasma that would emit at the same intensity.  Note that this definition can be used even when the assumptions (like thermal emission) are not valid, as it is still a useful quantity for comparison.  The brightness temperature is defined as:
\begin{equation}
    T_B(\nu) \equiv \frac{I_\nu c^2}{2 k_B \nu^2}
    \label{eqn:TB}
\end{equation}
where $I_\nu$ is the specific intensity (measured in ${\rm erg\,s}^{-1}\,{\rm cm}^{\red{-2}}\,{\rm Hz}^{-1}\,{\rm sr}^{-1}$ or equivalent), the frequency is $\nu$, $k_B$ is Boltzmann's constant, and $c$ is the speed of light. For  a source with flux density $S_\nu$ and solid angle $\Omega$, we have $I_\nu=S_\nu/\Omega$.  Note that this form assumes radiation is in the Rayleigh-Jeans limit, with $h\nu \ll k_B T$, even if that is not actually true.  For non-thermal sources the brightness temperature, $T_B$, is different from the actual electron temperature $T_e$, and typically $T_B \gg T_e$.  

We should also distinguish between \textit{incoherent} and \textit{coherent} emission \citep{melrose_coherent_2017}.  Most radiation sources are incoherent, such that each electron radiates independently and the total emission is found by summing over a distribution of electrons.  For an incoherent source, $T_B$ is limited by two mechanisms: $k_B T_B$ should be less than the energy of the emitting electrons, and $T_B$ should be $< 10^{12}\,$K because above this point radiation by inverse Compton scattering will increase significantly, leading to rapid energy loss and cooling back down to $10^{12}\,$K \citep{kellermann_spectra_1969}.  Note that the upper limit on $T_b$ could be even lower if set by equipartion between magnetic and particle energies \citep{readhead_equipartition_1994}.  In general it can be difficult to directly measure or constrain $T_B$. However, sources that vary on a timescale $\tau$, have their angular size limited to $\theta < c\tau/d$  at a distance $d$, which limits the solid angle  to $\Omega < (c\tau/d)^2$.  This then implies for the brightness temperature \citep[e.g.][]{readhead_equipartition_1994,cordes_dynamic_2004,miller-jones_zooming_2008,bell_murchison_2019}:
\begin{equation}
    T_B \gtrsim \frac{S_\nu d^2}{2 k_B \nu^2 \tau^2}
\end{equation}
with flux density $S_\nu$, ignoring relativistic beaming or cosmological effects.  While the latter can be easily estimated for a source with a known redshift, relativistic beaming is an important contributor to sources having apparent brightness temperatures much higher than their intrinsic values \citep{readhead_equipartition_1994}: depending on the speed of the relativistic motion and the geometry, the luminosity can  be boosted by the Doppler factor to some power (depending on the intrinsic spectrum) which can be 10 or more, with the inferred brightness temperature from variability boosted by the Doppler factor cubed since the timescale is also modified \citep{readhead_equipartition_1994,lahteenmaki_total_1999}, so the overall boost can exceed a factor of 1000.

When beaming is not a factor, or when corrections have been done, the intrinsic (excluding variability caused by propagation, discussed in Section~\ref{s_scint}) brightness temperature can then be estimated.  For brighter sources (large $S_\nu$) and faster variability (small $\tau$), sources can have $T_B >10^{12}\,$K, sometimes by many orders of magnitude, requiring an alternate  mechanism for the emission \citep{cordes_dynamic_2004}.  These coherent mechanisms, where the contributions of individual particles can add up in-phase, relate to plasma instabilities or `collective plasma radiation processes' \citep{melrose_collective_1991,melrose_coherent_2017}, where the emission from $N$ particles is $\propto N^2$ rather than $\propto N$ for incoherent emission.  A full description of these mechanisms is beyond the scope of this review, and we point the reader to helpful reviews such as \citet{melrose_coherent_2017}. In this review we will discuss each briefly and associate the mechanisms with the sources where they operate.

\subsection{Synchrotron emission}
\label{s_synch}
The majority of the extragalactic radio sources that are bright enough for us to observe, are either active galactic nuclei (AGN, both `radio-loud' and `radio-quiet'; \citealt{kellermann_vla_1989}) or star-forming galaxies, with the former dominating at higher flux densities \citep{condon_radio_2002,matthews_cosmic_2021}. 
For AGN the emission is primarily  via \textit{synchrotron radiation} \citep{rybicki_radiative_1985}, a (typically) non-thermal, incoherent process,  from a combination of their cores/nuclei and jets \citep{fanaroff_morphology_1974} produced by accretion onto central supermassive black holes \citep{blandford_relativistic_2019}.  For star-forming galaxies the emission is a combination of synchrotron emission from supernova remnants (SNRs) and diffuse gas as well as thermal emission from {H\sc{\,II}} regions \citep{helou_thermal_1985,chevalier_radio_1982}. Since  sources like SNRs and {H\sc{\,II}} regions are usually too extended to show variability we will not consider them further, although in some cases young SNRs can show secular evolution (e.g. \citealt{sukumar_radio_1989}), and their study can connect back to  supernovae \citep{milisavljevic_supernova_2017}.

Synchrotron emission is the result of relativistic electrons spiralling around magnetic field lines \citep{rybicki_radiative_1985,condon_essential_2016}.  The emission from a single electron depends on its energy, the magnetic field strength, and the angle between the electron's motion and the magnetic field, with a spectrum peaking near the critical frequency which is the gyrofrequency (depending on the magnetic field strength) modified by the electron's Lorentz factor.  But for a power-law distribution of electron energies, the summed emission results in a power-law with spectral index $\alpha$, defined as:
\begin{equation}
    S_\nu \propto \nu^\alpha
\end{equation}
depending on the underlying distribution of the electrons (note that some authors define the spectral index with the opposite sign, $S_\nu \propto \nu^{-\alpha}$, and care must be taken to correctly interpret values).  

Most synchrotron sources
have a steep spectrum ($\alpha<0$) with $\alpha\approx -0.7$ \citep[][and indeed most non-thermal sources have negative spectral indices in general]{sabater_lotss_2019,franzen_gleam_2021}.  At lower frequencies, the brightness temperature approaches the effective temperature of the relativistic electrons and this emission is modified by synchrotron self-absorption, which causes the steep spectrum to turn over into one with $\alpha \approx 2.5$.  

Synchrotron radiation is ubiquitous in radio astronomy, and most sources are steady emitters. However, there are two physical scenarios that can lead to transient or variable emission: changes in accretion rate and jet launching in accreting compact objects on different mass scales; and explosions interacting with the surrounding medium.

\subsubsection{Accretion and jet ejection}
\label{s_jets}
The process of disk accretion and jet ejection is ubiquitous across astrophysics \citep[e.g.][]{degouveia_astrophysical_2005}, and where compact objects are involved those jets are typically relativistic \citep[e.g.][]{bromberg_propagation_2011} with similar characteristics that depend primarily on the accretor mass, accretion rate, and ejection velocity.  In the context of radio transients, we focus on relativistic jets from AGN and X-ray binaries (XRBs), which we discuss below.  

AGN dominate the radio sky at flux densities above a few millijanskys (at centimetre wavelengths), 
and most show little variability \citep{hovatta_statistical_2007,hodge_millijansky_2013,stewart_lofar_2016,mooley_caltech-nrao_2016,bell_murchison_2019,
murphy_askap_2021}, although fluctuations of a few to few tens of percent are not uncommon \citep[e.g.][]{falcke_radio_2001,mundell_radio_2009} and variations are both more significant and faster at higher radio frequencies
\citep[e.g.][]{ackermann_radiogamma-ray_2011,hovatta_statistical_2007}, especially for sub-classes of AGN such as blazars \citep{max-moerbeck_ovro_2012,richards_blazars_2011,max-moerbeck_time_2014}.  

Aside from extrinsic causes (see Section ~\ref{s_scint}) most of these changes are thought to be due to modest changes in the jets \citep[e.g.][]{nyland_quasars_2020}, such as the propagation of shocks  along the jets \citep{marscher_models_1985,hughes_synchrotron_1989}, disk instabilities \citep{czerny_accretion_2009,janiuk_different_2011}, or changes in accretion power \citep{koay_parsec-scale_2016,wolowska_changing-look_2017}.  These effects can be magnified by the jet orientation, with jets directed close to the line of sight having larger variability on shorter timescales due to relativistic effects \citep{lister_relativistic_2001}: changes in orientation can therefore dominate any intrinsic changes in their effects on variability.

However, a subset of these sources show higher variability \citep[e.g.][]{barvainis_radio_2005}.  This may be due to large-scale changes in the jet orientation from accretion instabilities or interaction within a binary super massive black hole \citep{palenzuela_dual_2010,an_periodic_2013}, or large-scale changes in the jet power from changes in the accretion flow \citep{mooley_caltech-nrao_2016,nyland_quasars_2020}.  These sources apparently transition from radio-quiet (often undetectable) to radio-loud \citep[although the distinction may not be so simple, e.g.][]{kellermann_radio-loud_2016}, with the radio luminosity apparently increasing by more than an order of magnitude over several decades, potentially as a result of  newly-launched jets.

Such changes in the jet behaviour can be replicated on much smaller scales (both physical and mass) through accreting X-ray binaries hosting neutron stars or black holes \citep{fender_towards_2004,fender_transient_2006,fender_disc-jet-wind_2016}.  Here, small-scale changes in the jet power during the hard state correspond to normal AGN variability, while significant launching of superluminal jet components in the transition between hard and soft states  correspond to the radio-quiet to radio-loud transitions of AGN.

\subsubsection{Explosions and shocks}
\label{s_afterglow}
In contrast to the scenarios above, where accretion leads to jets that can vary as the accretion rate or direction varies, the radio emission from transient events such as classical novae \citep{chomiuk_new_2021,chomiuk_classical_2021}, supernovae \citep{weiler_radio_2002}, long and short gamma-ray bursts \citep{piran_physics_2004,berger_short-duration_2014}, magnetars \citep{frail_outburst_1999}, neutron star mergers \citep{hallinan_radio_2017}, tidal disruption events \citep{levan_extremely_2011,bloom_possible_2011} and similar phenomena lead more naturally to variable radio emission.  These sorts of explosions/ejections have been among the most anticipated targets for large-area radio surveys \citep[e.g.][]{metzger_extragalactic_2015}.  Note that even though some of these sources, like gamma-ray bursts, may be powered by very short-lived relativistic jets, we discuss them separately from Section~\ref{s_jets} as the jets themselves are not seen at radio wavelengths, but only their impact on the surrounding medium.

The basic scenario here is where an explosion leads to a rapid ejection of material.  This outflow can be Newtonian or relativistic, and it can be collimated or spherical.  Eventually the ejecta will impact the surrounding medium (typically circumstellar or interstellar material) leading to shock waves that amplify any magnetic fields present and accelerate electrons to relativistic energies, leading to synchrotron emission \citep{chevalier_radio_1982}.  This so-called `afterglow' model is present in many variants in many different physical scenarios, where the mass, velocity, and angular profile of the ejecta and the radial profile of the surrounding medium can all change \citep{sari_spectra_1998,chevalier_wind_2000,chandra_radio-selected_2012,mooley_mildly_2018,mooley_strong_2018,troja_outflow_2018,lazzati_late_2018}.  

Normally, for very energetic events like gamma-ray bursts the ejecta are observed to be ultrarelativsistic and jetted \citep{rhoads_how_1997,meszaros_optical_1997}, such that the initial afterglow emission is beamed into a solid angle of $\sim 1/\Gamma^2$ (where $\Gamma$ is the bulk Lorentz factor) centred on the jet axis, which is also where the $\gamma$-rays themselves are visible.  At later times the Lorentz factor will decrease so the solid angle will increase, leading to increasing visibility of the late-time afterglow emission. Eventually, when $1/\Gamma$ approaches the intrinsic jet size the entire jet will be visible. This tends to happen at the same time as the jet material will begin to expand sideways  \citep{sari_jets_1999}.  Together these lead to an achromatic `jet break' in the lightcurve \citep{sari_jets_1999,rhoads_dynamics_1999}. Measurement of jet breaks can then be used to infer the jet size and hence intrinsic energetics of collimated explosions \citep{frail_beaming_2001}.

This also gives rise to the concept of an `orphan afterglow', where only the late-time radio emission is visible from our line of sight \citep{rhoads_dirty_2003,ghirlanda_grb_2014}.  Such an event would have a steeper and shorter rise than an on-axis GRB, but merge onto the same power-law in the decline (see Figure~\ref{f_grb}).

The general behaviour for all of these afterglows is a series of power-law increases and declines \citep{sari_spectra_1998}, where the increase reflects the expanding shock front as well as the changing level of self-absorption.  This increase happens first at higher frequencies because the self-absorption is less significant there.  Eventually the entire afterglow becomes optically thin, at which point it transitions into a decline phase as the ejecta expand and slow down.  
During this phase a standard non-thermal steep spectrum is present \citep{weiler_radio_2002}.  This is illustrated in Figure~\ref{f_grb}, which shows lightcurves with and without the effect of self-absorption. If absorption is not an issue, the peak on-axis flux density  is expected to occur when the shock has swept up a mass comparable to its own material and starts to decelerate, and it will scale with the energy of the explosion \citep{nakar_detectable_2011,metzger_extragalactic_2015}. The time until the peak will scale with the energy of the explosion to the $\frac{1}{3}$ power (neglecting cosmological effects).

We can also see in Figure~\ref{f_grb} the lightcurve from a potential orphan afterglow.  One clear difficulty with all of these lightcurves is that they have the same late-time behaviour, so without a lucky early detection or robust limits on prompt high-energy emission (complicated when the explosion time is also weakly constrained), many different phenomena can be fit with the same data.

\begin{figure}
    \centering
    \includegraphics[width=1.0\linewidth]{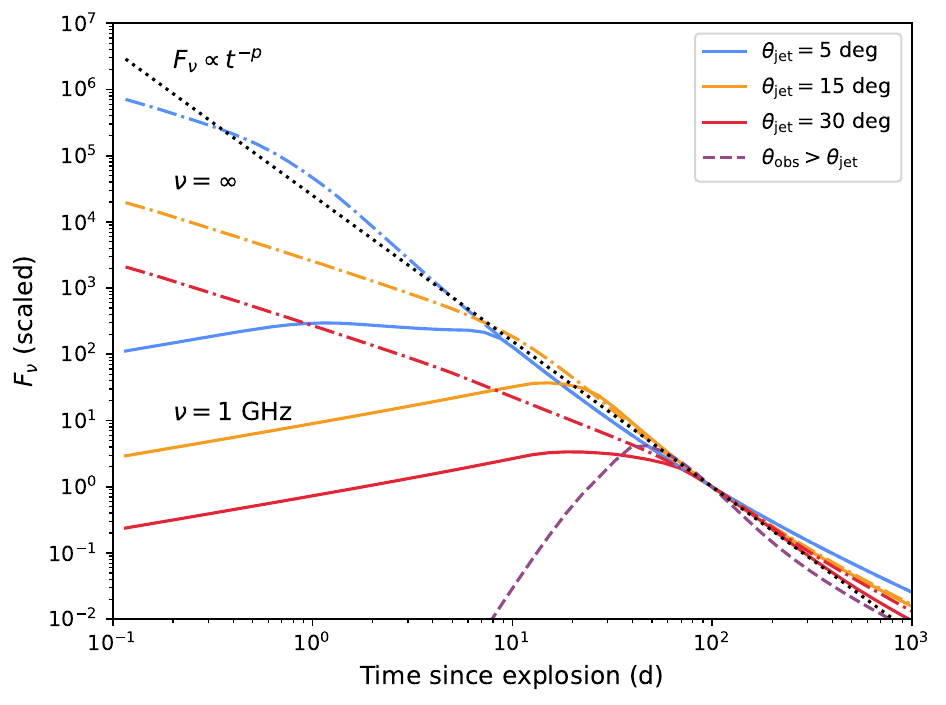}
    \caption{Model GRB lightcurves, computed using \citet{ryan_gamma-ray_2020} and inspired by \citet{piran_physics_2004}.  These models use the simplest possible model for a relativistic jet.  The top dash-dotted curves are for infinite frequency (ignoring any effects of self-absorption), while the lower solid curves are for a frequency of 1\,GHz.  All curves are normalised at 10\,d.  For both frequencies, we show jet opening angles of $\theta_{\rm jet}=5\deg$ (blue), $15\deg$ (orange), and $30\deg$ (green) with an on-axis observer ($\theta_{\rm obs}=0$).  The infinite frequency models show jet breaks when the jet has decelerated to a bulk Lorentz factor of $1/\theta_{\rm jet}$, after which they have a similar power-law behaviour with $F_\nu \propto \nu^{-p}$ (black dotted line), where $p$ is the power-law index of the electron distribution.  We also show a jet seen by an  off-axis observer, potentially an `orphan afterglow', since the observer would miss any high-energy emission (red dashed curve).  This has similar late-time behaviour but is much fainter at early times.}
    \label{f_grb}
\end{figure}

\subsection{Free-free emission}
\label{s_freefree}
A small fraction of variable sources emit primarily via free-free (or bremsstrahlung) emission \citep{condon_essential_2016,osterbrock_astrophysics_2006}, typically in a thermal plasma.  This is produced by electrons accelerated electrostatically around ions.  At higher frequencies the emission typically has $\alpha\approx -0.1$ \citep{condon_essential_2016,chomiuk_classical_2021}, which then turns over into an opaque source with $\alpha \approx 2$ once it becomes self-absorbed at lower frequencies.

The primary sources of variable free-free emission are classical novae \citep[CNe;][]{chomiuk_classical_2021,chomiuk_new_2021}, where a thermonuclear explosion on an accreting white dwarf has expelled material into the interstellar medium.  These are frequently detected through optical and infrared time-domain surveys\footnote{\url{https://asd.gsfc.nasa.gov/Koji.Mukai/novae/novae.html}.} although their emission spans the electromagnetic spectrum.  At radio wavelengths, the traditional emission  from CNe is free-free emission from a slowly expanding sphere of ionised ejecta \citep{hjellming_radio_1979,seaquist_thick_1977}, which evolves over timescales of weeks--years as the emission first increases  due to decreasing self-absorption and then declines due to decreasing density (since emission is proportional to the square of the electron density).  While more complex sources are now observed to have non-thermal synchrotron emission from interactions between the ejecta and the surrounding medium \citep[][and see Section~\ref{s_afterglow}]{krauss_expanded_2011,weston_non-thermal_2016,weston_shock-powered_2016,finzell_detailed_2018}, even the thermal emission itself can be more complex both spatially and temporally \citep{nelson_2011_2014,chomiuk_2011_2014,chomiuk_classical_2021}.  

\subsection{Stellar emission: gyrosynchrotron, plasma, and electron cyclotron maser emission}
\label{s_stellarmech}
In a number of stellar systems, especially for cooler, later-type stars we can detect \textbf{gyrosynchrotron emission,} which is an incoherent, (usually) non-thermal process related to synchrotron emission  \citep{dulk_radio_1985,gudel_stellar_2002}.  Here, however, the electrons are only mildly relativistic.  This leads to broad-band emission that has significant (but not 100\%) circular polarisation at higher frequencies, where it is optically thin.  At lower frequencies we see a typical optically-thick spectral index of $\alpha=+2.5$ and a steep  spectrum on the higher frequency side, with the power-law index depending on the underlying distribution of electron energies.

In contrast, \textbf{plasma emission} is a \textit{coherent} process \citep{melrose_coherent_2017} observed in stellar flares 
\citep{gudel_stellar_2002,bastian_radio_1990}, converting plasma turbulence into radiation via non-linear plasma processes.  It can account for brightness temperatures up to $10^{18}\,$K, with emission at the fundamental or harmonic of the plasma frequency (determined by the local electron density), typically at lower radio frequencies ($\lesssim 1\,$GHz): at higher frequencies the emission suffers from free-free absorption.

Finally, \textbf{electron cyclotron maser emission} (ECME) is also a coherent process observed in stellar flares \citep{dulk_radio_1985,gudel_stellar_2002}.  Instead of the plasma frequency, the emission is at the fundamental or (low) harmonic of the cyclotron frequency (determined by the local magnetic field).  ECME dominates over plasma emission in diffuse highly-magnetised plasma such that the plasma frequency is much less than the cyclotron frequency \citep{treumann_electron-cyclotron_2006}, so measurement of frequency structure or cutoffs can be used to constrain the magnetic field strength.  It can result in brightness temperatures up to $10^{20}\,$K and circular polarisation of up to 100\%.  The emission can also be elliptically polarised, with a combination of linear and circular polarisation.  This is typically only seen in pulsars \citep{melrose_coherent_2017} and planets \citep{zarka_auroral_1998}, although it has been seen in a small number of late-type stars \citep{spangler_four-stokes_1974,lynch_154_2017,villadsen_ultra-wideband_2019}.

\subsection{Pulsar-related variability}
\label{s_pulsarcause}
Pulsar radio emission as seen from Earth is inherently variable, with regular modulation on the timescale of the rotational period (although see \citealt{basu_detection_2011}).  Despite decades of study the pulsar emission mechanism still eludes detailed understanding \citep{melrose_coherent_2017,lorimer_handbook_2012,philippov_pulsar_2022}, but it can achieve extremely high brightness temperatures of above $10^{20}\,$K.  Typically the emission has a steep spectral index ($\alpha\approx -1.5$; \citealt{bates_pulsar_2013,jankowski_spectral_2018,anumarlapudi_characterizing_2023,karastergiou_thousand-pulsar-array_2024}), although some highly-magnetised pulsars known as magnetars can have much flatter spectra \citep{camilo_transient_2006}.  It can be highly polarised --- both linear and circular --- with rotations of the linear position angle and sign changes of the circular polarisation occurring during the pulse.

However, these properties do not explain why pulsars (especially millisecond pulsars) can be identified in traditional image domain transient and variable searches. The rotational modulation, on timescales of 1\,ms to 10's of seconds, is typically far too fast to appear as a variable source for an imaging survey (although this is now changing with short timescale imaging --- see Sections~\ref{s_lpts} and \ref{s_fast}). In addition to extrinsic modulations that can be very significant for pulsars (Section~\ref{s_scint}), there are a number of intrinsic mechanisms that can lead to variability on longer timescales.  These range from `nulling' \citep{backer_pulsar_1970}, where the pulsar turns off for tens to hundreds of pulses, to intermittency \citep{kramer_periodically_2006}, where the pulsar turns off for days to months, to eclipses where the pulses or even the continuum emission disappear for part of a binary orbit (timescales of minutes to hours; \citealt{fruchter_millisecond_1988,broderick_low-radio-frequency_2016,polzin_study_2020}).  It is also possible that rather than distinct variable sub-classes, there may be a continuum of variability across the population \citep{lower_ubiquity_2025,keith_thousand-pulsar-array_2024}  with ties between the rotational, pulse profile, and flux density variations that can be seen broadly with sufficient precision.  Like the underlying emission mechanism, the theoretical basis for these changes is also not well understood.

Finally, for magnetars \citep{kaspi_magnetars_2017}, presumed reconfiguration of their magnetic fields during bursts or giant flares can lead to the sudden appearance of pulsed radio emission \citep{camilo_transient_2006}, in addition to the ejection of relativistic plasma leading to synchrotron afterglows discussed in Section~\ref{s_afterglow}.  This pulsed emission can have different spectral properties to typical pulsars, and fades on timescales of months.

\subsection{Extrinsic variability}\label{s_extrinsic}
\subsubsection{Diffractive and refractive  scintillation}
\label{s_scint}
When the radio waves from a source propagate through an inhomogeneous ionised medium, the waves can bend and diffract to form spatial variations in the wavefront.  If the observer is moving relative to the wavefront, this can result in temporal variability known as {\it scintillation} \citep{rickett_radio_1990,narayan_physics_1992}.  Such scintillation has been observed coming from the heliosphere and interplanetary medium (IPS; \citealt{hewish_interplanetary_1964}) as well as  the interstellar medium (interstellar scintillation or ISS; \citealt{scheuer_amplitude_1968,rickett_frequency_1969}).  Interstellar scintillation is only detectable for the most compact radio sources, such as compact cores of AGN \citep{lovell_first_2003}, distance relativistic explosions  \citep{goodman_radio_1997,frail_radio_1997}, or pulsars and fast radio bursts. Moreover, scintillation can have a range of properties depending on the source size and observing frequency.  

Relevant for scintillation is the Fresnel scale, $r_{\rm F} \equiv \sqrt{\frac{\lambda d}{2\pi}}$, for a source at distance $d$ observed at wavelength $\lambda$.  Also relevant is the diffraction scale $r_{\rm d}$ over which the phase variance is 1\,rad.  The ratio of these then defines the scintillation strength, $ r_{\rm F}/r_{\rm d}$.
At higher frequencies and for closer sources the scintillation is typically in the `weak' regime ($r_{\rm F}/r_{\rm d} \ll 1$), with small ($\ll 1$) rms phase perturbations on the Fresnel scale \citep{rickett_radio_1990,cordes_interstellar_1991,narayan_physics_1992,hancock_refractive_2019}, leading to fractional variability $\ll 1$.  

More interesting is the `strong' regime ($r_{\rm F}/r_{\rm d} \gg 1$), where the phase fluctuations over the Fresnel scale are large.  In this regime  we  can observe both diffractive \citep{scheuer_amplitude_1968} and refractive \citep{sieber_causal_1982,rickett_slow_1984}  effects.

Diffractive scintillation results from multipath propagation \citep{goodman_effects_1987} from many independent patches which can add constructively or destructively, leading to large intensity fluctuations in a frequency-dependent manner.  The patches each scatter radiation into an angle $\theta_{\rm r}=r_{\rm r}/d$, with the refractive scale $r_{\rm r}\equiv r_{\rm F}^2/r_{\rm d}$.  
This phenomenon requires very small sources, with angular sizes $\theta_{\rm d}\equiv r_{\rm d}/d \ll \theta_{\rm r}$, and so is usually limited to pulsars,   although compact relativistic explosions like GRBs can also scintillate diffractively in their early phases \citep{goodman_radio_1997}. For diffractive scintillation, the timescale and bandwidth  both increase with observing frequency, and the modulation can saturate at $\sim 100$\% although it can also appear to be lower when averaging over multiple `scintles' (time-frequency maxima).

Refractive scintillation, on the other hand, is due to large scale focusing and defocusing over scales $r_{\rm r}$, and where the sources need to be compact compared to $\theta_{\rm r}$.  Refractive scintillation corresponds to lower modulation over longer timescales and wider bandwidths.  

\subsubsection{Extreme scattering events}
Beyond the stochastic variations caused by the turbulent interstellar medium discussed above, systematic monitoring of compact extragalactic radio sources discovered discrete high-amplitude changes in source brightness with a characteristic shape, called `extreme scattering events' (ESEs; \citealt{fiedler_extreme_1987}).  Subsequently also seen in Galactic pulsars \citep{cognard_extreme_1993}, ESEs are believed to be due to coherent lensing structures within the interstellar medium (ISM)  \citep{fiedler_summary_1994}.  However, significant questions remain regarding the natures of those structures that seem much denser and with higher pressure than the average ISM \citep{clegg_gaussian_1998}.  Whether the structures are one dimensional (filaments), two dimensional (sheets), or three dimensional is still debated 
\citep{romani_radio_1987,pen_refractive_2012,henriksen_hydrogen_1995,walker_extreme_1998}, as the models need to reproduce the temporal and frequency structure of ESEs \citep{walker_extreme_2007,vedantham_extreme_2018}.  

There is considerable work ongoing to improve real-time detection of ESEs \citep{bannister_real-time_2016}, which would lead to better observations of their environments.  At the same time there are studies of
the models of the structures that cause ESEs \citep{jow_cusp_2024,dong_extreme_2018,rogers_dual-component_2019}, as well as attempts to identify the underlying causes of the ISM inhomogeneities \citep{walker_extreme_2017}.  

\subsubsection{Gravitational lensing}
Separate from variability induced by scintillation, extragalactic radio sources may exhibit variability induced by gravitational lensing along the line of sight.  This has been predicted to appear in several ways.  Radio sources that already showed multiple lensed images could have substructure lensed by small compact objects (typical masses $\sim M_\odot$) within the lensing galaxy \citep{gopal-krishna_gravitational_1991,koopmans_microlensing_2000,biggs_vla_2023}, showing correlated but not identical variations between the separate images on timescales of days.
However, this can be hard to distinguish from scintillation \citep{koopmans_extrinsic_2003,biggs_time_2021} --- indeed there have been many reports of microlensing in the radio that have been difficult to verify \citep{koopmans_microlensing_2000,vernardos_microlensing_2024}.  While gravitational lensing should be achromatic, and hence show different behaviour from scintillation, the varying size of the radio emission region --- assumed to be a knot of synchrotron-emitting material within a Doppler-boosted relativistic jet ---  with frequency can cause apparent frequency dependence.  

A related phenomenon could be seen with larger lensing masses.  In this case \citep{vedantham_symmetric_2017,peirson_new_2022} the lens masses would be $10^{3-6}\,M_\odot$, causing achromatic variability on timescales of months--years in the lightcurve of a single source (the lensed images are not separable).  This would be similar to lensing observed from high-energy blazars in $\gamma$-rays \citep[e.g.][]{barnacka_first_2011}, although the details of the lensed sources are likely different \citep{spingola_radio_2016}.

\section{Classes of radio transients}\label{s_classes}
Now we have presented the main physical mechanisms that cause radio variability, in this section we discuss each of the main classes of radio transient and variable objects. Our focus is on intrinsic variability (propagation effects were covered in Section~\ref{s_extrinsic}) and what we can learn from radio observations. We also make the case for why radio transient surveys are important, in addition to targeted observations of individual objects. Since there is a limit to the amount of detail we can go into in this paper, where possible we point the reader to relevant review papers.  
The following subsections are roughly ordered from highly energetic extragalactic objects, through to stars, binaries with compact objects, and Galactic transients. We summarise the different variability timescales for the different classes of transients in Figure~\ref{f_time}.

\begin{figure*}
    \centering
    \includegraphics[width=1.0\textwidth]{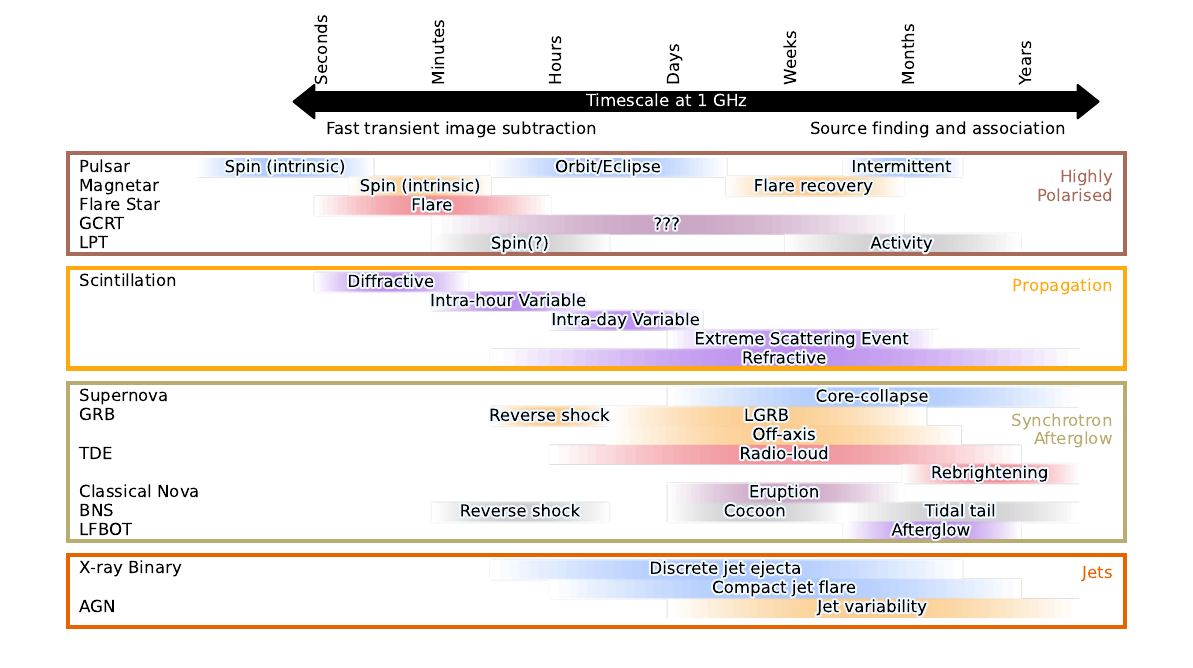}
    \caption{Plot showing the relevant timescales of different classes of radio transients.  Approximate limits of variability timescales are shown for different sources and different mechanisms.  We also separate the highly-polarised largely coherent transients in the top  box from the synchrotron afterglow (Section~\ref{s_afterglow}) in the bottom two boxes.  We  roughly delineate the timescales for traditional transient searches that find sources individually in each epoch and associate them across epochs ($\gtrsim\,$hours, e.g. \citealt{swinbank_lofar_2015,rowlinson_identifying_2019,pintaldi_scalable_2022}) and those that use image-subtraction or related techniques to find shorter-timescale variability at a reduced computational cost ($\lesssim\,$hours, e.g. \citealt{wang_radio_2023,fijma_new_2024,smirnov_tron1_2025}).}
    \label{f_time}
\end{figure*}

\subsection{Gamma-ray bursts}\label{s_grbs}
Detecting radio afterglows from gamma-ray bursts (GRBs), unbiased by higher frequency detections, was a major motivation for widefield radio transient surveys (as discussed in Section~\ref{s_motivation}). Because of their particular importance in this context, we give a brief background to radio observations of GRBs, before summarising our current understanding of their physics.

\subsubsection{Radio detection of gamma-ray bursts}
The detection of gamma-ray bursts was first reported by \citet{klebesadel_observations_1973}, and their origin remained the subject of debate for over two decades.   
Observations from the Burst And Transient Source Experiment provided a key breakthrough, showing that the sky distribution of GRBs was isotropic \citep{meegan_spatial_1992,briggs_batse_1996}; and hence ruling out progenitors such as Galactic neutron stars. The focus then shifted to identifying a counterpart at other wavelengths so that a distance scale could be established. The detection of a counterpart, now called an `afterglow,' to GRB 970228 in X-ray \citep{costa_discovery_1997} and optical \citep{van_paradijs_transient_1997} wavebands, in a host galaxy at $z = 0.695$, established that GRBs were cosmological. A review of early gamma-ray burst afterglow observations, and our derived understanding, is given by \citet{van_paradijs_gamma-ray_2000}.

The first GRB for which a radio afterglow was detected was GRB 970508 
\citep{frail_radio_1997}. High variability was observed in the first month after the explosion, which was attributed to diffractive scintillation. Since diffractive scintillation occurs only when the source is smaller than a characteristic size, the transition from the diffractive to refractive regime could be used to measure the angular size of the fireball, determining that it was a relativistically expanding explosion. The subsequent monitoring of this event in radio \citep{frail_450_2000} demonstrated the utility of radio observations in measuring the calorimetric properties of GRBs. 

The next decade and a half of radio follow-up campaigns is summarised by \citet{chandra_radio-selected_2012}. They collate observations of 304 afterglows at frequencies between 0.6\,GHz and 660\,GHz. 
Of the 306 GRBs with radio follow-up observations, 95 were detected in radio (31\%). Of these, 65 had radio lightcurves (three or more detections in the same band). A key issue is that if a source was not detected early in a follow-up program, it was generally not observed subsequently. This is one factor motivating untargeted searches for GRB afterglows.

In the rest of this section we discuss our current understanding of long and short GRBs, and the connection between short GRBs and neutron star mergers. 

\subsubsection{Long gamma-ray bursts}
Once GRB studies became more fruitful with the launch of dedicated satellites, it was quickly determined \citep{kouveliotou_identification_1993} that there appeared to be two separate populations of GRBs delineated mostly by burst duration but also somewhat by spectral characteristics, short-hard GRBs (SGRBs) with durations $<2\,$s and long GRBs (LGRBs) with longer durations, although there are other subtleties as well (e.g. low-luminosity long GRBs, GRBs with extended emission; \citealt{campana_association_2006,norris_short_2006}).  Long GRBs are generally understood in the `collapsar' model \citep{woosley_gamma-ray_1993,paczynski_are_1998,macfadyen_collapsars_1999,piran_physics_2004}. In this model the GRB originates from the collapse of a massive star (confirmed through observations of associated supernovae; \citealt{galama_unusual_1998}), leading to a jetted, relativistic outflow and then an afterglow when the jet hits the surrounding interstellar medium and/or the remains of the progenitor's winds (Section~\ref{s_afterglow}).  

Early radio observations made important contributions to the studies of relativistic expansion \citep{frail_radio_1997,taylor_angular_2004}, the nature of the progenitors \citep{li_wind_2001,frail_radio_2000}, the effects of viewing geometry on observables \citep{meszaros_viewing_1998}, the structure of the jet \citep{van_der_horst_radio_2005-1,panaitescu_multiwavelength_1998}, the passage of the reverse shock \citep{kulkarni_discovery_1999}, and the basic energetics of GRBs \citep{frail_beaming_2001,frail_complete_2003,frail_accurate_2005}.  

More recent observations and larger samples have refined and complicated this picture, providing constraints on the environment and related selection effects at multiple wavelengths \citep{schroeder_radio-selected_2022,osborne_are_2021}; information on the shock structure \citep{horesh_unusual_2015,laskar_reverse_2016,mcmahon_reverse_2006} and the nature of the central engine \citep{li_radio_2015}; and revealing a diversity in progenitors \citep{lloyd-ronning_lack_2017}.  See \citet{van_paradijs_gamma-ray_2000} and \citet{piran_physics_2004} for early reviews of GRBs, and \citet{chandra_gamma-ray_2016}, \citet{resmi_radio_2017} and \citet{anderson_ami_2028} for more recent discussions focusing on  radio observations.

Current sensitive, wide-band, surveys have enabled searches for larger samples of afterglows at later times, building on earlier results like \citet{seaton_possible_2001} to move beyond the biases present in other studies.  For instance, \citet{peters_observational_2019} did a dedicated search for late-time radio emission from three supernovae associated with GRBs, while \citet{leung_search_2021} used ASKAP data to do a similar search. This approach helps explore late-time afterglow behaviour, which is affected by the progenitor's wind output many years before the explosion.  

However, those searches still targeted GRBs known from high-energy triggers.  One of the longstanding goals of radio surveys \citep[e.g.][]{murphy_vast_2013,metzger_extragalactic_2015,mooley_caltech-nrao_2016,lacy_karl_2020} is understanding the true rate of cosmic explosions through the detection of orphan afterglows, which are not viewed along the jet axis.  While early results were encouraging \citep[e.g.][]{levinson_orphan_2002}, they were hard to validate as the candidates were discovered many years after the putative events, making followup impossible.  More recent surveys have come some way to rectifying this, with much better candidates that are amenable to followup \citep{law_discovery_2018,mooley_late-time_2022} or robust limits \citep{leung_matched-filter_2023}, although the numbers are still small and it can be hard to definitively determine the cause of a synchrotron transient due to their  similarity across many source types (e.g. tidal disruption events and supernovae).  

\subsubsection{Short gamma-ray bursts}
Neutron star mergers have long been believed to be the origin of short gamma-ray bursts (\citealt{eichler_nucleosynthesis_1989,kochanek_gravitational_1993}; although it can be more complicated, as in \citealt{troja_nearby_2022,rastinejad_kilonova_2022}).  Like with long GRBs there is a collimated, relativistic jet that impacts the surrounding medium, and so many of the phenomena related to  long GRBs could also take place in  short GRBs \citep{granot_gamma-ray_2014,berger_short-duration_2014}.

This means that we expect similar relativistic afterglows for short GRBs (Section~\ref{s_afterglow}). However, in contrast to long GRBs, the radio emission is considerably fainter \citep{panaitescu_observational_2001} and hence the first detections were not made until years later \citep{berger_afterglow_2005,soderberg_afterglow_2006}. This is due to both the difference in the energetics of the GRBs themselves, as well as considerably lower circumburst medium densities (\citealt{berger_short-duration_2014,fong_decade_2015,oconnor_constraints_2020}; with the flux density scaling as the density to the 0.5 power).  This environmental difference likely relates to the different underlying model (merger of compact objects instead of an exploding massive star) as well as the resulting change in delay time between formation and burst, which can be much longer in the cases of short GRBs (Gyr vs. Myr; \citealt{zevin_observational_2022}).  For a full review, see \citet{granot_gamma-ray_2014} and \citet{berger_short-duration_2014}.

After years of meagre (but important) returns at radio wavelengths \citep[e.g.][]{fong_decade_2015}, the advent of more sensitive radio instruments and surveys has improved the prospects for radio detections of short GRBs \citep{schroeder_long-lived_2025}.  Combined with X-ray observations, radio afterglow measurements constrain the energetics, geometry, and environment of short GRBs \citep[e.g.][]{jin_short_2018,dichiara_short_2020,rouco_escorial_jet_2023}.  Radio observations are especially important for measuring jet collimation \citep{fong_short_2014,troja_achromatic_2016}, and deviations from a simple shock model that could indicate prolonged energy injection and/or passage of the reverse shock \citep[e.g.][]{lloyd-ronning_reverse_2018,lamb_short_2019,troja_afterglow_2019,schroeder_radio_2024,anderson_early_2024}.  

In the future, observations with more sensitive instruments \citep{lloyd-ronning_science_2017} and untargeted surveys should detect a larger number of sources in a wider range of environments, exploring whether the conclusions drawn so far are biased and how they may relate to the related populations of Galactic neutron star binaries and extragalactic neutron star mergers (e.g. \citealt{chattopadhyay_modelling_2020,gaspari_galactic_2024}).

\subsubsection{Neutron star mergers}
\label{s_nsm}
While short GRBs have been studied with the understanding that they are most likely the result of neutron star mergers, this was largely untested empirically until the first (and so far still only) multi-messenger detection of a neutron star merger, GW170817/GRB170817A \citep{abbott_gw170817_2017,abbott_multi-messenger_2017}.  In this case a mildly relativistic off-axis afterglow was detected at radio wavelengths  \citep{hallinan_radio_2017} after first being detected at X-ray wavelengths \citep{troja_x-ray_2017,margutti_electromagnetic_2017}.  The proximity of this event allowed detailed study \citep[e.g.][]{mooley_mildly_2018,alexander_electromagnetic_2017,dobie_turnover_2018,troja_outflow_2018,corsi_upper_2018,mooley_strong_2018} giving constraints on the structure, opening-angle, and viewing geometry \citep[e.g.][]{lazzati_late_2018}.  Further observations, including very long baseline interferometry (VLBI) astrometry, allowed more robust constraints on the geometry \citep{mooley_superluminal_2018,ghirlanda_compact_2019} that in turn allowed direct constraints on the Hubble parameter \citep{hotokezaka_hubble_2019}.

While a burst of gamma-rays was seen from GW170817/GRB170817A \citep{goldstein_ordinary_2017}, it was considerably lower luminosity than that from most GRBs.  The distinction is that the observer needs to be on-axis to see a GRB, while the more general neutron star merger phenomena could be observed off-axis, as was the case with GW170817/GRB170817A. 
Most subsequent searches for neutron star mergers at radio wavelengths  still   rely on multi-messenger triggers.  Given the uncertain nature of many of those triggers and the wide sky areas they point to \citep[e.g.][]{petrov_data-driven_2022}, even triggered searches still retain some aspects of wide-field radio surveys, with many sources to consider and tight control over false positives required.  With these triggered searches, non-detections can still be used to constrain the geometry, energetics, and environments of neutron star mergers \citep{gulati_constraints_2025}.
But even without such a trigger, the collimated afterglow from a SGRB/neutron star merger could still be detectable by wide-field radio surveys \citep{metzger_extragalactic_2015,hotokezaka_radio_2016,lin_detectability_2020}, although as with many similar events it may be a challenge to distinguish them from other related synchrotron transients.  Nonetheless, they are promising targets to understand the diversity in jet launching and how neutron star mergers are tied to their environments.  If detected, followup observations including astrometry and searches for scintillation can then be used to determine the jet geometry and energetics \citep{dobie_constraining_2020}, especially with next-generation radio facilities \citep{dobie_radio_2021,corsi_radio_2024}.

Beyond the collimated, relativstic ejecta there are also predictions for an additional component, with mildly-relativistic dynamical ejecta interacting with the surrounding medium and causing a second afterglow-like component \citep{nakar_detectable_2011,piran_electromagnetic_2013,hotokezaka_mass_2015,hotokezaka_synchrotron_2018}.  Unlike the jet launching discussed above, detecting this dynamical ejecta could allow studies of the neutron star equation of state and merger dynamics.  So far there has not been a definitive radio detection of this emission from GW170817/GRB170817A \citep{hajela_two_2019,balasubramanian_gw170817_2022}, and while there have been claims of a second emission component at X-ray wavelengths \citep{hajela_evidence_2022}, this is currently a subject of debate \citep{troja_accurate_2022,ryan_modeling_2024,katira_late-time_2025}.  Nonetheless, for future events these transients should be detectable by wide-field radio surveys \citep{metzger_extragalactic_2015,hotokezaka_radio_2016,corsi_radio_2024,bartos_radio_2019}.

Finally, there are models of neutron star mergers where the gravitational wave signal is expected to be accompanied by a coherent burst of low-frequency radio waves, potentially even before the gravitational wave signal (e.g. \citealt{usov_low_2000,totani_cosmological_2013,zhang_fast_2020,yamasaki_repeating_2018,wang_pre-merger_2018,cooper_pulsar_2023}, and for candidate detections see \citealt{moroianu_assessment_2023,rowlinson_candidate_2024}).  Identification of prompt emission would aid in prompt localisation of the gravitational wave event (thus aiding further electromagnetic followup), potentially provide a dispersion measure-based distance \citep[e.g.][]{palmer_radio_1993}, and help understand the physical conditions at the site of the merger.  These signals would not generally be detectable by most of the imaging surveys discussed here, but could be identified by beam-forming surveys, especially wide-area/all-sky surveys (Section~\ref{s_allsky}).

\subsection{Supernovae}
\label{s_sne}
Core-collapse supernovae (SNe) represent the end point in the evolution of massive stars.  They are responsible for injecting mass, energy and enriched material into the ISM \citep{alsabti_handbook_2017}.  The many different observational classifications/manifestations of SNe come from differences in the evolutionary state and mass-loss history of the progenitor \citep{alsabti_handbook_2017,smith_mass_2014}.   Radio emission from SNe largely comes from interaction between the SN ejecta and the circumstellar material (\citealt{chevalier_interaction_1981,chevalier_radio_1982,chevalier_synchrotron_1998}; see \citealt{weiler_radio_2002}).  Observing the evolution of the radio emission allows one to unwind the mass-loss history of the progenitor, allowing studies of the latest stages of mass-loss right before explosion (which are unconstrained empirically; \citealt{chandra_circumstellar_2018}).  This can then be used to study the peculiarities of mass-loss for individual SNe (e.g. \citealt{sfaradi_dense_2024}) and for large samples \citep{bietenholz_radio_2021,sfaradi_observed_2025}, which can then be correlated with the optical classification.

The simple picture of steady spherical mass-loss is, in many cases, contradicted by detailed radio observations.  This has been studied in detail in the case of SN~1987A in the Large Magellanic Cloud \citep{staveley-smith_structure_1993,mccray_remnant_2016,petruk_polarized_2023}, through multi-wavelength modeling of the evolving morphology and spectrum, but usually this level of detail is not possible.  Instead observations can usually only constrain the flux density, and in small numbers of cases the size and rough shape (with very long baseline interferometry; e.g. \citealt{demarchi_radio_2022}).  But even the basic lightcurve information is complex, 
and can show multiple radio peaks \citep{anderson_peculiar_2017,palliyaguru_vlbi_2021,rose_late-time_2024} indicating episodic mass-loss, or broadened peaks \citep{soderberg_radio_2005,sfaradi_dense_2024} showing aspherical ejecta. In particular, we highlight the use of unbiased surveys \citep{stroh_luminous_2021,rose_late-time_2024} in systematically probing for unexpected emission of a large, diverse sample of SNe, and further searching for (likely) SNe that were not identified previously \citep{gal-yam_radio_2006,dong_transient_2021}.

Beyond the simple blast-wave model, sub-classes of SNe such as the stripped-envelope  types Ibc or Ic-BL can probe additional particle acceleration and/or central engines that may be present in some SNe 
\citep{corsi_search_2023,soderberg_relativistic_2010,soderberg_late-time_2006}. Studying the radio emission properties of these SNe can help constrain the energetics and geometries of relativistic jets \citep{shankar_proto-magnetar_2021,barnes_grb_2018}.  Some type Ic-BL supernovae also show long GRBs \citep{woosley_supernova_2006,cano_observers_2017}, presumably from a jet associated with a central engine.  Other observational manifestations can arise from partially successful jets that either lead to Ic-BL SNe without GRBs or with low-luminosity GRBs \citep{eisenberg_observational_2022,bromberg_are_2011,pais_velocity_2023,piran_relativistic_2019}.

Aside from core-collapse SNe, thermonuclear SNe (Type Ia; \citealt{blondin_type_2024}) are of considerable interest for their elemental enrichment patterns \citep{dwek_iron_2016} and their use in cosmology \citep{riess_observational_1998,perlmutter_measurements_1999}, but the underlying natures of their progenitors are still poorly understood \citep{maoz_observational_2014,liu_type_2023}.  Radio observations can in principle distinguish between progenitor scenarios \citep{horesh_early_2012,perez-torres_constraints_2014}, constraining the presence of a non-degenerate companion.  However, despite considerable investment of observational resources (e.g. \citealt{panagia_search_2006,hancock_visibility_2011,chomiuk_evla_2012,margutti_inverse_2012,horesh_early_2012,perez-torres_constraints_2014}) no type Ia SNe were detected for many years.  This changed with the detection of  SN2020eyj by \citet{kool_radio-detected_2023}, whose radio properties show signs of interaction between a blast wave and the circumstellar medium (CSM), suggesting a non-degenerate companion.  However, the optical/infrared signature of this SN is also somewhat unusual, showing signs of CSM interaction (Type Ia-CSM; \citealt{silverman_type_2013}), so it remains to be seen whether this signature is borne out by future observations of a larger sample of Ia SNe.

\subsubsection{Luminous fast blue optical transients}
\label{s_fbots}
Sitting at the intersection of core-collapse SNe and GRBs, there are expected to be a variety of relativistic explosions that would lack high-energy emission (e.g. \citealt{meszaros_tev_2001}).  Such sources have begun to be discovered in increasing numbers by large optical surveys.  The first of these from an emerging new class is AT2018cow \citep{prentice_cow_2018,perley_fast_2019,ho_at2018cow_2019,margutti_embedded_2019,nayana_ugmrt_2021}, which was notable for having bright, fast, blue optical emission along with very bright and long-lasting (sub)millimetre emission transitioning, along with quickly declining centimetre emission whose properties are not easily reconciled with the millimetre emission \citep{margutti_embedded_2019,ho_at2018cow_2019,ho_luminous_2022} in the usual way for synchrotron explosions.  These sources are called `(luminous) fast blue optical transients' or (L)FBOTs, with LFBOTs those with the highest peak luminosities ($10^{44}\,{\rm erg\,s}^{-1}$) that are achieved in the shortest times (few days), although FBOTs can  span a range of lower luminosities and slower timescales   \citep[e.g.][]{drout_rapidly_2014}.

While the model for these sources is not clear \citep[e.g.][]{kuin_swift_2019,lyutikov_fast-rising_2019,lyutikov_nature_2022,mohan_nearby_2020,pellegrino_circumstellar_2022,chen_radiative_2022,pasham_evidence_2021,vurm_gamma-ray_2021}, with models including tidal disruptions of white dwarfs, ongoing central engines, accretion, and circumstellar interactions in unusual environments, many observations suggest that they are related to relativistic explosions from massive stars although this is far from certain \citep{metzger_luminous_2022,migliori_roaring_2024}.  Increasing numbers of LFBOTs are being identified through optical surveys \citep[e.g.][]{ho_search_2023}, which are now able to optimise their selection procedures \citep{bright_radio_2022,ho_koala_2020,coppejans_mildly_2020,perley_real-time_2021,ofek_at_2021,ho_minutes-duration_2023,yao_x-ray_2022,chrimes_at2023fhn_2024}, and radio followup is playing a crucial role in understanding their properties and hence progenitor models.   What is still lacking, though, are discoveries beginning with radio or millimetre wavelengths, although this may change with the next generation of millimetre surveys \citep{eftekhari_extragalactic_2022}. Like the supernova cases above, such discoveries can help understand the biases in the optical selection functions and what the true underlying rates are.

\subsection{Active galactic nuclei}
\label{s_agn}
As discussed in Section~\ref{s_jets}, AGN are the dominant radio source by number in most widefield surveys to date\footnote{Narrower, deeper surveys show that star forming galaxies dominate at lower flux densities \citep[see][]{padovani_faint_2016A}.}  \citep{condon_radio_2002}, and also the dominant source of variability \citep[e.g.][]{falcke_radio_2001,mundell_radio_2009}.  Some of this is extrinsic (Section~\ref{s_scint}) but much of the variability is due to intrinsic changes in the sources themselves.  For most variable sources this manifests as modest stochastic variability at the few percent to few tens of percent level on timescales from days to years.  This low-level intrinsic variability is typically ascribed to changes in the accretion rate onto the central super-massive black hole, propagation of shocks along the jets, or similar phenomena \citep[e.g.][]{bignall_time_2015,marscher_models_1985}.

However, there are also examples with much more extreme variability, exceeding 100\%  \citep{barvainis_radio_2005}.  Causes for this can include  reorientation of AGN jets, potentially from a binary supermassive black hole \citep{palenzuela_dual_2010,an_periodic_2013}, or dramatic changes in the accretion rate \citep{kunert-bajraszewska_caltech-nrao_2020,wolowska_caltech-nrao_2021,nyland_quasars_2020}.  Both of these are of considerable interest.  Binary supermassive black holes are likely related to the model for hierarchical growth of these black holes through mergers \citep{volonteri_formation_2010}, and contribute to the low-frequency gravitational wave background \citep{agazie_nanograv_2023-1} detected through pulsar timing experiments \citep{agazie_nanograv_2023,epta_collaboration_second_2023,miles_meerkat_2025,reardon_search_2023,xu_searching_2023} and potentially  the expected signal at higher frequencies that would be detected through forthcoming space-based interferometers \citep{huang_impacts_2024}.  In contrast, significant changes in jet activity and accretion rate may give clues to the duty cycle and intermittency of AGN activity, which is important for understanding black hole growth \citep{volonteri_formation_2010} as well as feedback processes in galactic environments \citep[e.g.][]{czerny_accretion_2009}.

\subsection{Tidal disruption events}
\label{s_tdes}
Tidal disruption events (TDEs; \citealt{hills_possible_1975,rees_tidal_1988,komossa_tidal_2015}; see \citealt{gezari_tidal_2021} for a recent review) occur when a star is disrupted while passing within the tidal radius of a supermassive black hole (note that other TDE scenarios involving compact objects and lower-mass black holes have been proposed, e.g. \citealt{krolik_swift_2011}).  This leads to emission potentially across the electromagnetic spectrum, including at radio wavelengths (\citealt{van_velzen_radio_2011,giannios_radio_2011};  see \citealt{alexander_radio_2020} for a recent review).  

The best studied cases of TDE emission in the radio are where relativistic jets are launched as in Swift~J1644+57 \citep{bloom_possible_2011,zauderer_birth_2011,levan_extremely_2011}.  However, there is a large diversity in the radio properties of TDEs, with many lacking relativistic jets and only 20--30\% with detectable prompt (within the first few months after an optical/UV/X-ray transient) radio emission that instead originates from synchrotron emission from fast but sub-relativstic outflows and a wide range in luminosity \citep{alexander_radio_2020}.  There are a number of proposed models for this radio emission, ranging from debris streams or winds from the accretion disk interacting with the surrounding medium, to self-interaction within debris streams \citep{bower_late-time_2013,krolik_asassn-14li_2016,alexander_discovery_2016,pasham_discovery_2018}.

Perhaps even more interesting are the recent discoveries of significantly-delayed radio emission \citep{horesh_delayed_2021,horesh_are_2021,cendes_mildly_2022}, years after the original event.  The origin of this --- either a delay in mass ejection \citep{cendes_mildly_2022} or a viewing angle effect \citep{sfaradi_off-axis_2024} --- is still debated, but it seems to be relatively common (30--40\%) among TDEs discovered at other wavelengths \citep{horesh_are_2021,cendes_ubiquitous_2024,anumarlapudi_radio_2024}.  This points to a further diversity in TDE properties, based on the nature of the disrupted star, the characteristics of the surrounding medium, the mass/spin of the black hole, or additional factors.  Understanding the underlying cause of this full range of manifestations can help with studies of accretion and the environments of supermassive black holes.

The results above are from TDEs discovered at (primarily) optical wavelengths and then followed up in the radio. However, new radio surveys have led to discoveries of TDEs (and TDE candidates) starting with their radio emission \citep{anderson_caltech-nrao_2020,ravi_first_2022,somalwar_candidate_2023,somalwar_vlass_2025-1,somalwar_vlass_2025,dykaar_untargeted_2024}. Such discoveries, along with those at infrared wavelengths \citep[e.g.][]{mattila_dust-enshrouded_2018}, have the potential to more robustly probe the radio properties of TDEs without the underlying biases in optical surveys, such as being able to determine a rate independent of any host obscuration \citep{bower_constraining_2011}.  This is echoed by the use of unbiased surveys to reveal unexpected flares in late-time emission \citep{ravi_first_2022,anumarlapudi_radio_2024}.

We show a compilation of a diverse set of synchrotron transient lightcurves in Figure~\ref{f_explosions}.  This highlights the extreme range in luminosity (at least 6 orders of magnitude for the peak) and timescale ($>3$ orders of magnitude).  Such empirical analyses are often helpful when classifying new events \citep[e.g.][]{coppejans_mildly_2020,ho_koala_2020}.

\begin{figure}
    \centering
    \includegraphics[width=1.0\linewidth]{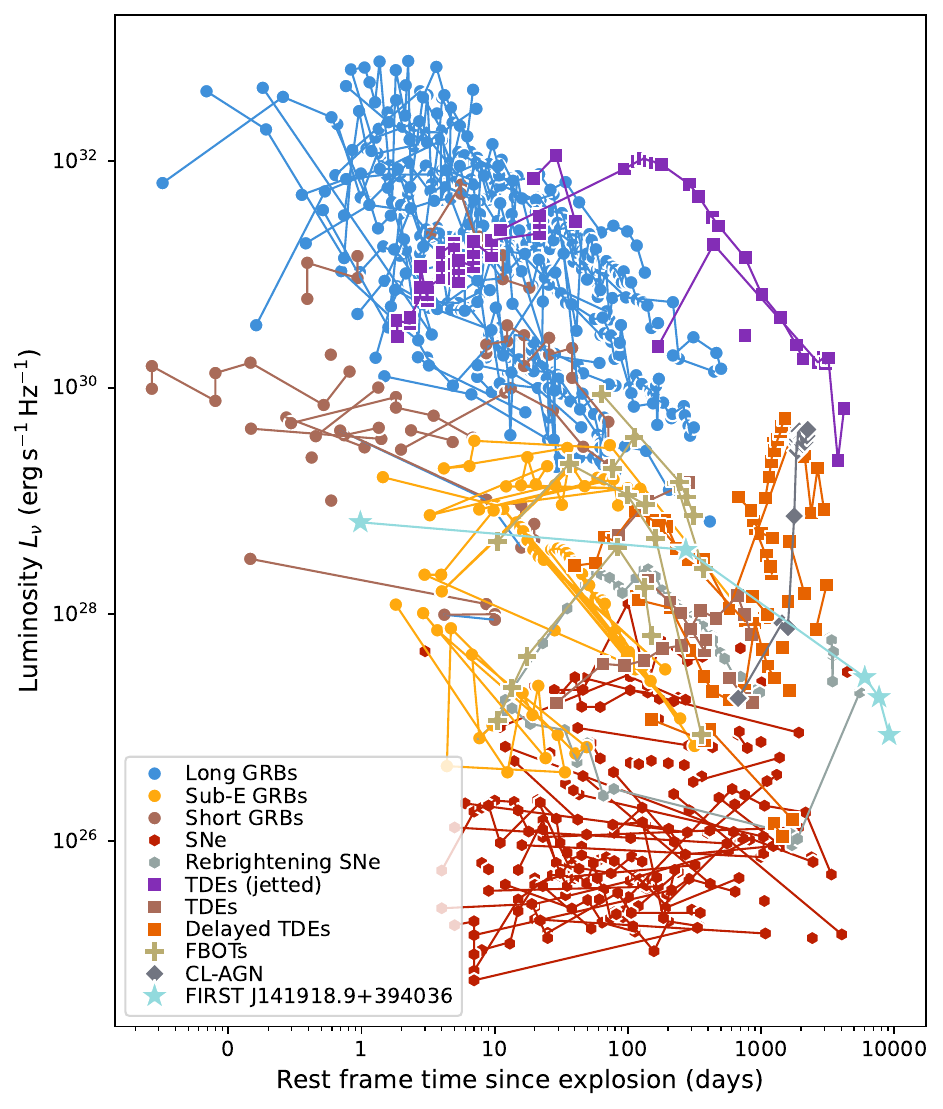}
     \caption{Radio lightcurves of a diverse set of synchrotron transients, following examples like \citet{ho_koala_2020} and \citet{coppejans_mildly_2020}.
     The sources include: long GRBs,  sub-energetic GRBs, and short GRBs (circles; Section~\ref{s_grbs}), supernovae (hexagons; Section~\ref{s_sne}), TDE (squares; Section~\ref{s_tdes}), FBOTs (pluses; Section~\ref{s_fbots}), changing-look AGN (diamonds; Section~\ref{s_agn}), and the potential orphan afterglow FIRST~J141918.9+394036.  Data are primarily at 7--10\,GHz, except FIRST~J141918.9+394036 (1.4\,GHz, but which was scaled following \citealt{mooley_late-time_2022}), a few GRBs at 5\,GHz, and a few TDEs at 15\,GHz; in those cases no spectral correction or $K$-correction has been applied.
     Data were compiled by A.~Gulati and are from \citet{alexander_discovery_2016,alexander_radio_2017,anderson_early_2024,andreoni_very_2022,berger_grb_2001,berger_host_2001,berger_jet_2000,berger_radio_2003,berger_afterglow_2005,bietenholz_radio_2021,bright_radio_2022,brown_late-time_2017,cendes_mildly_2022,cendes_radio_2021,cendes_ubiquitous_2024,cenko_afterglow_2011,cenko_multiwavelength_2006,cenko_swift_2012,chandra_comprehensive_2008,chandra_discovery_2010,chrimes_multi-wavelength_2024,coppejans_mildly_2020,djorgovski_afterglow_2001,eftekhari_associating_2018,fong_short_2014,fong_decade_2015,fong_broadband_2021,frail_450_2000,frail_complete_2003,frail_accurate_2005,frail_energetic_2006,frail_enigmatic_2000,frail_radio_1999,galama_bright_2000,galama_continued_2003,goodwin_radio_2023,goodwin_systematic_2025,greiner_unusual_2013,hajela_eight_2025,hancock_grb111209a_2012,harrison_broadband_2001,harrison_optical_1999,ho_at2018cow_2019,ho_koala_2020,horesh_are_2021,horesh_delayed_2021,horesh_unusual_2015,kulkarni_radio_1998,lamb_short_2019,laskar_radio_2023,laskar_reverse_2016,laskar_vla_2018,laskar_first_2022,law_discovery_2018,leung_search_2021,levan_heavy-element_2024,margutti_embedded_2019,margutti_signature_2013,mattila_dust-enshrouded_2018,meyer_late-time_2025,moin_radio_2013,mooley_late-time_2022,oconnor_structured_2023,pasham_multiwavelength_2015,perley_afterglow_2014,perley_grb_2008,price_grb_2002-1,rhodes_rocking_2024,rol_grb_2007,rose_late-time_2024,schroeder_long-lived_2025,schroeder_radio_2024,sfaradi_off-axis_2024,soderberg_afterglow_2006,soderberg_constraints_2004,soderberg_redshift_2004,soderberg_relativistic_2006,stein_tidal_2021,taylor_discovery_1998,van_der_horst_detailed_2008,zauderer_radio_2013}.  See \url{https://github.com/ashnagulati/Transient_Comparison_Plots}
    \label{f_explosions} 
    }
\end{figure}

\subsection{Stars}
\label{s_star}
Radio emission has been detected from stars at all stages of their evolution \citep{gudel_stellar_2002}: pre-main sequence and main sequence stars; ultracool dwarfs and supergiants; and a range of unusual stellar types such as magnetically chemically peculiar stars, RS~CVn binaries and Wolf-Rayet stars. Figure~\ref{f_cmd} \citep[adapted from][]{driessen_sydney_2024} demonstrates this diversity, showing the radio luminosity of a sample of known radio stars, spanning the entire colour-magnitude diagram, including the supergiant and white dwarf branches.

Radio emission gives us insight into a range of stellar phenomena that can not be fully understood with observations in other wavebands. This includes questions about: (i) stellar mass loss throughout the different stages of stellar evolution; (ii) how young stars and planets evolve in their common disk environment; (iii) the magnetic field structure of stars; and (iv) how the flaring activity of stars affects the habitability of planets they host \citep{matthews_radio_2019}.   

However, radio emission has only been detected from a tiny fraction of stars. Of the $\sim1.8$~billion stars detected by Gaia \citep{gaia_collaboration_gaia_2023}, fewer than 1\,500 have been detected at radio frequencies. Widefield radio transient surveys (and circular polarisation surveys), notably with ASKAP and LOFAR, are now transforming this area of research, with the number of known radio stars increased substantially from the well-established \citet{wendker_catalogue_1987,wendker_radio_1995} catalogue, which contains 228 radio stars detected at $<3$~GHz (and which was last updated in 2001), to the Sydney Radio Stars Catalogue which contains confirmed detections of 839 unique stars \citep{driessen_sydney_2024}. 

Radio stars exhibit a wide range of variability behaviours, driven by a range of emission mechanisms (see Section~\ref{s_stellarmech}). Transient and highly variable radio emission is an indication of strong magnetic fields, shocks, or interactions between stars in binary systems.
In this section we can only briefly summarise our current understanding of stellar radio emission (with a focus on variability), and point the reader to suitable review papers on each stellar class. We have summarised the key characteristics of each class in Table~\ref{t_stars}. Binary systems involving a compact object are addressed separately in Sections~\ref{s_novae} and \ref{s_xrb}. The following subsections are presented in (roughly) decreasing order of temperature. 

\begin{figure*}
    \centering
    \includegraphics[width=1.0\textwidth]{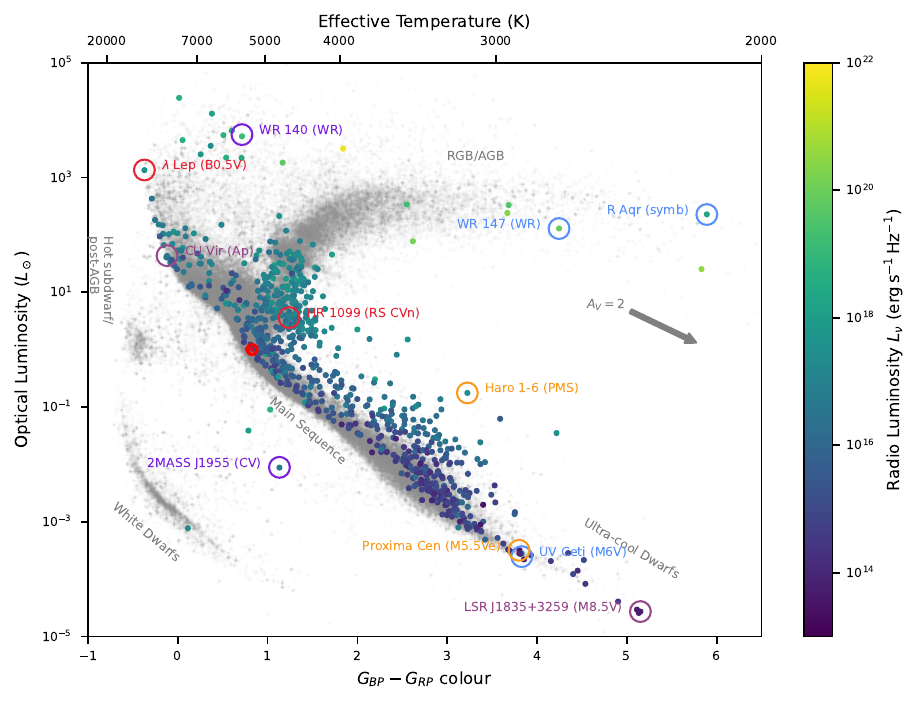}
     \caption{Gaia DR3 colour-magnitude diagram showing the stars in an updated version of the Sydney Radio Stars Catalogue. The colour scale shows the radio luminosity based on the maximum flux density of each star in the SRSC and the Gaia \texttt{rgeo} distance. The grey background points show the Gaia DR2 CMD for reference \citep{pedersen_diverse_2019}, and we annotate some of the major features.  The Sun is indicated by the red $\odot$ symbol.  Individual stars are indicated by the coloured circles and are labeled with their types.  We show a rough translation between Gaia $G_{BP}-G_{RP}$ colour and effective temperature determined from synthetic photometry (based on \citealt{stsci_development_team_pysynphot_2013}) and an extinction vector \citep{zhang_empirical_2023}. Even cooler sources such as brown dwarfs are located off the right edge of the figure, and are not included as they do not have Gaia measurements. Figure is adapted from \citet{driessen_sydney_2024}.}
    \label{f_cmd}
\end{figure*}

\subsubsection{O \& B type main sequence stars}\label{s_obstars}
The dominant source of radio emission from hot ($ \sim 10\,000$~K to $>50\,000$~K), massive ($>10\,{M_\odot}$), OB stars is their stellar winds \citep{gudel_stellar_2002}. As the stars shed their strongly ionised winds, they produce thermal (free-free) radio emission (Section~\ref{s_freefree}); and indeed radio observations are a well-established way of measuring stellar mass loss rates in OB stars \citep{leitherer_radio_1995,scuderi_radio_1998}. 
The associated radio emission is predicted to have a positive spectral index ($\alpha \sim+0.6$) and not be variable \citep{wright_radio_1975}. However, these theoretical models assume a steady, spherically symmetric wind. In cases where the wind is inhomogeneous, filaments and clumps in the wind can cause variations in the radio flux density. This has been observed in, for example, P~Cygni, a luminous blue variable star, and the first OB star detected in radio \citep{wendker_detection_1973}. VLBI observations by \citet{skinner_compact_1997} and \citet{exter_further_2002}  demonstrated that the stellar wind of P Cyg is highly inhomogeneous and as a result its integrated flux density changes on timescales of days.

Both blue and red supergiants (the highly evolved counterparts of massive OB stars) exhibit thermal radio emission from their stellar atmospheres, and where present, their stellar winds, via the same mechanisms as main sequence OB stars \citep{scuderi_radio_1998}. See, for example, radio studies of $\alpha$ Ori (Betelguese) \citep{lim_large_1998,ogorman_alma_2020}.

Up to $25-40\%$ of radio-detected OB stars have been found to have signatures of non-thermal radio emission \citep{bieging_survey_1989}, and in some cases this dominates the thermal component \citep{abbott_detection_1984}. This emission, due to synchrotron radiation (Section~\ref{s_synch}), has a characteristic flat or negative spectral index, and can be variable on short timescales \citep[e.g.][]{bieging_survey_1989,persi_thermal_1990}. Synchrotron emission implies the presence of strong magnetic fields and shocks. Initially there were a range of hypotheses for the source of non-thermal emission, considering both single stars and binary systems. However, it is now established that most (if not all) OB stars that show non-thermal emission are binary systems \citep[see Table 2 in][]{de_becker_non-thermal_2007}. The emission is caused by shocks in the colliding-wind region of binary OB systems, where the winds from both stars in the binary interact (see also Section~\ref{s_rscvn}). This has been demonstrated in a number of systems through the observed periodicity of the radio variability \citep[see the series of papers studying stars in the Cyg OB association for examples][]{blomme_non-thermal_2005,van_loo_non-thermal_2008,blomme_non-thermal_2010}. A review of non-thermal radio emission in massive stars is given by \citet{de_becker_non-thermal_2007}.

\subsubsection{Wolf-Rayet stars}\label{s_wrstars}
Wolf-Rayet (W-R) stars are the helium-burning remnants of massive OB stars. They are highly evolved, having lost their hydrogen-rich outer layers through the processes of Roche lobe overflow to a companion (in close binaries),  mass loss by a strong stellar wind (in single stars), or a combination of both \citep{crowther_physical_2007}. W-R stars have extremely hot, dense stellar winds, and so exhibit strong thermal radio emission, via the same mechanisms as OB stars, and are some of the most luminous stellar radio sources \citep{abbott_radio_1986}. As with OB stars, this radio emission can also be used to measure stellar mass loss rates, which, for W-R stars, are very high \citep{nugis_mass-loss_2000}.

A subset of W-R stars exhibit non-thermal synchrotron radiation, and as with OB stars, most (if not all) of these are in binary systems \citep[see Table 1 in][]{de_becker_non-thermal_2007}. The archetypal example is WR 140, a W-R$+$O star binary, with a 7.9~year orbit. Extensive monitoring with the VLA by \citet{white_eight-year_1995} showed variability in the non-thermal emission, clearly correlated with the orbital phase. Further work by \citet{dougherty_high-resolution_2005} showed that the spectral index is also correlated with the orbital phase, suggesting that the relative importance of emission, absorption and plasma processes changes with stellar separation (and hence electron and ion density) throughout the orbit. For more information we refer the reader to the review of non-thermal radio emission in massive stars (including W-R stars) by \citet{de_becker_non-thermal_2007}.

\subsubsection{A \& early F type main sequence stars}
A and early F-type stars ($\sim 6\,500$~K to $<10\,000$~K) have neither the strong ionised stellar winds of OB type stars, nor the dynamo-generated magnetic fields of later type stars \citep[][ and see below]{gudel_stellar_2002}. Hence they are generally expected to have weak thermal emission from their atmospheres, and mostly not be detectable at radio wavelengths \citep{brown_stringent_1990}. A small number of nearby objects have been detected in radio, for example Sirius A \citep[A0/1;][]{white_mesas_2018,white_mesas_2019} and Procyon \citep[F5;][]{drake_detection_1993}. 
For Procyon, the observed quiescent emission is thermal (free-free) chromospheric emission from the entire stellar disk. The variable emission is potentially due to either (a) active regions on the corona, analogous to our Sun; or (b) flares or bursts due to localised magnetic fields, again analogous to our Sun. See also the discussion of thermal radio emission detected from the late F9 star $\eta$~Cas~A by \citet{villadsen_first_2014}.

A subset of A-type (and B-type) stars that do show strong radio emission are the
 magnetic chemically peculiar (Ap and Bp type) stars. We discuss these separately in  Section~\ref{s_mcp}.

\subsubsection{Magnetic chemically peculiar stars}\label{s_mcp}
Magnetic chemically peculiar stars (MCPs, also referred to as Ap and Bp type stars), are a subset of main sequence stars that have strong (kilogauss), organised magnetic fields, and unusual chemical abundances on their surfaces \citep{preston_chemically_1974,landstreet_magnetic_1992}.  

There were many searches for radio emission from these stars, carried out over several decades \citep[e.g.][]{trasco_radio_1970,turner_12_1985}, until \citet{drake_discovery_1987} successfully detected five systems with the VLA: $\sigma$~Ori E, HR 1890, and $\delta$~Ori C, IQ Aur, and Babcock's star (HD 215441). Subsequent work by \citet{linsky_radio_1992} showed the radio emission was consistent with gyrosynchrotron radiation from particles in the magnetosphere, accelerated due to interactions with the stellar wind. 

The geometry of magnetic fields in MCP stars is generally dipolar, with the magnetic axis inclined with respect to the rotational axis. Hence the observed magnetic field of the star is modulated by its rotational period \citep{landstreet_magnetic_1992}. Consistent with this, \citet{leone_periodic_1993} showed that the observed radio emission of two MCPs (HD 37017 and $\sigma$~Ori~E) was periodic, and varied with the rotational period of the star.

However, the gyrosynchrotron mechanism can not explain the observed radio emission of all MCPs. Observations of CU~Vir by \citet{trigilio_coherent_2000} detected bright, highly directional, and highly circularly polarised emission, that required a coherent mechanism to explain it. They showed this was consistent with electron cyclotron maser emission, a result supported by long-term monitoring observations of CU Vir \citep[e.g.][]
{trigilio_radio_2008,ravi_radio_2010,lo_observations_2012}. 

The current generation of radio telescopes (and in particular widefield radio surveys) has enabled the detection of an increased number of magnetic chemically peculiar stars, including at lower radio frequencies. See \citet{hajduk_radio_2022}, \citet{das_discovery_2022} and \citet{das_vast-memes_2025} for recent results from LOFAR, GMRT (the Giant Metrewave Radio Telescope), and ASKAP respectively. 

\subsubsection{Late F, G \& K type main sequence stars}
Late F-type to early K-type main sequence stars are often referred to as `solar-type’ stars, as they are broadly similar to our Sun in terms of temperature, mass, and other physical processes. The Sun is extremely well-studied in radio: see \citet{dulk_radio_1985} and \citet{bastian_radio_1998} for earlier reviews of solar radio emission, and \citet{benz_physical_2010} for a more recent review of solar flares across the electromagnetic spectrum. \citet{kowalski_stellar_2024} provides an equivalent review of stellar flares. 

Like our Sun, F, G and K-type stars are observed to have quiescent thermal radio emission \citep{villadsen_first_2014}. If the Sun’s thermal radio emission was scaled to the distance of stars in our local neighbourhood, it would only reach \ujy\ levels, which until recently was below the detection threshold of most radio telescopes. Hence the stars detected to date tend to be more luminous than the Sun. 

F to K-type stars (and many other main sequence stars) have a magnetic corona like our Sun, and hence have magnetic fields generated by a dynamo at the interface between the radiative core and the convective envelope. They therefore show stellar flares, with both thermal and non-thermal components \citep[e.g.][]{gudel_radio_1992, lim_radio_1995,budding_radio_2002}.  

Solar-type stars have also been observed to flare with much higher energies than our own Sun \citep[so-called `superflares’;][]{maehara_superflares_2012}. Superflares are caused by magnetic reconnection, and often linked to starspot activity. Superflares have not been definitively detected at radio wavelengths to date. 

Finally, some F, G \& K stars exhibit coherent bursts from a range of mechanisms \citep[e.g.][]{matthews_radio_2025} --- although in general the stellar equivalents of solar radio bursts have been primarily studied on K \& M dwarfs due to their extremely strong magnetic fields, since these result in stronger and more frequent flare activity (see Section~\ref{s_mdwarfs}).

\subsubsection{RS CVn and Algol binaries}
\label{s_rscvn}
RS Canum Venaticorum (RS CVn) and Algol stars are two subtypes of chromospherically active binary star systems. 

RS CVn binaries consist of a subgiant or giant star of spectral type F--K with a companion main sequence star of spectral type G--M. They have typical orbital periods of 1--30 days, and typical separations of $\sim R_\odot$ \citep{hall_rs_1976}. They have active chromospheres, high rotational velocities and strong magnetic fields. 

Algol binaries consist of two semi-detached stars, in which one is a hot (B or A-type) main sequence star, and the second is a more evolved (G or K-type) giant or subgiant. They have typical orbital periods of 1--20 days and typical separations of $\sim 10-40R_\odot$ \citep{giuricin_general_1983}. 

At gigahertz frequencies, these stellar systems exhibit quiescent emission, due to mildly relativistic electrons radiating gyrosynchrotron emission in magnetic fields \citep{gudel_stellar_2002}. 
Both RS CVn and Algol-type binaries also show variability in this gyrosynchrotron emission, on timescales of minutes to hours. This activity can be present for a significant fraction ($\sim30\%$) of the time \citep{lefevre_variability_1994}. In addition, they show significant radio flaring activity
\Citep{drake_survey_1989,lefevre_variability_1994,umana_radio_1998}. 

At low radio frequencies, coherent emission mechanisms dominate in RS CVn binaries. \citet{toet_coherent_2021} propose that this electron cyclotron maser emission could be generated by enhanced chromospheric flaring due to tidal forces or the interaction between the two binary stars, analogous to the Jupiter-Io interaction. 

Two well-studied active binary systems are the RS CVn binary HR~1099 \citep[e.g.][]{owen_detection_1976,feldman_discovery_1978,osten_multiwavelength_2004,slee_coherent_2008} and the canonical Algol binary, $\beta$~Persei \citep{mutel_dual_1985,mutel_radio_1998}. 

\subsubsection{M dwarf and ultracool dwarf stars}\label{s_mdwarfs}
M-dwarfs are very low mass (typically $0.1 - 0.6$ M$_\odot$), cool ($T < 3\,800$~K) stars. They are long lived, and extremely abundant, accounting for roughly 75\% of the stars in our local neighbourhood \citep{winters_solar_2019}. M-dwarfs show bright radio bursts across all wavebands, associated with their extremely strong \citep[up to kilogauss;][]{shulyak_strong_2017} magnetic fields.

The observed radio emission comes from a range of different emission mechanisms. Intense and highly polarised electron cyclotron maser emission often dominates \citep[e.g.][]{hallinan_confirmation_2008,lynch_154_2017,villadsen_ultra-wideband_2019}. This likely originates both from magnetic active regions, and from auroral emission \citep[e.g.][]{hallinan_magnetospherically_2015}, analogous to solar system planets \citep{zarka_auroral_1998}. M-dwarfs also show quiescent and flaring emission due to gyrosynchrotron radiation \citep[e.g.][]{berger_discovery_2001,lynch_radio_2016}.

Ultracool dwarfs (later than M7-type) are the coolest ($T < 2\,900$~K) stellar and substellar objects. Only ten L \& T-type dwarfs have been detected in radio to date (see Table 2 in \citealt{rose_periodic_2023}, and in addition the recently published detection by \citealt{guirado_dwarf_2025}), but these observations have provided evidence for kilogauss magnetic fields even in the coolest of substellar objects \citep{kao_strongest_2018}. Given that ultracool dwarfs are fully convective (without a radiative core), this requires an alternative dynamo mechanism \citep[e.g.][]{browning_simulations_2008}.

Like M-dwarfs, ultracool dwarfs can show quiescent, flaring and auroral radio emission \citep[e.g.][]{route_arecibo_2012,kao_auroral_2016}. Because radio observations allow us to probe the magnetic field strength, they are critical for understanding energy generation in ultracool stars and substellar objects.

Radio observations complement X-ray and optical observations by providing direct measurements of the magnetic field strengths and plasma densities associated with stellar activity \citep{benz_physical_2010}. Because coherent emission mechanisms often dominate, many M-dwarf and ultracool dwarf stars do not fit the empirically established (over 10 orders of magnitude) Gudel-Benz relation between quiescent gyrosynchrotron radio and soft X-ray luminosity.  

Untargeted searches with current telescopes are enabling detection of larger samples of M dwarf and ultracool dwarf stars than were previously available \citep[e.g.][]{callingham_population_2021}. They also enable searches that are unbiased by the properties of the stars in other wavebands (previous samples were often selected on the basis of their known chromospheric activity \citep[e.g.][]{lynch_154_2017,crosley_low-frequency_2018}.

\subsubsection{Star-planet interactions and exoplanetary systems}
Radio emission is expected from exoplanetary systems, in analogy to what we observe from the magnetised planets in our own solar system \citep{zarka_plasma_2007}. This predicted emission originates from several different mechanisms: (i) auroral emission from the interaction between the host stellar wind and the exoplanetary magnetosphere, as seen in magnetised planets in our solar system \citep{zarka_auroral_1998,farrell_possibility_1999}; (ii) electron cyclotron maser emission from the magnetic interaction between a so-called `hot Jupiter' and its host star (referred to as star-planet interactions, or SPI; \citealt{zarka_plasma_2007}); (iii) emission from the interaction between an exoplanet and its exomoon/s, analogous to the Jupiter-Io system \citep{noyola_detection_2014}.

There has been substantial theoretical and empirical work in predicting the flux densities expected \citep[e.g.][]{lazio_radiometric_2004,griesmeier_predicting_2007,hess_modeling_2011,lynch_detectability_2018}.
For (i) the radio emission has a cutoff frequency determined by the strength of the exoplanetary magnetic field, and hence is only likely to be detectable at low radio frequencies ($< 300$~MHz). For (ii), the emission frequency can be higher (up to gigahertz), as it is determined by the magnetic field strength of the host star rather than the planet. 

Over the past two decades there have been many searches for radio emission from exoplanetary systems, from targeted observations of individual exoplanets considered likely to be detectable \citep[e.g.][]{bastian_search_2000,lazio_magnetospheric_2007,vidotto_stellar_2012} to large-scale searches for emission from known exoplanetary systems in both targeted \cite[e.g.][]{ortiz_ceballos_volume-limited_2024} and untargeted radio surveys \citep[e.g.][]{murphy_limits_2015}. In recent years there have been a number of tentative detections \citep[e.g.][]{vedantham_coherent_2020,turner_search_2021}, although none have yet been definitively confirmed as originating from star-planet interactions. 

The improved sensitivity of current and future telescopes, and the capability of widefield surveys to target many systems at once means detecting exoplanetary systems at radio wavelengths is increasingly likely \cite[see][for a discussion of SKAO capabilities]{zarka_magnetospheric_2015}. However, a significant observational challenge is that the emission is beamed, and dependent on the orbital period of the planet. Although beaming can make the emission brighter, the high directionality reduces the probability of detection. In addition, once radio emission has been detected from an exoplanetary system, it is challenging to distinguish between an exoplanet-related cause, and emission from the host star \citep[see, for example, the discussion in][]{pineda_coherent_2023}.   
See \citet{callingham_radio_2024} for a recent review of radio emission from star planet interactions and space weather in exoplanetary systems.


\begin{table*}[t!]
    \caption{Key characteristics of radio emission from different stellar types. For each type we give a couple of key example objects, and the associated references. The types are roughly ordered in decreasing temperature, following the subsections in the main text.} 
    \label{t_stars}

{\tablefont\begin{tabular}{>{\raggedright\arraybackslash}p{3cm}
    >{\raggedright\arraybackslash}p{4.5cm}
    >{\raggedright\arraybackslash}p{4.3cm}
    >{\raggedright\arraybackslash}p{2.9cm}
    >{\raggedright\arraybackslash}p{1.2cm}} \toprule
{\bf Stellar type} & {\bf Emission mechanisms} &	{\bf Key features} & {\bf Example sources} & {\bf Citations} \\
\hline 
Main seq. O \& B & (i) Thermal free-free emission from ionized stellar winds & Low variability; unpolarised; $\alpha ~+0.6$   & $\zeta$~Pup, P Cyg & \hyperlink{sref1}{1}, \hyperlink{sref2}{2}, \hyperlink{sref3}{3} \\
 & (ii) Non-thermal synchrotron emission in binaries & Variable; $\alpha \le 0.0$  & \\
Wolf-Rayet Stars & (i) Thermal free-free emission from ionized stellar winds & Strong, some of the most luminous radio stars; low variability & WR140, WR147	& \hyperlink{sref4}{4}, \hyperlink{sref5}{5}, \hyperlink{sref6}{6} \\
 & (ii) Non-thermal synchrotron emission in binaries	& Variable in flux and spectral index; can be aligned with orbital phase  &  \\
Main seq. A \& early-F & (i) Thermal free-free emission from the chromosphere & Weak, rarely detected; quiescent emission & Sirius A, Procyon & \hyperlink{sref7}{7}, \hyperlink{sref8}{8} \\
 & (ii) Thermal free-free emission from the corona & Low variability; weak flares or bursts  & & \\
Magnetic chemically peculiar (Ap \& Bp) & (i) Gyrosynchrotron radiation from the interaction between stellar wind and magnetosphere &  Quiescent; sometimes variable with rotation period; moderately to highly polarised & CU Vir, $\sigma$~Ori E & \hyperlink{sref9}{9}, \hyperlink{sref10}{10}, \hyperlink{sref11}{11} \\ 
 & (ii) Coherent auroral emission (ECMI) &  Variable and periodic; moderately to highly polarised & & \\ 
Main seq. late-F, G \& K & (i) Thermal free-free emission from the corona & Low variability; weak; unpolarised & $\alpha$ Centauri, AB Dor & \hyperlink{sref12}{12}, \hyperlink{sref13}{13}, \hyperlink{sref14}{14} \\
& (ii) Gyrosynchrotron (and thermal) radiation from stellar flares & Variable; can be periodic; can be polarised & &  \\
& (iii) Coherent bursts & Variable; highly polarised & &  \\
RS CVn and Algol binaries & (i) Gyrosynchrotron radiation (quiescent and flaring) & Both quiescent and variable & HR 1099, $\beta$ Persei & \hyperlink{sref15}{15}, \hyperlink{sref16}{16}, \hyperlink{sref17}{17} \\
& (ii) Electron cyclotron maser emission & Short strong bursts; highly polarised &  & \\

M dwarfs \& \hspace{10mm} ultracool dwarfs & (i) Electron cyclotron maser emission and coherent auroral emission & Strong, highly polarised, highly variable, can be periodic & Proxima Cen; LP944-20; WISE J062309$-$045624  & \hyperlink{sref18}{18}, \hyperlink{sref19}{19}, \hyperlink{sref20}{20}\\
 & (ii) Gyrosynchrotron radiation (quiescent and flaring) & &  & \\

\botrule
\end{tabular}}\begin{tabnote}
Citations: 
\hypertarget{sref1}{1} --- \citet{wendker_detection_1973}; 
\hypertarget{sref2}{2} --- \citet{morton_radio_1978}; 
\hypertarget{sref3}{3} --- \citet{bieging_survey_1989}; 
\hypertarget{sref4}{4} --- \citet{dougherty_high-resolution_2005};
\hypertarget{sref5}{5} --- \citet{churchwell_wolf-rayet_1992}; 
\hypertarget{sref6}{6} --- \citet{abbott_radio_1986};
\hypertarget{sref7}{7} --- \citet{white_mesas_2019};
\hypertarget{sref8}{8} --- \citep{drake_detection_1993}
\hypertarget{sref9}{9} --- \citet{leto_stellar_2006}; 
\hypertarget{sref10}{10} --- \citet{leone_periodic_1993};
\hypertarget{sref11}{11} --- \citet{drake_discovery_1987};
\hypertarget{sref12}{12} --- \citet{gudel_radio_1992} 
\hypertarget{sref13}{13} --- \citet{trigilio_alpha_2018}
\hypertarget{sref14}{14} --- \citet{lim_first_1992}
\hypertarget{sref15}{15} --- \citet{feldman_discovery_1978};
\hypertarget{sref16}{16} --- \citet{mutel_radio_1998};
\hypertarget{sref17}{17} --- \citet{drake_survey_1989}; 
\hypertarget{sref18}{18} --- \citet{zic_flare-type_2020};
\hypertarget{sref19}{19} --- \citet{berger_discovery_2001};
\hypertarget{sref20}{20} --- \citet{rose_periodic_2023};
\end{tabnote}

\end{table*}

\subsection{Classical novae}\label{s_novae}
Classical and recurrent novae --- forms of cataclysmic variables (highly-variable binary systems involving a white dwarf), with the distinction between classical and recurrent novae coming from comparing the recurrence time to observational spans --- are thermonuclear explosions on the surfaces of white dwarfs in binary systems (\citealt{gallagher_theory_1978,bode_radio_1987}; for a recent review see \citealt{chomiuk_new_2021}).  A thin layer of accreted material undergoes unstable nuclear burning, driving off a modest amount of mass (more or less than the accreted amount) at a speed of hundreds to thousands of km\,s$^{-1}$.  These are among the most common explosions in the Milky Way, although the number of detections (largely from wide-field optical surveys) is limited by available monitoring and dust obscuration.  While classical novae (CNe) are common, and broadly understood, further studies of CNe can help probe binary interactions, chemical enrichment, shocks, and high-energy acceleration \citep{chomiuk_new_2021}.

The traditional view of radio emission from CNe (\citealt{seaquist_thick_1977,hjellming_radio_1979}; and see \citealt{chomiuk_classical_2021} for a recent synthesis of radio emission properties) is where an expanding sphere of ionised gas emits via free-free radiation, with the temporal/spectral evolution depending on the expansion of the gas and the changes in the emission measure and self-absorption that result.  Even in this simple model, measurements of the free-free emission can help constrain the distances to CNe, the masses of ejected material, and their energetics \citep{bode_radio_1987,gulati_classical_2023}.

As more observations accumulated, the picture grew more complex.  Suggestions that the brightness temperature required non-thermal emission \citep{snijders_nova_1987,taylor_unusual_1987,krauss_expanded_2011,weston_non-thermal_2016,weston_shock-powered_2016} emerged as radio observations sampled more CNe at earlier times.  This emission --- presumably synchrotron emission from shocks between the ejecta and surrounding material --- may tie to the  GeV $\gamma$-ray emission \citep{ackermann_fermi_2014} that may have other signatures across the electromagnetic spectrum \citep{chomiuk_new_2021}.  Even the thermal emission itself may be more complex, with evidence for multiple, aspherical components \citep{chomiuk_2011_2014} or prolonged/delayed ejection \citep{nelson_2011_2014}, which can best be probed through detailed radio followup of a wide range of CNe.

Finally, while the vast majority of novae discovered come from optical surveys, these will miss many sources that are obscured by dust in the Milky Way \citep{chomiuk_new_2021}.  The advent of near-infrared \citep{lucas_vvv-wit-01_2020} surveys offers a new opportunity to discover obscured novae deep in the Galactic plane.  Even more than that, radio surveys with sufficient sampling can identify new nova candidates on their own, probing a different aspect of the eruption process (tied more to ejecta mass and velocity than bolometric luminosity).

\subsection{X-ray binaries}
\label{s_xrb}
Low-mass X-ray binaries, where neutron stars or black holes accrete from stellar-mass companions, are highly variable sources seen across the electromagnetic spectrum \citep{fender_overview_2014,van_den_eijnden_new_2021,fender_jets_2006}.  The jets produced by these systems inject energy into the surrounding medium \citep{gallo_dark_2005}, regulating star formation and ionising the surrounding gas \citep{fender_energization_2005}, and can be used to study the formation of jets and accretion across many orders of magnitude in accretor mass (from stellar-mass systems to supermassive black holes; \citealt{gallo_universal_2003,gallo_radiox-ray_2014,van_den_eijnden_new_2021}), giving insight into the fundamental mechanisms of jet launching \citep{mcclintock_black_2014} and synchrotron-emitting regions \citep[e.g.][]{chauhan_broadband_2021}.  Detailed imaging of the radio jets can also show direct evidence for relativistic jet launching and superluminal motion \citep{mirabel_superluminal_1994,fender_ultra-relativistic_2004}.

Radio emission from X-ray binaries depends on the jet type, which can change over time.  Discrete blobs of ejected material have steeper radio spectra ($\alpha\approx -0.7$), emitting as they move away from the compact object.  In contrast, steady jets have much flatter spectra coming from a superposition of  regions at different radii dominating at different frequencies.  

Observing such systems over time and across wavelengths, there is evidence for correlation between the radio  and  X-ray luminosities that extends across many orders of magnitude in mass \citep{merloni_fundamental_2003,migliari_jets_2006,gallo_universal_2003} when the X-ray emission is in the `hard' state, but which may also break into multiple tracks with different levels of radio emission \citep{soleri_nature_2011} whose origin and significance are not understood \citep{gallo_hard_2018}.  However, when the X-ray emission is in other spectral states (`intermediate' or `soft'; \citealt{belloni_states_2010}) this correlation disappears as steady jet emission stops and discrete blobs may be  ejected \citep{fender_towards_2004}.  The transitions between these states likely represent disc instabilities changing the rate at which material moves from the disc onto the compact object \citep{coriat_revisiting_2012}.  Such transitions can give rise to transient radio jets, either going from the `soft' X-ray state to the `hard'  or the other way, when newly-launched jets can impact the material from previous ejection cycles \citep{koljonen_2006_2013}.

Historically, most radio observations of X-ray binaries have been triggered by state transitions observed at X-ray wavelengths (e.g. \citealt{plotkin_2015_2017,russell_accretion-ejection_2014}), where the prevalence of all-sky X-ray monitors (e.g. \citealt{barthelmy_burst_2005,matsuoka_maxi_2009}) allows for easy identification of sources requiring follow-up (although see e.g. \citealt{miller-jones_time-sequenced_2004}).  However, this leads to time-lags between when X-ray behaviour occurs and radio monitoring commences, and may overlook radio variability that is less tied to X-ray state changes (e.g. \citealt{wilms_correlated_2007}).  Moreover, X-ray sources can be  subject to significant absorption by intervening Galactic material.  Therefore future radio monitoring of the Galactic plane can be used to identify transient/variable X-ray binary sources across a much wider range of radio luminosity, potentially independent of high-energy behaviour, to help understand if the standard paradigm actually holds in all cases \citep[e.g.][]{corbel_incoherent_2015}.

\subsection{Pulsars and magnetars}
\label{s_pulsar}
Rotation-powered pulsars (spinning, highly-magnetised neutron stars; \citealt{lorimer_handbook_2012}) and magnetars (neutron stars powered by decay of ultra-strong magnetic fields; \citealt{kaspi_magnetars_2017}) are most commonly identified by single-dish radio surveys and X-ray surveys, respectively. They are of broad astrophysical interest as probes of matter at extreme densities (e.g. \citealt{demorest_two-solar-mass_2010,miller_radius_2021}) and magnetic fields \citep{kaspi_magnetars_2017,philippov_pulsar_2022}, as well as related topics like how supernova asymmetries lead to high kick velocities (e.g. \citealt{cordes_guitar_1993}), searches for planets \citep{wolszczan_planetary_1992}, and tests of gravity (e.g. \citealt{archibald_universality_2018}); networks of pulsars can also be used to search for low-frequency gravitational waves \citep{agazie_nanograv_2023,reardon_search_2023,epta_collaboration_second_2023,xu_searching_2023,miles_meerkat_2025}.

Intrinsically, individual pulses from pulsars are highly variable (magnetospheric `weather', in the language of \citealt{philippov_pulsar_2022}), but they tend to add together to form highly-stable averages (magnetospheric `climate'; \citealt{philippov_pulsar_2022}; also see \citealt{lorimer_handbook_2012}) with stable flux densities \citep{kumamoto_flux_2021}. Hence, there must be some specific reason for pulsars to appear variable, and be detectable in image domain transient surveys; these mechanisms are discussed in Section~\ref{s_pulsarcause}. 

Many cases of variability are due to extrinsic effects, such as diffractive or refractive scintillation (Section~\ref{s_scint}) or interactions in tight binary systems.   The former results from the fact that neutron stars are extremely compact and hence are among the very few sources to show diffractive scintillation.
The latter can be either periodic, with eclipses and modulations during the orbits of so-called `spider' pulsars (pulsars with low-mass non-degenerate companions in tight orbits; \citealt{roberts_surrounded_2013}), or secular, with radio variability and optical state changes caused by the transition from an accretion-powered low-mass X-ray binary phase to a rotation-powered millisecond pulsar \citep{archibald_radio_2009}.  Both of these can cause pulsars to appear and disappear \citep[e.g.][]{bond_first_2002,zic_discovery_2024} during imaging observations, depending on the timescale of those observations compared to the orbital period and eclipse duration (for spiders) or state changes.  In all cases these effects can  be exploited to discover new pulsars \citep{backer_millisecond_1982,dai_detecting_2016,dai_prospects_2017}, while scintillation itself can help understand the properties of the turbulent interstellar medium \citep[e.g.][]{kumamoto_flux_2021}, the pulsar velocity distribution \citep[e.g.][]{cordes_diffractive_1998}, and the origin of extreme scattering events \citep{zhu_pulsar_2023}.

However, some types of pulsars also show intrinsic variability. In some cases, the time between seeing individual pulses is sufficiently long compared to the observation duration that normal pulse-to-pulse variations show up as significant variability between images.  This can occur when the rotation period is very long (say $\gtrsim 10\,$min; e.g. \citealt{hurley-walker_radio_2022,hurley-walker_long-period_2023}) although it is not yet clear if these objects are indeed neutron stars (see Section~\ref{s_lpts}) or when the neutron stars do not emit a pulse every rotation period, as with rotating radio transients (RRATs; \citealt{mclaughlin_transient_2006}), where pulses are only detected once per hundreds or thousands of rotation periods.  In fact, the RRAT phenomenon appears to be at least partly based on observational limitations, where the sources likely form a continuum of behaviour between pulsars that emit every pulse; pulsars that show short-term nulling behaviour \citep{backer_pulsar_1970}; intermittent pulsars \citep{kramer_periodically_2006}; and those that only emit sporadically \citep{keane_rotating_2011,keane_rotating_2011-1}.  Broadly, this is likely to be related to the dynamics of the pulsar magnetosphere as pulsar emission weakens and then (in most cases) ceases, moving toward the so-called `death line' \citep{philippov_pulsar_2022}.   

Finally, larger-scale magnetospheric changes can also lead to changes in the radio properties of magnetars \citep{kaspi_magnetars_2017}, which are powered not by rotation like most pulsars but by an overall magnetic energy store. Most magnetars do not emit radio pulses, but following outbursts there can be radio emission lasting months \citep{camilo_transient_2006} to years \citep{liu_86_2021}, although emission can also turn on absent any high-energy transient \citep{levin_radio-loud_2010,levin_spin_2019,dai_wideband_2019}.  While in some ways this emission is pulsar-like, the spectral and polarisation properties can differ from those of most pulsars, with flatter spectra \citep{camilo_transient_2006,torne_simultaneous_2015} and higher polarisation fractions \citep{camilo_magnetar_2007,philippov_pulsar_2022}.

As with the extrinsic effects, these intrinsic variations can be used to identify new pulsars through imaging surveys \citep{wang_discovery_2022,mcsweeney_discovery_2025} or new activations of known sources \citep[e.g.][]{driessen_radio_2023}, as well as gather robust statistics of nulling and intermittency.  

Beyond these phenomena, which relate to the pulsed emission from the neutron star, there are several related phenomena that likely tie to the birth and energetics of pulsars.  We discuss these below.

\subsubsection{Nascent pulsar/magnetar wind nebulae}
When a supernova leaves behind a compact object, the luminosity of this object (both particle and electromagnetic) can power a new pulsar wind nebula (PWN; \citealt{gaensler_evolution_2006,olmi_dawes_2023}) or magnetar wind nebula, depending on the nature of the compact object.  This represents a transition from the supernova stage to a supernova remnant, a phase that may span a decade or so and is poorly probed observationally (e.g. \citealt{milisavljevic_evidence_2018,milisavljevic_supernova_2017}) but can help constrain the population of newly formed compact objects and tie them to the properties of their parent supernovae (e.g. \citealt{metzger_diversity_2015,fransson_emission_2024}).  This is especially relevant for sub-classes of supernovae (including Type Ic-BL and superluminous supernovae) that may be powered by ongoing emission from a central engine, often thought to be a magnetar \citep{thompson_magnetar_2004,kasen_supernova_2010,dessart_long-term_2024}.

So far, direct radio searches for transitional sources have been mixed.  Broad and targeted radio surveys of supernovae (e.g. \citealt{law_search_2019,eftekhari_radio_2019}) have only yielded a few detections of new wind nebula candidates.  Results from untargeted radio surveys have also only discovered a small number of sources \citep{dong_flat-spectrum_2023,marcote_resolving_2019}, and the interpretation of these discoveries is often unclear \citep{mooley_late-time_2022} since the nature of the synchrotron emission is degenerate between many classes.  Nonetheless, continued deep radio observations of large samples of supernovae from diverse classes, whether done as targeted searches or just parts of broad surveys, will help to further resolve this question and establish links between different supernova types and potential central engines.

\subsubsection{Fast radio burst persistent radio sources}
While we are not in general discussing FRBs in this review, we will briefly discuss the related phenomenon of `persistent radio sources' (PRSs). First identified following the first FRB localisation \citep{chatterjee_direct_2017}, PRSs are long-lived compact ($\sim$mas; \citealt{marcote_repeating_2017,bhandari_constraints_2023}) non-thermal radio sources that are co-located with (so far) repeating FRBs \citep{chatterjee_direct_2017,marcote_repeating_2017,niu_repeating_2022,bruni_discovery_2024,bruni_nebular_2024}.  They are spatially offset from the nuclei of their host galaxies and have properties inconsistent with star-formation such as very high degrees of magnetisation \citep{michilli_extreme_2018,niu_repeating_2022} that appears to be tied to their luminosities \citep{yang_are_2020}. This suggests a common origin with FRBs and perhaps an interpretation as a magnetar-powered nebula  \citep{margalit_concordance_2018}, although other models are possible (e.g. \citealt{zhang_physical_2020,sridhar_radio_2022}, \citealt{ibik_search_2024} and references therein).  If confirmed, this interpretation would allow persistent radio sources to be used as FRB calorimeters and place constraints on the origin and evolution of FRBs and their progenitors.

Note that, while called `persistent', PRSs do in fact have significant changes in both their Faraday rotation measure 
\citep{michilli_extreme_2018,niu_repeating_2022}  and in their flux densities \citep{rhodes_frb_2023,zhang_temporal_2023}, likely from intrinsic causes.  This then open an avenue for discovery of new PRSs: searching for compact, non-thermal, potentially variable radio sources either associated with known FRBs \citep{ibik_search_2024,eftekhari_associating_2018} or not \citep{ofek_search_2017,dong_flat-spectrum_2023}.  Such searches can help constrain the rate and lifetimes of repeating FRBs and PRSs, although care must be taken to not be confused by galactic nuclei, chance alignments, or thermal radio emission (e.g. \citealt{ravi_host_2022,dong_mapping_2024}).

\subsection{Long period transients and Galactic Centre Radio Transients}\label{s_lpts}
Imaging surveys with the latest generation of radio telescopes  have begun to discover apparently periodic radio sources that lie at or well beyond the traditional pulsar `death line' where radio emission is expected to stop (Figure~\ref{f_ppdot}).  Called  long-period transients (LPTs), their periods range from  minutes \citep{hurley-walker_radio_2022,hurley-walker_long-period_2023,caleb_emission-state-switching_2024,dong_chime_2025,wang_detection_2025} to hours \citep{hurley-walker_29_2024,lee_interpulse_2025}. We summarise the properties of the currently known LPTs in Table~\ref{t_lpts}. 

LPTs emit very bright, highly polarised radio pulses that last for several minutes (exhibiting peak flux densities of up to tens of Jy), with polarisation position-angle swings like those observed in pulsars or FRBs \citep{mckinven_frb_2025}, but on timescales hundreds of times longer \citep{dobie_two-minute_2024,men_highly_2025}.  By using sub-structure within the pulse, we can determine a dispersion measure (DM) and infer the distance using a Galactic model \citep{cordes_ne2001i_2002,yao_new_2017}; most LPTs appear to be at  several kpc and largely within the plane of the Galaxy (which may be an observational bias). Some sources are only radio-loud for a few weeks to months, while others remain radio-loud for decades \citep{hurley-walker_long-period_2023}. Many show a range of different emission `modes' reminiscent of pulsar behaviour \citep{caleb_emission-state-switching_2024}, while others show `interpulses', again much like pulsars \citep{lee_interpulse_2025}.  Even among those that do not show discrete modes, there is extremely large variability in the intensity and morphology of individual pulses \citep{lee_interpulse_2025}.
However, unlike other populations of very intermittent radio sources \citep{mclaughlin_transient_2006,surnis_discovery_2023,anna-thomas_unidentified_2024}, LPT energetics computed from spin-down limits suggest that these sources \textit{cannot} be rotation-powered neutron stars \citep{hurley-walker_radio_2022,wang_detection_2025}, unlike some of the long-period radio pulsars also discovered recently \citep[e.g.][]{wang_discovery_2025}.  

These sources were discovered only once it became possible to explore widefield radio surveys with short-timescale imaging (sometimes called `fast imaging'). Prior to that, objects with periods of this length were largely overlooked by both pulsar surveys (often only sensitive to signals with periods $<10\,$s, although there are some exceptions often related to detection of individual pulses rather than periodicity, such as \citealt{caleb_discovery_2022}) and radio transient imaging surveys.   They are therefore filling in an under-populated area in the transient phase space (Figure~\ref{f_phase}). Short-timescale imaging approaches are discussed more in Section~\ref{s_fast}.

The initial discoveries lacked any detections at optical/IR or X-ray wavelengths, despite some extensive searching \citep[e.g.][]{rea_constraining_2022,lee_interpulse_2025}, although given their locations in the Galactic plane with significant extinction/absorption, not all of the upper limits were equally constraining.  We also note that many of those searches were not contemporaneous with periods when the radio source was active, raising questions about variability.  Nonetheless, recent discoveries have uncovered sources with a wider range of multi-wavelength properties that include binary systems containing a white dwarf and late-type star \citep{hurley-walker_29_2024,rodriguez_spectroscopic_2025,de_ruiter_sporadic_2025} where it is clear that the radio periodicity is tied to the orbit, instead of the spin.  These  sources were discovered as radio transients like the other LPTs, but may represent a different underlying population (see below).  Even more intriguing is ASKAP~J1832$-$0911, which was serendipitously detected in soft X-rays by \textit{Chandra} \citep{wang_detection_2025}. 
The X-ray luminosity exceeds the radio luminosity by a factor of $\sim 10$, making the standard pulsar energetic argument even more challenging, and the X-ray flux is seen to pulse on the radio period and then vary secularly together with the radio flux.  This likely points to a shared magnetospheric origin (not accretion or intra-binary shocks), but detailed models are still lacking.

The characteristics of LPTs were immediately reminiscent of the so-called Galactic Centre Radio Transients (GCRTs) that had been discovered two decades earlier (\citealt{hyman_low-frequency_2002,hyman_powerful_2005};  also see \citealt{davies_transient_1976,zhao_transient_1992}) The most well known of these is GCRT~J1745$-$3009, the so-called `cosmic burper,' which showed bursts of up to 1\,Jy every 77\,min, but then long spans in which it was undetectable. Long-term follow-up \citep[e.g.][]{hyman_new_2006,hyman_faint_2007,roy_circularly_2010} showed intermittent behaviour (which contributes to the difficulty of monitoring these sources) and different burst properties (flux density, spectral index, burst duration and periodicity).   Additional GCRT sources have been discovered \citep{wang_discovery_2021,wang_pilot_2022} recently, although it is unclear if they resemble the periodic emission of GCRT~J1745$-$3009, and if their location near the Galactic centre is a selection effect.

LPTs also share some radio properties with radio pulsating white dwarf binary systems (that consist of a white dwarf star with an M-dwarf companion)\footnote{Although these are sometimes referred to in the literature as `white dwarf pulsars', we find this terminology confusing as (a) it mixed both physical and phenomenological information, and (b) `pulsar' is a well established term for neutron stars, which are not present in these systems.}. There have been potentially six such systems detected to date: AR Sco, with a radio period of $\sim1.95$\,min in a 3.5\,h orbit \citep{marsh_radio-pulsing_2016};  J1912$-$4410, with a radio period of $\sim 5.3$\,min in a 4\,h orbit \citep{pelisoli_53-min-period_2023};  ILT J1101$+$5521, with a much longer period of 2.1\,h, matching the orbital period \citep{de_ruiter_transient_2024};  GLEAM-X J0704$-$37, with a period of 2.9\,h (again matching the orbit; \citealt{hurley-walker_29_2024,rodriguez_spectroscopic_2025}); SDSS~J230641.47+244055.8 with  pulsations at the 92\,s WD spin period and a 3.5\,h orbit \citep{castro_arsco_2025}; and CHIME/ILT~J163430+445010, with a 14\,min radio period and a 0.6 or 1.2\,h orbital period \citep{dong_chimefast_2025,bloot_strongly_2025}.  

We summarise the properties of all the sources that fit into this broad category in Table~\ref{t_lpts}\footnote{See \url{https://vast-survey.org/LPTs/} for an up-to-date version.}. It is highly likely there are multiple underlying source classes that produce the observed behaviour of long period transients: the two most promising being neutron stars and highly magnetised white dwarfs (either isolated or in binaries) \citep{rea_long-period_2024,qu_magnetic_2025}. In this table we have roughly categorised them based on how the authors report them in the discovery papers and subsequent papers.  There are further sources that have been recently discovered but not yet fully characterised, and which might align with any or none of these classes.

While the field is rapidly moving, at the time of writing the origin(s) of LPTs is unclear.
LPTs may be old  magnetars whose radio emission would ordinarily be expected to have ceased \citep{hurley-walker_radio_2022,rea_constraining_2022,zhou_formation_2024}, which were spun down to long periods through an initial `propeller' accretion phase (\citealt{ronchi_long-period_2022,fan_evolutionary_2024}; also see \citealt{ho_ejector_2017}); 
this would explain the long periods and energetic emission. Fully explaining the radio emission and evolution are still a challenge \citep{cooper_beyond_2024,suvorov_evolutionary_2023}, but work is ongoing.

However, other models such as isolated or binary (like some of the sources discussed above) white dwarf (WD) pulsars, or even other binaries, are all possible \citep{zhang_gcrt_2005,katz_gleam-x_2022,loeb_hot_2022,rea_long-period_2024,tong_discussions_2023,qu_magnetic_2025,xiao_apparently_2024,mao_binary_2025,zhou_nature_2025}, and may be relevant for a part of the population.  
For instance, \citet{rodriguez_spectroscopic_2025} suggest that the longer period sources ($\gtrsim 1.2\,$hr) are different classes, with the short period sources related to white dwarf or neutron star spins and the longer period sources related to (in at least some cases) orbits in cataclysmic variables, potentially magnetised systems like polars: disentangling orbital versus spin properties remains a challenge, especially in the absence of optical counterparts, but one that could also be solved by space-based gravitational wave observations \citep[e.g.][]{suvorov_lptgw_2025}.  
Recent long-term timing observations show that even among the `classical' LPTs in Table~\ref{t_lpts} there may be connections to binary systems, may place a larger number of similar sources on a continuum of magnetic WD binaries, although evolutionary challenges still remain \citep{castro_arsco_2025}.

Regardless of the specific model, both observationally and theoretically it appears  that there is a significant underlying population of similar sources \citep{rea_long-period_2024,beniamini_evidence_2023} which are only now being discovered.  Hopefully through continued identification of new sources and longer-term studies of known ones (e.g. radio timing to determine the spin-down rate) the origin of these sources can be elucidated.

\begin{table*}
\caption{Properties of long period transients and related Galactic sources. The sources are broadly grouped according to how they are reported in the literature. We have only included published objects here, but we know of a number of new discoveries currently in preparation.}
\label{t_lpts}
\setlength{\tabcolsep}{2pt}
{\tablefont\begin{tabular}{l|c c c c c ccc c >{\raggedright\arraybackslash}p{2.0cm} l}
\toprule
     \textbf{Source} & \textbf{Period} & \textbf{Duty} & \textbf{DM} & $|$\textbf{\textit{b}}$|$ & \multicolumn{2}{c}{\textbf{Polarisation}} & \textbf{Steep} & \textbf{Optical} & \textbf{X-ray} & \textbf{Comment} & \textbf{Citation}  \\
&     & \textbf{Cycle} & $\mathbf{(pc/cm^3)}$& &\textbf{Linear} & \textbf{Circular} & \textbf{Spectrum} \\ \hline
\multicolumn{1}{l}{\textbf{GCRTs:}\dotfill}\\
GCRT J1745$-$3009 & 77\,min & 13\%\phantom{.0} & -- & \phantom{0}$0.5\degr$ & \red{No} & Yes & Yes & No & No & ``cosmic burper" & \hyperlink{ref1}{1}, \hyperlink{ref2}{2}, \hyperlink{ref3}{3}, \hyperlink{ref4}{4}\\
GCRT J1746$-$2757 & -- & -- & -- & \phantom{0}$0.4\degr$ & ? & ? & ? & No & No &  & \hyperlink{ref5}{5}\\
GCRT J1742$-$3001 & -- & -- & -- & \phantom{0}$0.1\degr$ & ? & ? & Yes & No & No &  & \hyperlink{ref6}{6}\\
ASKAP J173608.2$-$321635 & -- & -- & -- & \phantom{0}$0.0\degr$ & Yes & Yes & Yes & No & No &  & \hyperlink{ref7}{7}\\

\multicolumn{1}{l}{\textbf{Unclassified LPTs:}\dotfill}\\
CHIME J0630+25 & \phantom{0}7\,min & \phantom{0}0.6\% & \phantom{0}$23.0$ & \phantom{0}$7\degr$\phantom{.0} & ? & ? & Yes & No & No & Glitch & \hyperlink{ref8}{8}\\
GLEAM-X J162759.5$-$423504.3 & 18\,min & \phantom{0}4\%\phantom{.0} & \phantom{0}$57.0$ & \phantom{0}$3\degr$\phantom{.0} & Yes & No & Yes & No & No &  & \hyperlink{ref9}{9}, \hyperlink{ref10}{10}\\
GPM J1839$-$10 & 21\,min & 10\%\phantom{.0} & $274.0$ & \phantom{0}$2\degr$\phantom{.0} & Yes & No & Yes & No & No & $P_b=8.8\,$h & \hyperlink{ref11}{11}\\
ASKAP J1832$-$0911 & 44\,min & \phantom{0}6\%\phantom{.0} & $458.0$ & \phantom{0}$0.1\degr$ & Yes & Yes & Yes & No & {Pulsed} &  & \hyperlink{ref12}{12}\\
ASKAP J1935+2148 & 54\,min & $<1$\% & $146.0$ & \phantom{0}$0.7\degr$ & Yes & Yes & Yes & No & No & Mode switching & \hyperlink{ref13}{13}\\
ASKAP J175534.9$-$252749.1 & 70\,min & \phantom{0}2\%\phantom{.0} & $710.0$ & \phantom{0}$0.1\degr$ & Yes & Yes & Yes & No & No &  & \hyperlink{ref14}{14}, \hyperlink{ref15}{15}\\
ASKAP J1839$-$0756 & 6.5\,h & \phantom{0}2\%\phantom{.0} & $188.0$ & \phantom{0}$1\degr$\phantom{.0} & Yes & Yes & Yes & No & No & Interpulse & \hyperlink{ref16}{16}\\

\multicolumn{1}{l}{\textbf{Magnetic WD Binaries:}\dotfill}\\
SDSS J230641.47+244055.8 & 1.5\,min & {?} & -- & $32\degr$\phantom{.0} & ? & ? & ? & {WD+M4} & No & $P_b=3.5$\,h & \hyperlink{ref17}{17}\\
AR Sco & \phantom{0}2\,min & 50\%\phantom{.0} & -- & $19\degr$\phantom{.0} & Weak & Yes & No & {WD+M5} & {Pulsed} & $P_b=3.6$\,h & \hyperlink{ref18}{18}, \hyperlink{ref19}{19}\\
J191213.72$-$441045.1 & \phantom{0}5\,min & $<1$\% & -- & $22\degr$\phantom{.0} & ? & ? & ? & {WD+M4.5} & {Pulsed} & $P_b=4.0$\,h & \hyperlink{ref20}{20}\\
CHIME/ILT J163430+445010 & 14\,min & \phantom{0}1.2\% & \phantom{0}$25.0$ & $42\degr$\phantom{.0} & Yes & Yes & ? & {WD+?} & No & $P_b=1.2$\,h or 0.6\,h; Spin-up & \hyperlink{ref21}{21}, \hyperlink{ref22}{22}\\
ASKAP J1448$-$6856 & 1.5\,h & 25\%\phantom{.0} & -- & \phantom{0}$8\degr$\phantom{.0} & Yes & Yes & Yes & {Hot} & {Yes} &  & \hyperlink{ref23}{23}\\
ILT J1101+5521 & 2.1\,h & \phantom{0}2\%\phantom{.0} & -- & $55\degr$\phantom{.0} & Yes & No & Yes & {WD+M4.5} & No & $P_b=2.1$\,h & \hyperlink{ref24}{24}\\
GLEAM-X J0704$-$37 & 2.9\,h & 0.004\% & \phantom{0}$36.0$ & $13\degr$\phantom{.0} & Yes & Yes & Yes & {(WD?)+M3} & No & Most likely interpretation & \hyperlink{ref25}{25}, \hyperlink{ref26}{26}\\

\multicolumn{1}{l}{\textbf{Long-period Pulsars:}\dotfill}\\
PSR J0311+1402 & 41\,s & \phantom{0}1\%\phantom{.0} & \phantom{0}$19.9$ & $37\degr$\phantom{.0} & Weak & Weak & Yes & No & {?} & Neutron star & \hyperlink{ref27}{27}\\
PSR J0901$-$4046 & \phantom{0}1\,min & \phantom{0}1\%\phantom{.0} & \phantom{0}$52.0$ & \phantom{0}$4\degr$\phantom{.0} & Yes & Yes & Yes & No & No & Neutron star & \hyperlink{ref28}{28}\\
\botrule
\end{tabular}
\begin{tabnote} 
Periods given are the primary radio periods, but these may be a mix of rotation and orbital periods or even beat frequencies, depending on the source; 
where both orbital and rotation periods are known they are given separately.  
Many sources show multiple modes with differing pulse widths and polarisation properties, and are highly variable.   Properties given as ``?" could not be determined from the available information. An up-to-date version of this table is available at \url{https://vast-survey.org/LPTs}.
Also see \citet{dong_chime_2025} and \url{https://lpt.mwa-image-plane.cloud.edu.au/published/tables/1} for related samples. 
\end{tabnote}\begin{tabnote}
Citations: \hypertarget{ref1}{1} --- \citet{hyman_powerful_2005}; \hypertarget{ref2}{2} --- \citet{hyman_faint_2007}; \hypertarget{ref3}{3} --- \citet{roy_circularly_2010}; \hypertarget{ref4}{4} --- \citet{kaplan_search_2008}; \hypertarget{ref5}{5} --- \citet{hyman_low-frequency_2002}; \hypertarget{ref6}{6} --- \citet{hyman_gcrt_2009}; \hypertarget{ref7}{7} --- \citet{wang_discovery_2021}; \hypertarget{ref8}{8} --- \citet{dong_chime_2025}; \hypertarget{ref9}{9} --- \citet{hurley-walker_radio_2022}; \hypertarget{ref10}{10} --- \citet{rea_constraining_2022}; \hypertarget{ref11}{11} --- \citet{hurley-walker_long-period_2023}; \hypertarget{ref12}{12} --- \citet{wang_detection_2025}; \hypertarget{ref13}{13} --- \citet{caleb_emission-state-switching_2024}; \hypertarget{ref14}{14} --- \citet{dobie_two-minute_2024}; \hypertarget{ref15}{15} --- \citet{mcsweeney_new_2025}; \hypertarget{ref16}{16} --- \citet{lee_interpulse_2025}; \hypertarget{ref17}{17} --- \citet{castro_arsco_2025}; \hypertarget{ref18}{18} --- \citet{marsh_radio-pulsing_2016}; \hypertarget{ref19}{19} --- \citet{stanway_vla_2018}; \hypertarget{ref20}{20} --- \citet{pelisoli_53-min-period_2023}; \hypertarget{ref21}{21} --- \citet{bloot_strongly_2025}; \hypertarget{ref22}{22} --- \citet{dong_chimefast_2025}; \hypertarget{ref23}{23} --- \citet{anumarlapudi_askap_2025}; \hypertarget{ref24}{24} --- \citet{de_ruiter_sporadic_2025}; \hypertarget{ref25}{25} --- \citet{hurley-walker_29_2024}; \hypertarget{ref26}{26} --- \citet{rodriguez_spectroscopic_2025}; \hypertarget{ref27}{27} --- \citet{wang_discovery_2025}; \hypertarget{ref28}{28} --- \citet{caleb_discovery_2022}.
\end{tabnote}
}
\end{table*}

\begin{figure}
    \centering
    \includegraphics[width=1.0\linewidth]{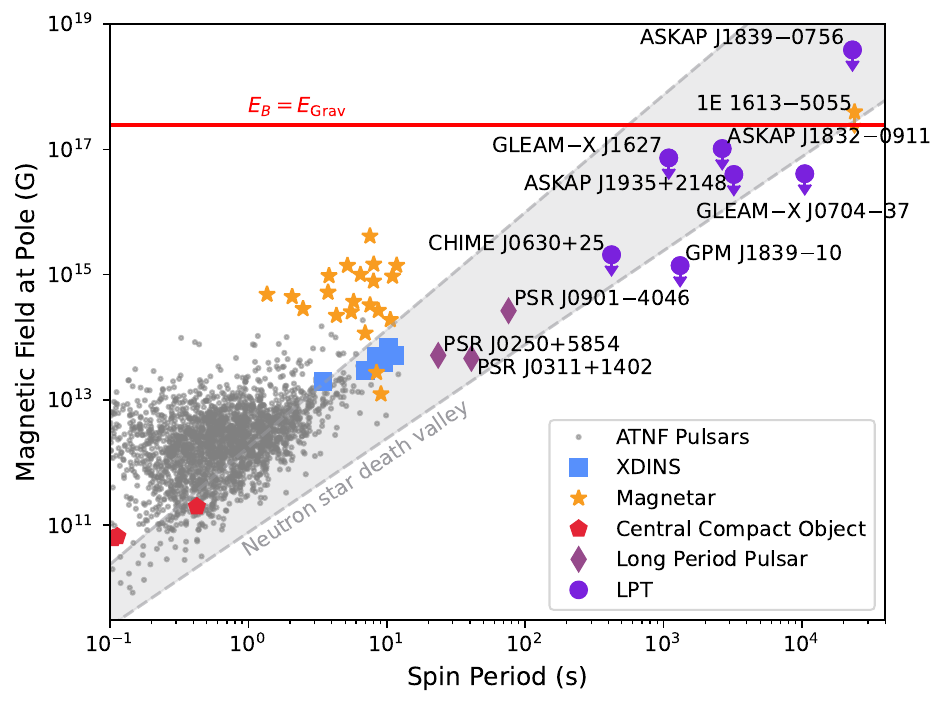}
    \caption{Pulse period versus polar magnetic field ($6.4\times 10^{19}\,{\rm G}\sqrt{P \dot P}$) diagram for pulsars, magnetars, LPTs, and related sources, following  \citet{rea_long-period_2024}.  We show radio pulsars as well as magnetars (stars), X-ray dim isolated neutron stars (XDINS; squares), central compact objects (CCOs, also some times referred to as `anti-magnetars'; pentagons) and select long-period radio pulsars (diamonds) from \citet[][version 2.6.0]{manchester_australia_2005}.  We also plot LPTs (large circles; \citealt{wang_detection_2025,lee_interpulse_2025,wang_discovery_2021,caleb_emission-state-switching_2024,hurley-walker_29_2024,hurley-walker_long-period_2023,hurley-walker_radio_2022,dong_chime_2025,hurley-walker_29_2024}), the radio pulsar J0311+1402 \citep{wang_discovery_2025}, and the long-period magnetar 1E~1613$-$5055 \citep{esposito_swift_2011}.  In those cases we are assuming that the measured period is the spin period, and that the moment of inertia is that of a neutron star.  We plot a range of possible `death lines' for purely-dipolar (top) and highly twisted (bottom) magnetospheres, based on \citet{rea_long-period_2024}, which delineate a `death valley' in which radio emission is expected to cease. Finally, we show the limit where the total magnetic energy is roughly the gravitational binding energy (solid red line).}
    \label{f_ppdot}
\end{figure}

\subsection{Unknown classes }
Serendipity (the fact of finding interesting or valuable things by chance\footnote{Definition from the Cambridge Dictionary \url{https://dictionary.cambridge.org/dictionary/}}) is a driving force in discovery-driven science. The importance of serendipity in astronomy has been discussed by many people, for example \citet{harwit_cosmic_1981} and \citet{fabian_serendipity_2009}. One of the most well known examples of serendipity in time-domain astronomy is the discovery of gamma-ray bursts using data from the Vela nuclear test monitoring satellites (this history is summarised by \citealt{bonnell_brief_1996}). 

There have been many serendipitous discoveries in radio astronomy, as discussed in the proceedings of an 1983 workshop “{\it Serendipitous Discoveries in Radio Astronomy}” held at NRAO\footnote{The full PDF is available online: \url{https://library.nrao.edu/public/collection/02000000000280.pdf}}. Serendipity was built into the science case for the Square Kilometre Array \citep{schilizzi_square_2024} and has been considered in the design of widefield transients surveys on SKAO pathfinder and precursor instruments. 

When new transient candidates are discovered in radio imaging surveys, there is usually an extensive process of identification. This typically involves using other radio data, archival multi-wavelength data, and conducting new follow-up observations in radio and other wavebands. See \citet{ofek_long-duration_2010} and \citet{wang_discovery_2021} for examples of this kind of analysis. It is often straightforward to identify the object as an already-catalogued object based on the available information. However, sometimes the source of the variable radio emission can not be identified. 

In early surveys this was often due to poor time sampling in radio, a lack of multi-wavelength information, and the archival nature of the searches meaning that the transient was discovered long after the event occurred. An example of a well-investigated radio transient that remains unidentified is the one reported by \citet{stewart_lofar_2016}. It is likely that this object will remain unidentified, but the improved specifications and availability of rapid multi-wavelength follow-up in current and future surveys will make the discovery of new classes of radio transients more likely. The most recent example of this is the emergence of a new class of long-period transients, as discussed in Section~\ref{s_lpts}.

\section{Radio transient surveys}\label{s_surveys}

In this section we outline the motivation for radio transient surveys (in contrast to triggered observations or targeted monitoring). We discuss the parameters of an ideal radio transient survey, and then give a brief history of how such surveys have evolved from the early days of radio astronomy to the present day.

\subsection{Motivation for radio transient surveys}\label{s_motivation}

There are a relatively small number of classes of astronomical objects that have been discovered and detected primarily at radio frequencies. Some key examples are pulsars (e.g. \citealt{hewish_observation_1968}), fast radio bursts (e.g. \citealt{lorimer_bright_2007}), Galactic Centre radio transients (e.g. \citealt{hyman_powerful_2005}) and the newly emerging class of long period transients (e.g. \citealt{hurley-walker_radio_2022}). We also observe radio variability caused by diffractive and refractive scintillation, that can either be observed as long-term stochastic behaviour in the case of scintillating quasars (e.g. \citealt{dent_quasi-stellar_1965}), or discrete occurrences such as extreme scattering events (e.g. \citealt{fiedler_extreme_1987}).

Much of what we have learned about other transient and variable radio phenomena has come from monitoring individual objects with targeted observations, often triggered by a discovery event detected in other wavebands. This is the case for explosive phenomena including long gamma-ray bursts (e.g. \citealt{frail_radio_1997}) short gamma-ray bursts \citep[e.g.][]{berger_afterglow_2005}, core-collapse radio supernovae \citep[e.g.][]{gottesman_first-epoch_1972} and recently, Type Ia supernovae (e.g. \citealt{kool_radio-detected_2023}).

For many classes of persistent radio variables, objects of interest have been selected for radio monitoring based on their variability at other wavebands. This is largely the case for radio stars \citep[e.g.][]{lovell_radio_1963}, X-ray binary systems \citep{andrew_detection_1968} and blazars \citep{hufnagel_optical_1992}. This targeted radio follow-up and monitoring has led to significant physical insights into these source classes (see Section~\ref{s_classes}), but remains limited by the selection effects present at other wavebands. Hence there are still gaps in our understanding of the phenomena themselves and also the overall populations as observed in the radio band.

Perhaps the best case study to demonstrate both the value of targeted follow-up and the need for untargeted surveys is radio follow-up of gamma-ray burst afterglows. As discussed in Section~\ref{s_grbs}, our knowledge of the radio properties of these objects (and the subsequent insights in their physical properties) comes from radio follow-up campaigns of individual objects, initially detected at higher energies. 

However, there are still some unanswered questions to address, for which untargeted surveys are needed. For example, as mentioned in Section~\ref{s_grbs} gamma-ray telescopes can not detect GRBs for which the jets are beamed away from our line of sight \citep{rhoads_how_1997}. In principle, widefield radio surveys allow us to detect such `orphan afterglows' and hence constrain the beaming fraction of GRBs --- a key quantity for understanding the true GRB event rate 
\citep{perna_constraining_1998}. They will also result in a significant increase in the number of radio afterglows detected, unbiased by selection criteria at other wavebands. These factors were key motivators in some of the early searches for radio transients in untargeted widefield surveys (e.g. \citealt{levinson_orphan_2002,gal-yam_radio_2006}) which we will discuss further in Section~\ref{s_history}.

\subsection{Ideal survey design}
The exploration of the unknown was a critical driver from the beginning of the SKAO project \citep[see the discussion in][which gives a detailed account of the history of the SKAO]{schilizzi_square_2024}. Time domain science has always been one of the key science areas, and as the SKAO project progressed, discussions of what an ideal time domain imaging survey would look like were initiated. 

\subsubsection{A figure of merit for transient surveys}\label{s_fom}

In this context, \citet{cordes_dynamic_2004} proposed a figure of merit for transient detection surveys:
\begin{equation}\label{e_merit}
{\rm FOM}_{\rm CLM} \equiv    A \Omega_{\rm epoch} \left(\frac{T}{\Delta T}\right)
\end{equation}
that should be maximised for a given survey (subject to various constraints regarding actual telescope designs etc.). In this figure of merit, $A$ is the collecting area of the telescope, $\Omega_{\rm epoch}$ is the solid angle coverage in a single epoch (not an instantaneous field-of-view like the usage in a classical \'{e}tendue), $T$ is the total duration of the observation, and $\Delta T$ is the time resolution. 
For continuum imaging surveys, the trade-off has typically been between sensitivity and area, spending a finite total time either on a smaller number of deeper pointings or a larger number of shallow ones (e.g. the LOFAR Two-metre Sky Survey (LoTSS) covering $20,000\,{\rm deg}^2$ with a rms of $2000$\,\ujybeam\ vs. LoTSS Deep covering $30\,{\rm deg}^2$ at $10$\,\ujybeam\ rms; \citealt{shimwell_lofar_2017,best_lofar_2023}).

For the figure of merit in Equation~\ref{e_merit}, increasing the collecting area, $A$, increases the instantaneous sensitivity of the observations, which means a fainter source population can be probed. Increasing the solid angle coverage, $\Omega_{\rm epoch}$, means the sky can be covered more efficiently, increasing the survey speed. Increasing the total duration of the observation, $T$, increases the sensitivity to phenomena on the timescale of the observations, and increasing the time resolution, $\Delta T$, broadens the range of timescales that can be explored by the survey. 

However, there are limitations to the metric in Equation~\ref{e_merit}, especially when comparing surveys with very different telescopes at very different frequencies over very different timescales.  

Firstly, $A$ is not an ideal way to fully parameterise the sensitivity of a telescope.  Better would be $A/T_{\rm sys}$ (where $T_{\rm sys}$ is the system temperature, including both the receiver temperature and the sky temperature, and $A$ could also include the antenna efficiency $\eta_A$ to define the effective area $A_{\rm eff}\equiv\eta_A A$).  Alternately, especially when considering surveys done at a range of timescales, more appropriate might be the delivered RMS noise $\sigma$, which can include contributions from the system-equivalent flux density (related to $A/T_{\rm sys}$; e.g. \citealt{crane_sensitivity_1989,wrobel_sensitivity_1999,lorimer_handbook_2012}) but also things like confusion noise, and which will depend on the angular resolution and the timescale probed ($\sigma(\Delta T) \propto \Delta T^{-1/2}$ typically, if not dominated by confusion).  This can be further modified to account for varying sensitivity as a function of observing frequency. Since most of the sources considered here have non-thermal spectra (the overall population will have a median spectral index of $\alpha\approx -0.7$ or lower) we will use $\sigma (\nu/\nu_0)^{-\alpha}$ comparing to some reference frequency $\nu_0$ (this is also done elsewhere, as in \citealt{best_lofar_2023}).  And unlike $A$ or $A/T_{\rm eff}$, which we seek to maximise, smaller values of $\sigma$ are better.

Secondly, the survey area is hard to compare between telescopes that have very different angular resolutions.  For example, some low-frequency dipole arrays (e.g. the Long Wavelength Array) have essentially instantaneous fields-of-view of $2\pi$\,sr, but might have limited resolutions of a few arcmin or worse changed (which can lead to a higher confusion limit), while others (e.g. LOFAR) have intrinsic fields-of-view of $2\pi\,$sr but are limited by beam-formed to focus on smaller areas.  This compares with traditional dish-based interferometers at higher frequencies (e.g. the VLA and the Australia Telescope Compact Array; ATCA) that have fields-of-view of arcmin$^2$ but resolutions of arcsec.  Therefore instead of just $\Omega_{\rm epoch}$ we use $\Omega_{\rm epoch}/\theta^2$, with $\theta$ the angular resolution.  This effectively normalises to the rough number of independent pixels (ignoring oversampling) or potential  sources in a pointing.  For a traditional interferometer this reduces to roughly $B^2/D^2$, where $B$ is the maximum baseline and $D$ the element diameter, but systems like phased-array feeds \citep{hotan_australian_2021} can increase this significantly.

Finally, rather than use $T/\Delta T$ separately we just use the number of epochs, $N_{\rm epochs}$, which has the same value.
Putting these together we choose:
\begin{equation}
    \label{e_merit2}
{\rm FOM} \equiv    N_{\rm epochs}(\Delta T) \left(\frac{\Omega_{\rm epoch}}{\theta^2}\right) \left(\frac{\nu}{\nu_0}\right)^\alpha \frac{1}{\sigma(\Delta T)} 
\end{equation}
as the figure-of-merit that a survey should maximise,
where we retain the explicit dependence of $N_{\rm epochs}$ and $\sigma$ on the time-scale probed since the same data can be processed in multiple ways, although this assumes that there are sufficient sources on the relevant timescales to make this reprocessing worthwhile.  This functions like an effective survey volume, multiplying the number of epochs times the number of sources divided by the noise. It could in principle be further modified to account for the total physical volume probed and various source number density scalings (such as $N(\geq S_\nu)\propto S_\nu^{-3/2}$ for standard candles in a Euclidean universe), which would adjust the relative importance of going wider versus deeper, but we do not do that here.  Such explorations can be done, but are best tied to individual source types and do not generalise to the entire field as well.

Our figure of merit here expands on the commonly used survey speed metric (the number of square degrees per hour needed to achieve a given sensitivity; e.g. \citealt{johnston_science_2007}) which is more relevant for design of a telescope than a particular survey (indeed we revisit it in Section~\ref{s_future} when discussing future facilities).  Firstly, we divide the survey speed by the synthesized beam solid angle $\theta^2$ to allow comparison of widely different telescopes.  Secondly, we do an explicit frequency correction assuming a spectral index $\alpha$ to allow comparison of widely different frequencies.  But most importantly we explicitly consider the number of individual epochs and the timescale over which that sensitivity is achieved, to make sure that we probe variability on the timescales of interest.

\subsubsection{Comparing transient surveys}
In the ideal case, a survey that optimised all of the parameters in Equation~\ref{e_merit2} would be suitable for detecting and monitoring radio time domain phenomena across all timescales (which, as Figure~\ref{f_phase} shows, span many orders of magnitude). In practice, in any given survey to date, some aspects of this ideal were significantly compromised, limiting the effectiveness of the survey for particular phenomena. 
In addition, in many of the surveys conducted to date, it has not been possible to control the time sampling or integration times, because they relied on archival data \cite[e.g.][]{levinson_orphan_2002,bower_submillijansky_2007, bannister_22-yr_2011}.

Over the past two decades there have been a large number of radio transient surveys, which we discuss in Section~\ref{s_history}. Figure~\ref{f_fom} shows how the effective survey `volume' has evolved over time, using the figure of merit in Equation~\ref{e_merit2}. It is important to note that while this way of representing the surveys does distinguish between different sampling times $\Delta T$, it does not look at how sources vary over such timescales, as discussed below.

\begin{figure*}
    \centering
    \includegraphics[width=1.0\textwidth]{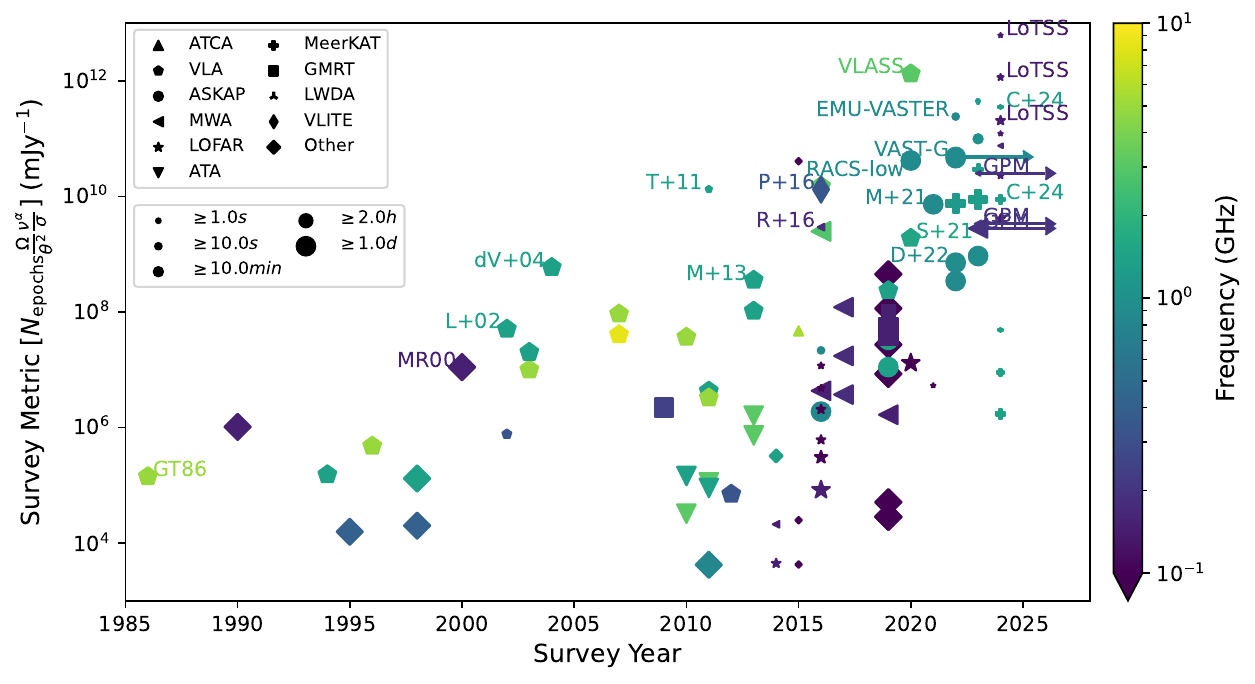}
    \caption{This plot shows how the effective survey volume, using the figure of merit in Equation~\ref{e_merit2} has evolved over time. Estimates are included for upcoming, but as yet unpublished, surveys. Some individual surveys of interest are labelled.  The surveys are coloured by observing frequency, the shape corresponds to the telescope, and the size corresponds to the survey timescale (if multiple results are reported for a single survey on different timescales, it is plotted multiple times).  Information has been updated from the survey compendium by Kunal Mooley and Deepika Yadav \url{http://www.tauceti.caltech.edu/kunal/radio-transient-surveys/index.html}.  Labelled surveys include: GT86 \citep{gregory_radio_1986}; MR90 \citep{mcgilchrist_7c_1990,riley_radio_1993}; MR00 \citep{minns_compact_2000}; L+02 \citep{levinson_orphan_2002}; dV+04 \citep{de_vries_optical_2004}; T+11 \citep{thyagarajan_variable_2011}; M+13 \citep{mooley_sensitive_2013}; R+16 \citep{rowlinson_limits_2016}; P+16 \citep{polisensky_exploring_2016}; S+21 \citep{sarbadhicary_chiles_2021}; M+21 \citep{murphy_askap_2021}; D+22 \citep{dobie_comprehensive_2022}; and C+24 \citep{chastain_commensal_2025}.  Additional surveys from Table~\ref{t_current} are RACS-low, VAST-G (for VAST-Galactic), VLASS, LoTSS, and GPM, as well as a projection for using the VASTER fast-imaging pipeline \citep{wang_radio_2023} on the EMU survey \citep{hopkins_evolutionary_2025}.} 
    \label{f_fom}
\end{figure*}

To compare the effectiveness of different surveys, a range of metrics have been used, most of which involve collapsing down the time dimension to form a `two-epoch equivalent' source density parameter, introduced by \citet{bower_submillijansky_2007}. For the two-epoch source density to provide a meaningful comparison, it requires the assumption that we do not know the characteristic timescale of variability of the underlying population/s, and hence a survey of 10 epochs spanning hours is equivalent to a survey of 10 epochs spanning months. Clearly this is not a valid assumption for any given, known source type: a survey with a total duration spanning hours would not be optimal for detecting gamma-ray bursts and supernovae, which evolve over months. However, the two-epoch source density has proved to be a useful tool in comparing surveys as the field has grown, from the four surveys listed by \citet{bower_submillijansky_2007} to the more than 50 that have been conducted to date, some of which are shown in Figure~\ref{f_fom}. In Section~\ref{s_history} we discuss the evolution of radio transient surveys over the past two decades. 

\subsubsection{Other survey properties}
Looking beyond the pure detection capabilities, survey design also affects the capacity for multi-wavelength follow-up, which is often critical for classification and identification of radio transients. A survey with two epochs separated by years \citep[such as the NVSS versus FIRST comparison conducted by][]{levinson_orphan_2002} can identify objects that have changed on those timescales, which is useful for estimating transient rates. However, confirming the nature of the individual objects (in that case, trying to distinguish between orphan afterglows, radio supernova and scintillating AGN) is made challenging by (a) the length of time after the transient event that the detection is made; (b) the lack of precision in identifying when the event occurred; and (c) the poor sampling of the event. So in addition to optimising the survey figure of merit, an ideal radio transients survey would also have:
\begin{itemize}
    \item A radio reference survey at a similar, or better, sensitivity;
    \item Sampling of a wide range of timescales;
    \item Rapid data processing to enable rapid identification; 
    \item Effective RFI and artefact rejection;
    \item Archival multi-wavelength survey data;
    \item Contemporaneous multi-wavelength observations;
    \item Capability to trigger follow-up with other telescopes.
\end{itemize}

Triggered follow-up is often important in identifying transients, particularly those identified at higher frequency wavebands (optical, X-ray) or with multi-messenger detectors. The most notable example of a comprehensive multi-wavelength follow-up campaign in recent years is that of the binary neutron star merger GW170817, detected in gravitational waves \citep{abbott_gw170817_2017}, with a counterpart identified in electromagnetic radiation \citep{abbott_multi-messenger_2017} and followed up in all wavebands from gamma-rays to radio, providing an incredibly detailed picture of the event \citep{kasliwal_illuminating_2017}. 

For many phenomena discovered in radio, follow-up is complicated by the fact that radio emission arrives later than X-ray, optical and other higher frequency emission due to the nature of the underlying physical processes. Taking GW170817 as an example, gamma-rays were detected 1.7s after the gravitational wave event \citep{goldstein_ordinary_2017}, optical/IR at 10.9 hours post-event \citep{coulter_swope_2017}, X-ray after 9 days \citep{troja_x-ray_2017}, and finally radio at 15 days after the event \citep{hallinan_radio_2017}. Had the radio been detected first, say as an orphan afterglow, triggers at other wavelengths would most likely have been too late to detect the event itself, or the subsequent time evolution of the emission.

For more short-lived phenomena, such as flares or bursts from radio stars, triggering from radio is also problematic, as it is rare that the burst is detected in real time, due to the data processing times. \citet{fender_prompt_2015} demonstrated that rapid triggering (in this case the Arcminute Microkelvin Imager Large Array (AMI-LA), operating at 15~GHz, responding to a Swift burst) can been successful in detecting stellar flares, but these examples are relatively uncommon. An ideal survey for detecting these phenomena would have simultaneous, or contemporaneous, multi-wavelength monitoring. 

There have been a number of successful coordinated monitoring programs. For example, \citet{osten_multiwavelength_2004} reports on a 4-year monitoring campaign of RS CVn system V711 Tau, spanning radio, ultraviolet, extreme ultraviolet, and X-ray; and \citet{macgregor_discovery_2021} reported an extreme flaring event on Proxima Centauri, detected during a joint monitoring campaign with radio, millimetre and optical telescopes. Of course several challenges of this approach are the resources required, and the difficulty of coordinating time availability on competitive-access telescopes. The ideal for future large scale projects would be dedicated facilities for conducting shadowing observations in other wavebands. A demonstrator of this approach is the MeerLICHT telescope \citep{bloemen_meerlicht_2016} that was built to shadow MeerKAT observations, hence providing optical lightcurves once a radio transient is identified \citep[e.g.][]{andersson_serendipitous_2022}.

\subsection{A brief history of radio transient surveys}\label{s_history}

Time variable behaviour has been observed since the early days of radio astronomy \citep[e.g.][]{hey_cosmic_1946,ryle_search_1951}. It was first observed in the closest, and brightest, radio sources: the Sun (see \citealt{wild_solar_1963}, for a review of early results on solar bursts, and \citealt{sullivan_cosmic_2009} for a more general history), Jupiter \citep{burke_observations_1955}, and supernova remnant Cassiopeia A \citep{hogbom_secular_1961}. 

There were a number of attempts to monitor supernova remnants and radio galaxies for longer timescale variability in the late 1950s and early 1960s. However, most found either no significant variability, or were inconclusive, due to limited sensitivity or poor time sampling. \citet{pauliny-toth_variations_1966} give a good summary of the research during this period. For example, the search for long period variability in radio `stars' (as most discrete radio objects were thought to be, before the discovery/identification of radio galaxies, and then quasars), by \citet{burke_observations_1955}, found that none of $\sim$100 sources monitored showed variability of greater than 10\% over 18 months. The discovery of quasars and the advent of synthesis imaging changed time domain radio astronomy significantly.

In this section we summarise the development of radio transient surveys in the image domain, as they evolved from discovery and monitoring of variable radio continuum sources, through to the current state of the field.  In doing this we acknowledge the very useful compilation of survey specifications originally provided by \citet{mooley_caltech-nrao_2016} and updated since, which we draw on here. We present the summary in largely chronological order, with some survey types (i.e. low and high frequency surveys, Galactic plane surveys, and circular polarisation surveys) treated separately for clarity (see Sections~\ref{s_low} to \ref{s_circ}).

\subsubsection{Targeted monitoring of stars and radio galaxies}

Early radio observations were not expected to be sensitive enough to detect radio stars (and here we mean actual stars), on the assumption that the stars were similar in brightness to our own Sun \citep{lovell_radio_1964}. However, the discovery that flare stars (for which UV Ceti is the prototype) had flares several orders of magnitude greater than the Sun in optical \citep{luyten_new_1949} saw renewed attempts to detect stellar radio emission. These attempts faced two main challenges: firstly the sensitivity of the telescopes, and secondly the difficulty of distinguishing flares from radio frequency interference \citep[see, for example, the discussion of stellar flare rates by][]{lynch_154_2017}. 

To unambiguously determine that observed flares were associated with a given star, combined optical and radio monitoring programs were carried out, resulting in simultaneous optical and radio flares detected on several stars including UV Ceti and V371 Ori \citep{slee_radio_1963, lovell_radio_1963}. Calculations from these observations implied a non-thermal emission mechanism, but the nature of this emission mechanism was not determined until much later. See \citet{hjellming_radio_1971} for a contemporaneous review of early radio star discoveries, and  \citet{dulk_radio_1985} for a review of stellar emission mechanisms.

The first quasar to be discovered and confirmed as extragalactic \citep[quasar 3C 273;][]{schmidt_3c_1963} was also one of the first extragalactic radio sources to have its variability characterised through monitoring \citep{dent_quasi-stellar_1965}. Over the next decade a number of teams conducted long-term repeated observations of radio galaxies and quasars from low to high radio frequencies \citep[e.g.][]{pauliny-toth_variations_1966, hunstead_four_1972, schilizzi_observations_1975}

The 1960s also saw the discovery of, and further searches for, interplanetary scintillation \citep[e.g.][]{hewish_interplanetary_1964,cohen_radio_1966}, which ultimately led to the discovery of pulsars \citep{hewish_observation_1968}. It was recognised from the start that IPS was a tool for measuring the angular size of extragalactic objects \citep[e.g.][]{cohen_ips_1967}.

In the 1980s, monitoring of extragalactic radio sources continued, including on shorter timescales \citep[e.g.][]{heeschen_rapid_1987} and lower frequencies \citep[e.g.][]{slee_survey_1988}. By the end of the decade it was a well established that a majority of the observed variability could be explained by interstellar scintillation \citep[e.g.][]{blandford_flicker_1986,rickett_radio_1990}. In other words, it was a propagation effect extrinsic to the source itself (Section~\ref{s_scint}).

Later work identified annual cycles in the observed variability due to interstellar scintillation \citep[e.g.][]{jauncey_intra-day_2001} caused by the relative velocity of the scintillation pattern as the Earth orbits the Sun. Eventually large surveys of compact extragalactic sources were conducted, most notably the Microarcsecond Scintillation-Induced Variability \citep[MASIV;][]{lovell_first_2003,lovell_micro-arcsecond_2008} survey. However, all of this work still involved targeted observations of individual objects. 

\subsubsection{The first imaging surveys for radio variability}

The advent of synthesis imaging meant it was possible to conduct untargeted surveys for variability, observing a (relatively) large sample of sources, with better flux stability and source localisation than what had previously been available. 

The first radio transient detected in a synthesis image was the X-ray binary Cygnus X-3, which was found in a Westerbork search for radio counterparts to X-ray sources \citep{braes_cygnus_1972} and which led to numerous followup studies \citep[e.g.][]{hjellming_radio_1972,hjellming_unusual_1972,gregory_discovery_1972,seaquist_highly_1974,molnar_low-level_1984}. About a decade later, in the 1980s, the first larger scale searches for radio transients and variability commenced.

At low frequencies, \citet{mcgilchrist_7c_1990}, followed by \citet{riley_radio_1993} and \citet{minns_compact_2000} conducted a $\sim$600\,\degsq\ survey over 10 years, at 151\,MHz, with the Cambridge Low Frequency Synthesis Telescope. \citet{riley_408-mhz_1995,riley_408-mhz_1998} conducted a 100~\degsq\ survey at 408\,MHz with the Dominion Radio Astrophysical Observatory Synthesis Telescope. In both cases, the sampling was roughly yearly, and these projects found that only a very small percentage of radio sources ($<1-2\%$) were highly variable at low frequencies, although that percentage was much higher for steep spectrum sources.  

The conclusions of these surveys were that most radio sources were not highly variable, and that most observed variability was consistent with scintillation. However, it is important to note that the sensitivity and limited time sampling meant that most timescales were not well explored.

\subsubsection{Widefield few-epoch searches}
The possibility of detecting orphan afterglows from gamma-ray bursts motivated the next phase of radio transient searches, including both dedicated surveys and uses of archival data. Rather than focusing on characterising variability, as the previous imaging surveys had done, this wave of surveys had the stated aim of finding explosive `transient’ objects. Here (and in Section~\ref{s_archival}), astronomers made use of existing single-epoch large-scale surveys. The properties of some of the key large-scale continuum surveys are summarised in Table~\ref{t_past}. By comparing sources between individual surveys, multi-epoch transient searches could be done. 

The first of these was \citet{levinson_orphan_2002}, who used two of the primary radio continuum reference surveys, Faint Images of the Radio Sky at Twenty centimeters \citep[FIRST;][]{becker_first_1995,white_catalog_1997} and the NRAO VLA Sky Survey \citep[NVSS;][]{condon_nrao_1998}, to search for isolated compact sources (with a flux density limit of 6\,\mjybeam\ at 1.4\,GHz) that were present in one catalogue but not detected in the other. They found nine candidate afterglows, and identified that the most likely confusing source types were radio-loud AGN and radio supernovae. Subsequent radio and optical follow-up of these candidates by \citet{gal-yam_radio_2006} found that five were not true variables, one was an AGN and one a known pulsar. One of the objects was likely to be a radio supernova, and the other remained unidentified, but the authors concluded it was unlikely to be a GRB afterglow. The analysis presented in this pair of papers provided a foundation for radio transient searches for the next decade.

Several 2--3 epoch VLA surveys followed, each exploring different timescales and sensitivities. \citet{carilli_variability_2003} explored timescales of 19 days and 17 months, at submillijansky sensitivities in the Lockman Hole region. \citet{de_vries_optical_2004} used a repeat observation of the FIRST Southern Galactic Cap zero-declination strip region to explore timescales of 7 years at millijansky sensitivities. Neither survey found any transients (objects that had appeared or disappeared between the epochs) and the authors set limits on the expected transient rate, as well as calculating the percentage of highly variable objects. 

In the following decade, \citet{croft_allen_2010} used the Allen Telescope Array Twenty-Centimetre Survey (ATATS) to do a two-epoch comparison with NVSS. They took the combined catalogue of 12 epochs of ATATS and compared to the NVSS, observed 15~years earlier. At the same time, \citet{bower_allen_2010} conducted a similar analysis with the ATA Pi GHz Sky Survey (PiGSS) with a smaller sky area but better sensitivity. Neither survey found any radio transients, and both put limits on the number of transients and variables expected in an all-sky survey. 

Most surveys in future years were much better sampled than these. However, there were a few more widefield few-epoch surveys, such as \citet{heywood_wide-field_2016} and \citet{murphy_search_2017}. Both of these found one candidate transient or highly variable object, but did not have enough information to identify the likely source class. 

These few-epoch transient surveys provided a good starting point for initial characterisation of the dynamic radio sky. However, it was evident that to do useful science, it would be necessary to have more observation epochs, better sensitivity, and timely analysis to enable follow-up.

\subsubsection{Exploring archival data and surveys}
\label{s_archival}
To get more information about the dynamic radio sky before future surveys became available, researchers delved deeper into archival datasets in order to exploit repeated observations that had been done for other purposes, for example calibration. The first example of this was by \citet{bower_submillijansky_2007} who searched 5 and 8.4\,GHz observations from the VLA data archives. The dataset consisted of 944 epochs of the same calibrator field, taken over 22 years. They found 10 transients, which had no optical counterparts \citep{ofek_long-duration_2010}, and were not able to be identified. Later re-analysis by \citet{frail_revised_2012} showed that most of these were likely to be imaging and calibration artefacts, but at the time the results motivated the exploration of a number of other archival searches.

\citet{bower_search_2011} built on this work to conduct a search using 1.4\,GHz observations of the same calibrator field (3C 286). Using more stringent criteria, they found no transient candidates, and calculated limits on the expected all-sky rate. \citet{bell_automated_2011} further expanded on the work, using an automatic pipeline developed for the LOFAR transients project \citep{fender_lofar_2008}. They analysed 5037 observations of seven calibrator fields, primarily at 5 and 8.4\,GHz. They also found no transient sources on timescales of 4--45 days, and reported new limits on the all-sky rate.

\citet{bannister_22-yr_2011} conducted an analysis of the Molonglo Observatory Synthesis Telescope \citep[MOST;][]{bock_sumss_1999} archive, consisting of 22 years of observations at 843\,MHz. Although most of the fields were only observed once as part of the Sydney University Molonglo Sky Survey \citep[SUMSS;][]{mauch_sumss_2003} and Molonglo Galactic Plane Surveys \citep{green_molonglo_1999,murphy_second_2007}, about $30\%$ of the fields were observed multiple times for calibration, data quality, or monitoring of specific objects. This search found 15 transients, some of which were identified with known objects (e.g. Nova Muscae 1991) and some of which remained unknown. 

\citet{thyagarajan_variable_2011} conducted a similar analysis of repeated observations from the FIRST survey. Each field was observed 2--7 times, and a large number of variable sources (but no transients) were identified. 
These projects demonstrated the challenges of working with an extremely inhomogeneous dataset in terms of time sampling, and the challenge of doing meaningful follow-up and analysis of transients that occurred many years before discovery. To make substantial progress, continuum surveys designed specifically for time domain science were required. 

\subsubsection{Multi-epoch radio transient surveys (2010s)}\label{s_multiepoch1}
The 2010s saw the emergence of designed-for-purpose multi-epoch radio transient surveys. Some of the first of these were conducted with the Allen Telescope Array \citep[ATA;][]{deboer_allen_2006}, an instrument designed for SETI searches. \citet{croft_allen_2011} published results from the 12-epoch, 690\,\degsq\ survey (ATATS) at 1.4\,GHz. \citet{bower_search_2011} and \citet{croft_allen_2013} presented results from the PiGSS survey at 3\,GHz. Neither of these surveys detected any transients (which they defined as varying by a factor of $>10$). However, they set more stringent limits on transient rates. Unfortunately, due to funding changes, radio imaging surveys did not continue after these initial results, and the full potential of the array as a transient machine was not realised. 

There were also a set of relatively small area (0.02 to 60~deg$^2$) multi-epoch surveys conducted with the VLA \citep[e.g.][]{ofek_very_2011,hodge_millijansky_2013,mooley_sensitive_2013,alexander_new_2015} and the ATCA \citep[e.g.][]{bell_search_2015,hancock_radio_2016}. Only one of these, \citet{ofek_very_2011}, detected a transient object, which, although not unambiguously identified, they state is consistent with being from a neutron star merger. These more sensitive surveys, with multiple epochs, were able to be used to start characterising the variability of the radio source population beyond a single numerical metric, for example looking at how variability changed with flux density, timescale and Galactic latitude. \citet{radcliffe_insight_2019} was notable for exploring the variability of radio sources at \ujy\ sensitivities (and finding the rates comparable to those at mJy sensitivities).

The re-evaluation of the \citet{bower_submillijansky_2007} transient results and the subsequent reduction of the derived rates by \citet{frail_revised_2012}, motivated \citet{aoki_reliability_2014} to revisit the rates that had been reported from the 1.4\,GHz Nasu Sky Survey. Although not strictly an imaging survey (transients were detected directly in drift scan fringe data), 11 extreme transients had been reported over a number of years \citep[see][and others]{kuniyoshi_strong_2007,niinuma_3_2007,matsumura_closely_2009}. The derived transient rate was inconsistent with other rates reported at the time. However, in the re-analysis, only a single transient passed the more rigorous thresholds applied by \citet{aoki_reliability_2014}, giving a rate more consistent with other projects.

The first results from the Australia SKA Pathfinder Boolardy Engineering Test Array \citep[ASKAP-BETA;][]{hotan_australian_2014} came in 2016, with \citet{heywood_wide-field_2016} and \citet{hobbs_pilot_2016}  publishing transient surveys covering timescales of days and minutes to hours, respectively. These surveys used the 6-dish ASKAP test array, operating at $\sim 1$\,GHz. This work was followed by \citet{bhandari_pilot_2018} who conducted a pilot survey at 1.4\,GHz using a 12-dish array, in the Early Science phase of ASKAP operations. None of these surveys detected any transient sources, but provided the first demonstration of ASKAP’s capabilities for transient science. In particular, the prospect of imaging of long integration observations on short timescales \citep{hobbs_pilot_2016}. 

A big step forward came with the Caltech-NRAO Stripe 82 Survey \citep[CNSS;][]{mooley_caltech-nrao_2016}, a multi-year, sensitive (rms $\sim 40\,$\ujybeam) survey of the 270\,\degsq\ Stripe 82 region. The survey was conducted with the VLA at 3\,GHz, and the Stripe 82 region was chosen for its existing multi-wavelength coverage, to improve the chances of identifying transients. By many measures this could be considered the most successful radio transient imaging survey to date, at that time. In addition to characterising the variability of the 3\,GHz source population, the initial pilot survey over $50\,{\rm deg}^2$ detected two transients serendipitously, which were both identified —- as an RS CVn binary and a dKe star.  Further studies discovered a tidal disruption event \citep{anderson_caltech-nrao_2020} and identified state changes in AGN \citep{kunert-bajraszewska_caltech-nrao_2020,wolowska_caltech-nrao_2021}. 

Another novel aspect of CNSS was a contemporaneous optical survey carried out with the Palomar Transient Factory \citep{law_palomar_2009}. This allowed a comparison of the dynamic sky in optical and radio wavebands; the main finding being that they are quite distinct, something that needs to be considered in planning for complementarity with, for example, the SKAO and Rubin Observatory \citep{ivezic_lsst_2019} transient surveys. 

\subsubsection{Multi-epoch radio transient surveys (2020s)}\label{s_multiepoch2}
By the start of the 2020s, roughly 50 untargeted radio transient imaging surveys had been conducted using both dedicated observations and commensal data obtained for other purposes, and yet only a handful of radio transients had been discovered via those surveys, most of which had not been conclusively classified. The main reasons for this were (i) the lack of surveys that were able to optimise the metrics presented in Equations~\ref{e_merit} and \ref{e_merit2} — all were limited in either sensitivity, sky coverage, or time sampling; and (ii) processing limitations meant that in many cases candidates were identified long after the event, limiting the potential for multi-wavelength follow-up.

The promise of large scale, well-sampled radio transient imaging surveys was finally (partially) realised, with the arrival of three SKAO pathfinder telescopes: ASKAP, MeerKAT and LOFAR, along with the VLA and MWA. The specifications for the major transient surveys on these telescopes are given in  Section~\ref{s_current}. In this section we will summarise some of the key results that have been published to date.

On ASKAP, the dedicated Variables and Slow Transients survey \citep[VAST;][see Section~\ref{s_vast}]{murphy_vast_2013} has both a Galactic and extragalactic component. The pilot survey, conducted at 888\,MHz \citep{murphy_askap_2021}, consisted of 12 minute observations (with typical sensitivity of 0.24\,\mjybeam), repeated with cadences of 1\,day to 8\,months. This survey detected 28 highly variable and transient sources including: known pulsars, new radio stars, sources associated with AGN and galaxies, and some sources yet to be identified. A pilot survey search of the Galactic Centre by \citet{wang_pilot_2022} resulted in the detection of a new GCRT (see Section \ref{s_lpts}).

\citet{dobie_askap_2019,dobie_comprehensive_2022} conducted 10 observations of a single ASKAP field (30\,\degsq) which covered 87\% of the localisation area of the gravitational event GW190814 localisation area. They reported a single synchrotron transient, which was not associated with the neutron star-black hole merger. Further multi-wavelength observations were conducted, but are so far inconclusive (D. Dobie, {\it private communication}). Reprocessing of a subset of this data on 15-minute timescales resulted in the discovery of 6 rapid scintillators in a linear arrangement on the sky \citep{wang_askap_2021}. These were used to constrain the physical properties of a degrees-long plasma filament in our own Galaxy. 

\citet{wang_radio_2023} extended this short-timescale (15\,min) imaging approach to 505 hours of data from several ASKAP pilot surveys, with typical frequency $\sim 1$\,GHz. Their search found 38 variable and transient sources, which is enough to start characterising the fraction of different source types on these timescales. They found 22 AGN or galaxies, 8 stars, 7 (known) pulsars and one unknown source, somewhat similar to the population found by \citet{murphy_askap_2021}. 

On MeerKAT there have also been several surveys that sample a range of timescales, often resulting from data focused on monitoring other sources \citep[e.g.][]{driessen_mkt_2020,driessen_detection_2022,andersson_serendipitous_2022}. \citet{rowlinson_search_2022} searched data from a monitoring campaign of MAXI J1820+070. There were 64 weekly observations with a typical integration time of 15 minutes and typical sensitivity of 1\,\mjybeam. This resulted in 4 AGN, a pulsar and one unidentified object. \citet{driessen_21_2022} searched data from weekly observations of GX~339$-$4 and identified 21 variable sources, most of which have low-level variability consistent with refractive scintillation from AGN.
On shorter timescales, \citet{fijma_new_2024} re-imaged 5.45 hours of MeerKAT data, centred on NGC 5068, on timescales of 8\,s, 128\,s, and 1\,hr, but did not find any transients.

On the VLA, the CHILES Variable and Explosive Radio Dynamic Evolution Survey \citep{sarbadhicary_chiles_2021} was designed to study radio variability in the extragalactic COSMOS deep field. It had 960\,h of 1--2\,GHz data, observed over 209 epochs, with typical sensitivity of $\sim10\,$\ujybeam. They found 18 moderately variable sources, all of which were identified as AGN or galaxies.

The VLA Sky Survey \citep[VLASS;][]{lacy_karl_2020} is underway and producing results (see Section~\ref{s_vlass}) but no general transient survey paper has been published yet. At low frequencies, the MWA Galactic Plane Monitor and LOFAR transient surveys are discussed in Sections~\ref{s_mwagpm} \& \ref{s_lotss}.

We have now reached a regime where radio transient imaging surveys are detecting significant numbers of transient and highly variable objects. Equally importantly, they have sufficient time sampling and resolution, along with multi-wavelength information, to allow the class of most of the objects to be identified. We discuss these populations further in Section~\ref{s_rates}. 

\subsubsection{Low frequency surveys}\label{s_low}
In the decades leading up to the mid-2010s, a majority of radio transient imaging surveys were conducted at gigahertz frequencies, predominantly with the VLA. There were a small number of exceptions, for example \citet{lazio_surveying_2010} conducted a survey at 73.8 MHz with the Long Wavelength Demonstrator Array (LWDA), finding no transients above 500\,Jy, and \citet{jaeger_discovery_2012}, who found one (unidentified) transient source in a search of 6 archival VLA fields observed at 325\,MHz. 

However, the advent of three new wide-field, low-frequency telescopes --- the Long Wavelength Array \citep[LWA1;][]{ellingson_lwa1_2013}, the Murchison Widefield Array \citep[MWA;][]{tingay_murchison_2013} and the Low Frequency Array \citep[LOFAR;][]{van_haarlem_lofar_2013} --- opened up the possibility of more fully characterising the time domain radio sky at $< 300$\,MHz. 

The LWA1 enabled very widefield surveys at low frequencies (10 -- 88\,MHz), but with relatively poor sensitivity (10s --100s of Jy). Surveys by \citet{obenberger_detection_2014,obenberger_monitoring_2015} detected fireballs from meteors, but no transients originating outside the Solar System. 

The next set of results were from early surveys with the MWA and LOFAR. \citet{bell_survey_2014} conducted a survey on timescales of 26\,minutes and 1\,yr with MWA at 154\,MHz. They found no transients above 5.5\,Jy. \citet{carbone_new_2016} reported on a LOFAR survey of 4 fields, observed on cadences between 15\,min and several months. The frequency of the observations were between 115 and 190\,MHz. They found no transients above 0.3\,Jy. \citet{stewart_lofar_2016} reported results from a 4 month monitoring campaign of the Northern Galactic Cap as part of the Multifrequency Snapshot Sky Survey. The survey consisted of 2149 11-min snapshots, each covering 175\,\degsq. They reported the discovery of one transient source, with no optical or high-energy counterpart. Despite an extensive investigation, the source class was not identified, and future attempts to discover more sources from this hypothetical population were unsuccessful \citep[e.g.][]{huang_matched_2022}.

The \citet{stewart_lofar_2016} transient increased the motivation for low frequency transient surveys. However, the next set of surveys yielded no new candidates. These included:
\citet{rowlinson_limits_2016,feng_matched_2017,tingay_multi-epoch_2019,kemp_image-based_2024} all with the MWA at 150--190\,MHz; \citet{polisensky_exploring_2016} with the Jansky Very Large Array Low-band Ionosphere and Transient Experiment (VLITE) survey at 340\,MHz; \citet{anderson_new_2019} taking advantage of the wide field of view of the Owens Valley Radio Observatory Long Wavelength Array (OVRO-LWA) at $<$100 MHz and \citet{hajela_gmrt_2019} with the GMRT at 150 MHz. Even with improved sensitivity, a significant population of low frequency transients remained elusive. 

The lack of radio transients at low frequencies is partly expected from predictions of synchrotron transient rates \citep[e.g.][]{metzger_extragalactic_2015}. As emission from, for example, GRBs evolves, the brightness decreases as the spectral peak moves to lower frequencies and the evolution slows, meaning they are less likely to be detectable given the sensitivity of low frequency telescopes \citep{ghirlanda_grb_2014}. Recently there has been a shift in focus towards searching for shorter timescale Galactic transients: these often have coherent emission mechanisms, leading to much faster variability. For example, \citet{de_ruiter_transient_2024} present an analysis of the LOFAR Two-metre Sky Survey \citep[LoTSS;][]{shimwell_lofar_2017} on 8-second timescales, which resulted in the detection of a new magnetic white dwarf M-dwarf binary system \citep{de_ruiter_sporadic_2025}. See Sections~\ref{s_mwagpm} and \ref{s_lotss} for work currently underway at low frequencies. 

\subsubsection{High frequency surveys}\label{s_high}
All of the surveys discussed so far have been conducted at or below frequencies of a few gigahertz. At higher radio frequencies (and in the millimetre regime), telescope fields of view are much smaller. For example, the field of view of the Australia Telescope Compact Array is $\sim2$\,arcminutes at 20\,GHz, and the field of view of the Atacama Large Millimeter Array \citep[ALMA;][]{wootten_atacama_2009} is $\sim19\arcsec$ at 300\,GHz. As a result, there are very few large scale continuum surveys at frequencies $>10$\,GHz. Two examples are the 9C survey \citep{waldram_9c_2003} covering 520\,\degsq\ at 15\,GHz and the AT20G survey \citep{murphy_australia_2010} covering 20\,626\,\degsq\ at 20\,GHz. Two conclusions from these projects were that 
(i) the overall properties of the source population at high frequencies can not be extrapolated from the $\sim1$\,GHz population;
(ii) the high frequency radio sky has a comparable level of variability to gigahertz frequencies: \citep[$\sim 5\%$ of bright sources were variable on $1-2$ year timescales;][]{sadler_properties_2006}.

The challenges of conducting large-scale high frequency radio surveys means that there have been a limited number of studies of high frequency radio variability. \citet{bolton_15-ghz_2006} selected 51 sources from the 9C survey for further monitoring. The motivation for 9C was to identify foreground contaminants for cosmic microwave background (CMB) maps; and variability could impact the completeness of such a sample. Monitoring over 3 years confirmed that about half of the steep spectrum objects (likely to be compact young radio galaxies), and none of the flat spectrum objects, were highly variable. There are various other high frequency monitoring programs in which sources have been selected from surveys \citep[e.g.][]{franzen_follow-up_2009}, however there have not been any untargeted multi-epoch imaging surveys, as such.

At even higher (millimetre) frequencies there have been a relatively small number of transient surveys using bolometric arrays on single-dish telescopes. Several of these are with the South Pole Telescope
\citep[SPT;][]{carlstrom_10_2011}. \citet{whitehorn_millimeter_2016} searched for transients in the SPT 100 Square Degree Survey at 95 and 150\,GHz, finding a single transient consistent with an extragalactic synchrotron source. \citet{guns_detection_2021} searched at 95, 150, and 220 GHz, resulting in the detection of 15 transient events including stellar flares and possible AGN activity. Other examples are \citet{chen_long-term_2013} who searched for transients using the combined Wilkinson Microwave Anisotropy Probe and Planck epochs at 4 bands between 33 and 94\,GHz; \citet{naess_atacama_2021}, \citet{li_atacama_2023} and \citet{biermann_atacama_2024} using the Atacama Cosmology Telescope at 77 to 277\,GHz. 

We note that the majority of these surveys have been designed for cosmological purposes and hence have focused on high Galactic latitudes,   where the transient targets are largely extragalactic synchrotron transients (although the actual discoveries often include Galactic stars).  Therefore they have excluded low-latitude sources like X-ray binaries, which can show high levels of variability at millimetre frequencies associated with state transitions \citep[e.g.][]{fender_cygnusx1_2000,russell_rapid_2020}. Future low-latitude surveys will doubtless identify more Galactic sources. 

We will not discuss millimetre transient surveys further in this review. An analysis of the expected populations of extragalactic transients detectable in future CMB surveys (including long and short GRBs, tidal disruption events and fast blue optical transients) is given by \citet{eftekhari_extragalactic_2022}.

\subsubsection{Galactic Centre and Galactic plane surveys}\label{s_gal}

Parallel to the efforts to characterise variability in the extragalactic radio sky, there were a number of surveys of the Galactic Centre, Galactic bulge and the Galactic plane more generally. These surveys were motivated by the dense stellar and compact object environment in the plane (and centre) of our Galaxy, making it a likely rich source of transient events \citep[e.g.][]{muno_deep_2003,morris_galactic_1996,calore_radio_2016}.   
However, radio imaging of Galactic regions comes with additional challenges, including (a) producing high quality images in regions of significant diffuse emission; and (b) identifying multi-wavelength counterparts when the optical and infrared source density is high, and there is substantial extinction. 

One of the first Galactic surveys was a 5-year monitoring program of the northern Galactic plane, conducted between 1977 and 1981 with the NRAO 91m transit telescope, operating at 6\,GHz \citep{gregory_radio_1981,gregory_radio_1986}. This survey explored timescales from days to $\sim1$ year. A key motivation was to find more sources like the high mass X-ray binary Cyg X-3, which had recently been discovered to have large radio outbursts \citep{gregory_discovery_1972}. About 5\% of the 1274 sources monitored showed significant variability on either short or long timescales. The authors were not able to identify any of these objects, but determined that most were likely to be extragalactic.

The next major long-term Galactic centre monitoring program was conducted by \citet{hyman_low-frequency_2002}. This project used archival and new widefield images of the Galactic centre taken with the VLA at 0.33\,GHz, with targeted follow-up of objects of interest at 1.4 and 4.8\,GHz. They introduced the nomenclature Galactic Centre Radio Transient (GCRT) to describe these sources, as their nature was unknown. Subsequent work by \citet{hyman_powerful_2005,hyman_new_2006, hyman_faint_2007,hyman_gcrt_2009} resulted in the discovery of three GCRTs. These sources shared some common properties (highly variable, steep spectrum, circularly polarised) but their identity has remained elusive (see Section~\ref{s_lpts}).

The challenges presented by imaging near the Galactic region, and difficulty of identifying potential transients has limited the number of surveys with a Galactic focus. 
\citet{becker_variable_2010} searched for transients in three epochs of the Multi-Array Galactic Plane Imaging Survey, a collection of VLA observations conducted at 4.8\,GHz. The observations spanned 15 years, with a sky area of $23.2$\,\degsq. They found no transients, and a number of highly variable sources (at a higher source density than found by extragalactic surveys at that time). However, they were not able to conclusively identify their object class.

\citet{williams_asgard_2013} conducted a two-year survey with the Allen Telescope Array at 3\,GHz, with weekly visits to 23\,\degsq\ in two fields in the Galactic plane. This was a substantial improvement on previous surveys, but did not detect any transient sources, pointing to lower frequencies potentially being more productive for Galactic transient discovery.   
The capabilities of SKAO pathfinder and precursor telescopes has seen a renewed interest in transient surveys of the Galactic centre and Galactic bulge \citep[e.g.][]{hyman_two_2021, wang_pilot_2022, frail_image-based_2024} as well as the re-exploration of archival data \citep[e.g.][]{chiti_transient_2016}. Some of the large-scale Galactic plane surveys currently underway are outlined in Section~\ref{s_current}.

\subsubsection{Circular polarisation surveys}\label{s_circ}
A vast majority of radio sources are driven by synchrotron emission and so do not show significant levels of circular polarisation \citep{saikia_polarization_1988}. For example, AGN, by far the most populous objects in the radio sky, are typically $<0.1\%$ circularly polarised \citep[e.g.][]{rayner_radio_2000}. However, some highly variable radio source classes do have significant fractions of circularly polarised emission, including pulsars (Section~\ref{s_pulsar}), radio stars (Section~\ref{s_star}), and unusual transients like GCRTs and LPTs (Section~\ref{s_lpts}). 

Circular polarisation surveys provide an alternative way of discovering and identifying these objects. Historically, most large radio surveys have only produced total intensity (Stokes I) data products (see Table~\ref{t_past}). For example the Westerbork Northern Sky Survey \citep{rengelink_westerbork_1997}, SUMSS \citep{mauch_sumss_2003} and the GMRT 150 MHz All-sky Radio Survey First Alternative Data Release \citep{intema_gmrt_2017}. The NRAO VLA Sky Survey \citep{condon_nrao_1998} produced linear polarisation products (Stokes Q and U) but did not produce circular polarisation products. 

When circular polarisation is available, it provides a method for easily identifying many interesting objects \citep[e.g.][]{kaplan_serendipitous_2019,wang_discovery_2021,wang_discovery_2022} while filtering out the non-variable sources.  In the cases where sensitivity in Stokes I is limited by confusion, Stokes V can also often have better sensitivity. Finally, unlike linear polarisation no correction for Faraday rotation \citep{burn_depolarization_1966} is needed, which eliminates the need for multiple frequency channels and hence provides better sensitivity (cf.\ \citealt{brentjens_faraday_2005}) and easier processing.

The first all-sky circular polarisation survey was conducted using the MWA at 200\,MHz \citep{lenc_all-sky_2018}. An untargeted search of this data resulted in detections of Jupiter, some known pulsars, several artificial satellites, and the tentative detection of two known flare stars. Several large survey telescopes now produce circular polarisation products by default, enabling searches for new objects via their circularly polarised emission. For example, \citet{pritchard_circular_2021} detected 33 radio stars in the ASKAP RACS survey at 888\,MHz; and \citet{callingham_v-lotss_2023} produced a catalogue of 68 circularly polarised sources from V-LoTSS (the circularly polarised LOFAR Two-metre Sky Survey). These sources consisted of radio stars, pulsars, and several that remain unidentified. This is an area that the SKAO pathfinder and precursor surveys are opening up for discovery.

\begin{table*}
   \caption{Specifications of past large scale single-epoch radio continuum surveys, in order of observing frequency.}
   \label{t_past}
   \setlength{\tabcolsep}{2pt}
   {\tablefont
    {\tablefont\begin{tabular}{@{\extracolsep{\fill}}l *{9}{>{\centering\arraybackslash}p{1.5cm}}}
   \toprule
Survey & GLEAM & GLEAM-X & VLSS & TGSS & WENSS & SUMSS & MGPS-2 & NVSS & FIRST\\ \hline
Telescope & MWA & MWA-II & VLA & GMRT & WSRT & MOST & MOST & VLA & VLA\\ 
Frequency (MHz) & 72--231 & 72--231 & 74 & 150 & 325 & 843 & 843 & 1400 & 1400\\ 
Bandwidth (MHz) & 31 & 31 & 1.56 & 16.7 & 5 & 3 & 3 & 100 & 100\\ 
Resolution (arcsec) & 150$^{\rm a,c}$ & 66$^{\rm c}$ & 80 & 25$^{\rm a}$ & 54$^{\rm a}$ & 43$^{\rm a}$ & 43$^{\rm a}$ & 45 & 5.4\\ 
Sky coverage (\degsq) & 29\,300 & 30\,900 & 30\,939 & 36\,900 & 10\,300 & 8\,000 & 2\,400 & 33\,800 & 10\,600\\ 
Sky region & $\delta \leq +25\degrees$ & $\delta \leq +30\degrees$ & $\delta \geq -30\degrees$ & $\delta \geq -53\degrees$ & $\delta\geq +30\degrees$ & $\delta \leq -30\degrees$ & $|b|<10\degrees$, $l>245\degrees$ & $\delta \geq -40\degrees$ & Galactic caps\\ 
Sensitivity (\mjybeam) & 25$^{\rm d}$ & 1.3$^{\rm e}$ & 100 & 5.0 & 3.5 & 1.0 & 1.0--2.0 & 0.45 & 0.2\\ 
Stokes products &  &  & I & I & IQUV & I & I & IQU & \\ 
Reference & \citet{wayth_gleam_2015} & \citet{hurley-walker_galactic_2022} & \citet{cohen_vla_2007} & \citet{intema_gmrt_2017} & \citet{rengelink_westerbork_1997} & \citet{bock_sumss_1999} & \citet{murphy_second_2007}$^{\rm b}$ & \citet{condon_nrao_1998} & \citet{becker_first_1995}

\botrule
\end{tabular}}
\begin{tabnote}
{$^{\rm a}$ The resolution depends on the declination as $\csc |\delta|$.}\tnp
{$^{\rm b}$ There was an earlier epoch of MGPS \citep{green_molonglo_1999}. However, only images were released, not a source catalogue.} \tnp
{$^{\rm c}$ At 154\,MHz.}\tnp
{$^{\rm d}$ At 154\,MHz, roughly.}\tnp
{$^{\rm e}$ For the combined 170--231\,MHz image.}\tnp
\end{tabnote}}
\end{table*}

\subsection{Current large-scale radio transient surveys}\label{s_current}
There are a number of large image domain radio transient surveys underway (or recently completed). In this section we briefly describe their properties in the context of the history we have presented in Section~\ref{s_history}, and the future plans for the SKAO, ngVLA, DSA-2000 and other instruments, that we will discuss in Section~\ref{s_future}. The specifications of these current surveys are summarised in Table~\ref{t_current}, in order of increasing frequency. The specifications of major single-epoch radio continuum surveys are summarised in Table~\ref{t_past}, for comparison.

Note that we are not including the many smaller-scale transient searches (e.g. of various archival datasets) that are underway but not-yet-published. Also, our focus remains on imaging surveys, so we do not include other ongoing radio transients surveys such as CHIME \citep[e.g.][which primarily looks for dispersed signals in beamformed data]{chimefrb_collaboration_first_2021} and MeerTRAP \citep[e.g.][]{turner_discovery_2025}.

\begin{table*}
   \caption{Specifications of current large scale radio transient surveys in the image domain. See the main text for the key reference papers.}
   \label{t_current}
    {\tablefont\begin{tabular}{@{\extracolsep{\fill}}lccccccccc}
   \toprule
Survey & \multicolumn{2}{c}{\dotfill LoTSS \dotfill} & MWA GPM$^a$ & \multicolumn{2}{c}{\dotfill VAST \dotfill} & \multicolumn{3}{c}{\dotfill RACS$^{\rm b}$\dotfill} & VLASS	   \\
       &  & Deep & & Extragal & Gal & RACS-low & RACS-mid & RACS-high & \\
\hline 
Telescope & \multicolumn{2}{c}{\dotfill LOFAR \dotfill} & MWA & \multicolumn{2}{c}{\dotfill ASKAP \dotfill} & \multicolumn{3}{c}{\dotfill ASKAP \dotfill} & VLA   \\ 
Frequency (MHz) & 144 & 144 & 200 & 887.5 	& 887.5 & 887.5$^{\rm c}$ & 1295.5 & 1655.5	& 3000	 \\
Bandwidth (MHz) & 48 & 48 & 30 & 288 & 288 & 288 & 144 & 288 & 2000    \\
Resolution  & $6\arcsec$ & $6\arcsec$ & $45\arcsec$ & $12-24\arcsec$	& $12-24\arcsec$ &$15-25\arcsec$ & $\gtrsim8\arcsec$ & $\gtrsim8\arcsec$	& $2.5\arcsec$ \\
Sky coverage (\degsq) & 20\,627 & 30 & 2\,700--4\,200 & 12\,279	& 2\,402  & 34\,240	& 36\,200 & 35\,955	& 33\,885\\
Sensitivity (\ujybeam)	&2\,000 (1\,h) & 10& 10\,000 (5\,min)$^{\rm d}$& 220	& 250& 250 & 200	& 195	& 70	   \\
Polarisation & IQUV & IQUV& IV & IV & IV& \multicolumn{3}{c}{\dotfill IQUV \dotfill} & IQU  \\
N repeats & 1$^{\rm e}$& ongoing & ongoing& 21$^{\rm f}$ &120$^{\rm f}$ & 3 & 1 & 1 & 3.5    \\
Typical cadence	&  8\,s, 2\,min, 1\,h & months & $\sim$2 weeks & $\sim$2\,months	& $\sim$2\,weeks& \multicolumn{3}{c}{\dotfill 12--24\,months \dotfill}	& 32\,months	 \\
Total time span & N/A & N/A & ongoing & 4\,years & 4\,years & \multicolumn{3}{c}{\dotfill 4.5 years \dotfill}  & 10\,years  
\botrule
\end{tabular}}
\begin{tabnote}
    {$^{{\rm a}}$At the time of writing the MWA GPM survey paper has not been published, but has been briefly described in the Methods section of \citet{hurley-walker_long-period_2023}. \tnp The other MWA GPM specifications listed here have been provided by N. Hurley-Walker (private communication). The sky coverage changes throughout the year.} \tnp
    {$^{{\rm b}}$RACS can be considered as a 5-epoch survey at 3 comparable frequencies, see Section~\ref{s_racs}.}\tnp
    {$^{{\rm c}}$RACS-low3 was observed at a slightly different frequency, 943.5\,MHz.}\tnp
     {$^{{\rm d}}$The estimated sensitivity is 7--10\,\mjybeam\ on 5 minute timescales and 70--100\,\mjybeam\ on 4 second timescales.}\tnp
    {$^{{\rm e}}$Note that although LoTSS was conceived as single-epoch survey, it can be reprocessed on short timescales to allow for transient searches see Section~\ref{s_lotss}.}\tnp
    {$^{{\rm f}}$At the time of writing VAST is $50\%$ complete.}\tnp
    \end{tabnote}
\end{table*}

\subsubsection{The LOFAR Two-metre Sky Survey}\label{s_lotss}
The LOFAR Two-metre Sky Survey \citep[LoTSS;][]{shimwell_lofar_2017} is a 150\,MHz survey of the sky north of  $\delta=0^\circ$. Each LoTSS image is an 8-hour observation, resulting in a typical sensitivity of $83\,$\ujybeam\ and a resolution of $6\arcsec$, with full Stokes parameters. The most recent data release covers $27\%$ of the sky \citep{shimwell_lofar_2022}, with an upcoming data release expected to cover $70\%$ of the sky.\footnote{\url{https://lofar-surveys.org/}}

Although LoTSS is essentially a single-epoch continuum survey, we have included it here because there is a substantial short-time scale imaging project underway that will search for transients on timescales of 8\,s, 2\,min and 1\,h \citep{de_ruiter_transient_2024}. The discovery of a white dwarf binary system from this survey was reported by \citet{de_ruiter_sporadic_2025}.

In addition, there are other LOFAR surveys underway that will be useful for transient searches, for example the LoTSS Deep Fields which consist of hundreds of hours of repeated observations of three fields (ELIAS-N1, Bo{\"o}tes and the Lockman Hole), with observing blocks scheduled months to years apart \citep{best_lofar_2023,shimwell_lofar_2025}, with eventual sensitivities of $\approx 10\,$\ujybeam.

\subsubsection{The MWA Galactic Plane Monitor}\label{s_mwagpm}
The MWA Galactic Plane Monitoring program (MWA GPM; project code G0080)\footnote{\url{https://mwatelescope.atlassian.net/wiki/spaces/MP/pages/24969577/2022-A+and+2022-B+Extended}} is a transient survey being conducted at 185--215\,MHz. 
It covers the Galactic plane south of $\delta < +15^\circ$ with $|b|<15\degrees$,  with longitude coverage changing through the year and 30\,min integrations on a bi-weekly cadence. It has a resolution of $45\arcsec$ and a typical sensitivity of 10\,\mjybeam\ on 5\,minute timescales (N.~Hurley-Walker, {\it private communication}).

The main scientific goal of the survey is to discover long-period transients. 
The survey parameters are briefly described in the Methods section of \citet{hurley-walker_long-period_2023}, which reports the discovery of GPM~J1839$-$10, a new long period transient, and will be fully described in a paper in preparation. 
\subsubsection{The ASKAP VAST Survey}\label{s_vast}
The ASKAP Variables and Slow Transients survey \citep[VAST;][]{murphy_vast_2013,murphy_askap_2021} is one of nine key survey projects on the ASKAP telescope. The survey is being conducted at 887.5\,MHz, and produces total intensity (Stokes I) and circular polarisation (Stokes V) data products. There are two components to the survey: an extragalactic component covering 10\,954 square degrees, with typical sensitivity of $220\,$\ujybeam, and a Galactic component covering 2\,402 square degrees with a typical sensitivity of $250\,$\ujybeam. Note that the Galactic survey also includes two fields that cover the Large and Small Magellanic Clouds. 
The extragalactic fields are sampled with a median cadence of 2 months, and the Galactic fields are sampled with a median cadence of 2 weeks. 

The survey was designed to cover a wide range of science goals, and these goals have evolved over time since the original proposal \citep{murphy_vast_2013}. Results from the VAST pilot survey, a 200 hour test of the observing strategy, data quality and data processing approach have confirmed the survey is effective for a number of these goals. Two examples are the multi-epoch sampling of the radio star population \citep{pritchard_multi-epoch_2024}, and the discovery of unusual Galactic transients \citep[e.g.][]{wang_discovery_2021}.  

\subsubsection{The Rapid ASKAP Continuum Survey}\label{s_racs}
The Rapid ASKAP Continuum Survey \citep[RACS;][]{mcconnell_rapid_2020} was designed to produce a reference image of the entire sky visible to ASKAP ($\delta < +50^\circ$), that would help in calibration of all ASKAP surveys. The first epoch (RACS-low1) was conducted at 887.5\,MHz, starting in April 2019, with a resolution of 15\arcsec\ and typical rms sensitivity of 250\,\ujybeam. There have been five subsequent epochs, two in the higher ASKAP bands: RACS-mid1 at 1367.5\,MHz starting December 2020 \citep{duchesne_rapid_2023,duchesne_rapid_2024}; and RACS-high1 at 1665.5\,MHz, starting in December 2021 \citep{duchesne_rapid_2025}; and two repeats of RACS-low starting in March 2022 and December 2023. See  Table~\ref{t_current} for the detailed specifications of these surveys. 

The RACS data products include total intensity, linear and circular polarisation (Stokes I, Q, U, V). The repeated epochs were designed to allow exploration of the time variable sky. Early results have demonstrated the effectiveness for RACS for this. For example, in the detection of late-time radio counterparts for optically detected tidal disruption events \citep{anumarlapudi_radio_2024}.

\subsubsection{The Very Large Array Sky Survey}\label{s_vlass}
The VLA Sky Survey \citep[VLASS; ][]{lacy_karl_2020} is an all-sky survey conducted at $2-4$ GHz, with 2.5\arcsec\ resolution. It covers the entire sky above $\delta > -40^\circ$, with typical rms sensitivity of 120\,\ujybeam. It was designed to have a time domain component: one of the four science goals is `{\it Hidden Explosions and Transient Events}'. Hence the survey has three repeated epochs separated by $\sim32$ months, with a fourth epoch partially completed. The VLASS data products include total intensity and linear polarisation images (Stokes I, Q, U) but, at the time of writing, no public circular polarisation (Stokes V). 

The science goals for VLASS are outlined in \citet{lacy_karl_2020}. In the time domain, the survey is best suited to search for objects that vary on timescales of $\sim$months, such as gamma-ray burst afterglows, radio supernovae and tidal disruption events, although it will also detect bursts from radio stars and Galactic transients. Indeed, early results have demonstrated the survey’s effectiveness for finding synchrotron transients, for example: a merger-triggered core collapse supernova \citep{dong_transient_2021}, and a sample of radio-detected tidal disruption events \citep{somalwar_vlass_2025}.  However, the sparse sampling has meant that extensive followup is needed to fully characterise the discoveries.

\subsection{Trends and developments in survey design}
\label{s_trends}
As we look at the progression of surveys over time (see Figure~\ref{f_fom} and Table~\ref{t_current}) and consider the developments in the last $\sim 5$\,years (see Sections~\ref{s_multiepoch2} and \ref{s_current}), some trends emerge.  These include:
\begin{itemize}
    \item {\bf Wider surveys:} Some surveys go deep \citep[e.g.][]{sarbadhicary_chiles_2021}, but more surveys are going as wide as feasible \citep[e.g.][]{murphy_askap_2021,lacy_karl_2020} at the expense of depth.  The advantages this offers in terms of transient rates can be debated, but having brighter sources certainly makes followup and classification easier. Planned future surveys will be able to go both deeper and wider than currently possible \citep{adams_radio_2019}.

\item {\bf Galactic transients:} More surveys are concentrating on  Galactic sources \citep{murphy_askap_2021,frail_image-based_2024,hurley-walker_long-period_2023}, especially in the Galactic plane.  This represents a shift from the extragalactic synchrotron transients targeted in previous surveys \citep[e.g.][]{mooley_caltech-nrao_2016}, and is related to instruments that focus on lower frequencies ($\lesssim 1.4\,$GHz) where the synchrotron transients are slow and faint.  Instead, searches for faster, often coherent, sources in the Galactic plane \citep{wang_pilot_2022,pritchard_circular_2021} are becoming prominent.

    \item {\bf Polarisation products:} Surveys are increasingly producing full polarisation products \citep[e.g.][]{shimwell_lofar_2017,murphy_askap_2021}, and in particular, circular polarisation.  This works together with the targeting of Galactic sources mentioned above.
    
    \item {\bf Many epochs:} Surveys are increasing from 2--3 epochs that are only sufficient for  identification/discovery, to multiple ($\gtrsim 10$) epochs \citep[e.g.][]{murphy_askap_2021}, which enables source classification and probing of different timescales.
    
    \item {\bf Short-timescale imaging:} Even when a survey only has a small number of observing epochs (including single-epoch surveys), it is now possible to reprocess the data on multiple timescales to reveal faster behaviour \citep[e.g.][]{wang_radio_2023,de_ruiter_transient_2024}.  This approach is most relevant to Galactic sources \citep{de_ruiter_sporadic_2025}, and a major early driver was detecting fast radio bursts \citep[e.g.][]{tingay_search_2015}, but it has also revealed some scintillating extragalactic sources \citep{wang_askap_2021}.
\end{itemize}
    
The VLA Sky Survey \citep{lacy_karl_2020} stands apart from some of these trends, drawing on the success of the Caltech-NRAO Stripe 82 Survey \citep[][and following papers]{mooley_caltech-nrao_2016}.  It operates at 3\,GHz, consists of 3--4 epochs, and has yet to release circular polarisation data.  Note that unlike some other surveys, VLASS was designed not solely as a transients survey, but with a broad set of science goals, even though much of the justification for the multiple epochs do come from the time domain.  Perhaps because of these aspects, it has been very successful in discovering extragalactic synchrotron transients that other current surveys do not probe as well \citep{somalwar_candidate_2023,stroh_luminous_2021,dong_flat-spectrum_2023,dong_transient_2021,somalwar_vlass_2025,somalwar_vlass_2025-1}.

Overall, the combination of newer instruments (e.g. ASKAP, MeerKAT), newer techniques \citep{wang_radio_2023,fijma_new_2024,smirnov_tron1_2025} and new targets along with increasing investment of telescope time has allowed both the scope of the surveys to increase together with their yield.  We expect that future developments in instrumentation (SKAO, ngVLA, DSA-2000, LOFAR2.0) will allow this to continue into the next decade. These telescopes, and the prospective surveys that may run on them, are discussed in Section~\ref{s_future}.

\section{Detection methods and approaches}\label{s_detection}

In optical astronomy, the well-established method of finding transient sources in the image domain is via image subtraction \citep[e.g.][]{zackay_proper_2016}. This approach has been used in major projects, for example, the Palomar Transient Factory \citep{law_palomar_2009} and the Zwicky Transient Facility \citep{bellm_zwicky_2019}. However, the limited sampling of the $uv$-plane makes this less straightforward in radio astronomy, and in general, a catalogue cross-matching approach has proved more successful. Catalogue cross-matching is also used in some major optical projects, for example the Catalina Real-Time Transients Survey \citep{drake_first_2009}.

Regardless of the approach, calibration and imaging artefacts are a significant issue in radio transient surveys, and, as in optical surveys, can result in false `transients’. For example, many of the transient candidates reported by \citet{bower_submillijansky_2007} from their analysis of archival VLA data were later found to be artefacts, or the result of calibration errors \citep{frail_revised_2012}. 

In this section we review the main transient detection methods used in radio transient surveys, as well as some of the main approaches to follow-up, monitoring and source identification. 

\subsection{Image subtraction and difference imaging}
\label{s_differencing}
In the optical domain, image subtraction is a core part of most transient detection pipelines. The typical approach used is to form a reference image from the first $n$ observations of a given field (or from dedicated deep observations). Each subsequent observation of that field is then astrometrically aligned with the reference image using a catalogue of extracted sources, and convolved to a common resolution. The new image is subtracted from the reference image following \citet{alard_method_1998,alard_image_2000} or similar \cite[e.g.][]{zackay_proper_2016}. Sources that are unchanged are effectively removed, leaving sources that have appeared, disappeared or changed substantially in brightness between the observations. These transient sources are detected in the subtracted images using source finding software such as SExtractor \citep{bertin_sextractor_1996}. This approach (and variations on it) has been used in many major optical transients surveys \citep[e.g.][]{wozniak_difference_2000, riess_type_2004,law_palomar_2009,bellm_zwicky_2019}.

In radio images, the sparse sampling of the $uv$-plane results in artefacts of a number of kinds, including ghost sources \citep[e.g.][]{grobler_calibration_2014} and sidelobes near imperfectly modelled bright sources \citep[e.g.][]{polisensky_exploring_2016}. 
Hence when image subtraction is used, the resulting images are dominated by these artefacts, rather than genuine transient sources. As a result, although there was a lot of experimentation with image subtraction  in early radio transient surveys, it has had very limited use in published results. One example is \citet{varghese_detection_2019} who detect a transient using two Long Wavelength Array stations, via image subtraction, as described by \citet{obenberger_monitoring_2015}.

A variation on image subtraction that is only possible in radio astronomy is visibility differencing and imaging, also just called `difference imaging'. In this case, the subtraction can be done using the model visibilities, hence removing the need for the (generally time-consuming) image deconvolution stage. There are a few examples of this \citep[e.g.][]{law_realfast_2018} but it has not be widely used. 

A further variation is making a deep reference image and subtracting this continuum image from  snapshot images with much shorter integration times from the same data-set (this is possible  since the correlator integration time of 1--10\,s is typically much smaller than the image integration time). This is appropriate only for transients with small duty cycles such that they do not appear in the deep image, or significant variability, and has been used in a number of recent projects, for example \citet{wang_discovery_2021, wang_radio_2023,fijma_new_2024,smirnov_tron1_2025}. See also the discussion in Section~\ref{s_fast}.

In current (and future) surveys with improved $uv$-coverage, image subtraction and difference imaging are becoming feasible approaches for radio transient detection. In addition, if there is full control over the observation strategy, some of the above problems mentioned in this section may be mitigated. For example, if the images to be subtracted are observed at the same local sidereal time, it reduces the image of residual synthesised beam sidelobes (although changes in beam shape due to ionospheric refraction or differences in flagging may still be present). In practice, this is difficult to achieve, but should be considered in the design of current and future automatic schedulers. 

\subsection{Other direct image analysis techniques}
\label{s_techniques}
A technique related to difference imaging is variance imaging. In this approach, an image is created from either a frequency (or time) cube by computing the variance of each pixel across all frequency (or time) channels. This has been proposed as a way of detecting new pulsars via their interstellar scintillation \citep[e.g.][]{crawford_variance_1996}. This technique has not been widely used, but has discovery potential with the increased sensitivity and frequency resolution of current imaging surveys \citep[e.g.][]{dai_detecting_2016,dai_prospects_2017,morgan_interplanetary_2018}.

Another approach is the use of matched filters --- which are in general an optimal statistic for detection \citep[e.g.][]{helstrom_elements_1995} directly on image pixels. One component of the rms noise in radio images is the classical confusion noise, which comes from the background of faint, unresolved sources \citep{condon_confusion_1974}. For telescopes with low angular resolution, this can become a dominant factor in the total noise. In this `confusion-limited’ regime, matched filter techniques might be useful. Applying a matched filter involves searching for a signal of a known or approximate functional form (e.g. a top-hat), taking advantage of the fact that nearly all of the background sources contributing to the confusion noise are static with time. This approach is used widely in signal processing in engineering, and in gravitational wave detection \citep[e.g.][]{allen_findchirp_2012}. It has not been used widely in radio transient detection, but one example implementation is given by \citet{feng_matched_2017} for a transient search with the MWA. 

Matched filters are also useful when searching for a particular class of transients in catalogue data (rather than image cubes); for example, searching for gamma-ray burst afterglows \citep{leung_matched-filter_2023}. 
In cases like this, slowly evolving lightcurves mean the sources of interest may not be highly variable using traditional metrics, and hence a matched filter is more effective for discovery.  See Section~\ref{s_methods} for further discussion of this topic.

\subsection{Catalogue cross-matching}
A vast majority of radio transient imaging surveys use a transient detection approach that follows these basic steps:
\begin{enumerate}
\item {\bf Image} each epoch of data;
\item Run {\bf source-finding} software on each image to extract sources above a noise threshold;
\item {\bf Associate} sources from different epochs with each other via positional cross-matching;
\item {\bf Measure} upper limits in cases where a previously detected object is not detected;
\item Create {\bf lightcurves} for each detected object;
\item Use various metrics to {\bf identify} transients and highly variable objects.
\end{enumerate}
There are many variations on how each of these steps is implemented (or whether the step is done at all). Two key examples of this approach that have been formalised into software pipelines are the LOFAR Transients Pipeline \citep[TraP;][]{swinbank_lofar_2015} and the VAST Pipeline \citep{pintaldi_scalable_2022}. In an ideal survey, transient detection would work in real time, as part of the imaging processing pipeline; to allow immediate identification and follow-up of transients. However, to date, the computational challenges of rapid imaging have meant that most projects have done their image-based transient detection in an offline or batch mode. See, for example, recent results from ASKAP VAST \citep{murphy_askap_2021}, LOFAR \citep{kuiack_aartfaac_2021} and MeerKAT \citep{rowlinson_search_2022}.   

In the rest of this subsection we briefly describe each of the steps in the catalogue cross-matching approach, pointing the reader to the relevant papers for more detailed explanations. Discussion of radio interferometric imaging is beyond the scope of this paper, so we will assume the imaging has already been done on the relevant timescale. Transient and variability metrics will be discussed in Section \ref{s_metrics}.

\subsubsection{Source finding}
The goal of source finding software is to identify astronomical objects in radio images. The typical approach to this is to identify statistically significant collections of pixels (`sources’) and `extract’ them by fitting elliptical Gaussians to measure their properties: position, angular size and flux density (note that the more general radio source finding approach can consider more complex source shapes, but transients are assumed to be point sources, although they can be in complex regions).  

In optical astronomy, one of the most well-used source finding packages is SExtractor \citep{bertin_sextractor_1996}, and many large projects use their own custom software \citep[e.g.][]{masci_zwicky_2019} that works in a similar way. However, the needs for radio source finding can be both more complex and simpler than in the optical.  For instance, in the radio the typical noise properties of the images are different (adjacent pixels in a restored image are correlated in the radio, but not in the optical); images are usually restored with a single beam shape that is not subject to seeing variations (although there can be ionospheric effects and variations due to RFI flagging); there are often multiple simultaneous images to search (e.g. as a function of frequency or polarisation), and imaging artefacts are usually deconvolution sidelobes rather than halos, satellite trails, and cosmic rays. These limitations and differences led to the development of radio-specific source finding packages for continuum surveys. 

Source finding is a significant focus for radio continuum surveys as we prepare for the SKAO. See, for example, the results of the ASKAP Evolutionary Map of the Universe \citep[EMU;][]{norris_emu_2011,hopkins_evolutionary_2025} Source Finding Data Challenge \citep{hopkins_askapemu_2015} and the Square Kilometre Array Science Data Challenge 1 \citep{bonaldi_square_2021}. However, many of the most challenging aspects (such as characterising sources that are diffuse or have a complex morphology) are not as relevant for radio transient surveys where the objects of interest are expected to be intrinsically compact\footnote{There are cases, particularly at lower frequencies, where a compact variable object is embedded in extended or diffuse emission. For example, a pulsar within a supernova remnant, or the core of a radio galaxy within extended lobes.}. 

Nevertheless, there are some aspects of source finding of particular importance to radio transient surveys. For example:
\begin{itemize}
    \item Speed: if source finding is running as part of real-time processing then it needs to be computationally efficient without reducing reliability and completeness; 
\item Identifying sources with negative flux density: for example in the case of circular polarisation (Stokes V) images in which sources can have either sign of polarisation; 
\item Source finding in the $uv$-plane: an alternative to image-plane source finding that can be useful when rapid processing is needed, but requires modelled sources to subtract.
\end{itemize}

Some of the main source finding packages used in current radio transients surveys are: Selavy \citep{whiting_source-finding_2012} used in the ASKAP workflow; AEGEAN \citep{hancock_compact_2012, hancock_source_2018}, used in GLEAM-X \citep{hurley-walker_galactic_2022}; and PyBDSF \citep[Python Blob Detector and Source Finder,][]{mohan_pybdsf_2015}, used in RACS \citep{hale_racs_2021} and LoTSS \citep{shimwell_lofar_2019}. A summary of the main aspects of radio source finding, for example background noise estimation, island-finding and deblending (shared by many of these packages) is given by \citet{lucas_efficient_2019}. A comparison of the performance of some of these packages (with a focus on continuum survey science applications) is given by \citet{boyce_hydra_2023}.

\subsubsection{Source association}
Once sources have been identified and fit in each individual epoch, the next step is associating detections of the same object throughout multiple epochs. This is done via positional cross-matching. Both the TraP and VAST pipelines cross-match using the `de Ruiter radius' \citep{de_ruiter_westerbork_1977}: the angular distance on the sky between a source and its potential counterpart in the next epoch, normalised by the positional error of both sources.

To achieve optimal results in source association there are a number of issues to consider. These include how to deal with variable image quality between epochs and how to ensure unique matches in cases were multiple potential matches are identified. There are also computational challenges when scaling this up to large datasets with many repeated fields. These issues are discussed in more detail by \citet{swinbank_lofar_2015}. 

\subsubsection{Lightcurve creation}
Source association produces a lightcurve of flux density measurements with time for all epochs in which the source was detected. To make a more complete lightcurve most pipelines measure either (i) limits or (ii) forced fit parameters for each source in each epoch in which it is not detected \citep[this is also done for optical transient surveys, e.g.][]{masci_new_2023}. Limits are typically based on the local rms noise in the region of the source (a more computationally efficient but less accurate alternative is to use the overall rms noise of each image). Forced fitting involves fitting an elliptical Gaussian at the known position of the transient source. In both cases, the implementation approach depends on whether the pipelines are running in real time (and hence cannot access previously observed images, beyond a given buffer) or in an offline or batch mode. 

Subsequent analysis, such as applying variability metrics, searching for periodic behaviour, or doing automatic classification, is performed on the resulting lightcurves. 

\subsection{Short-timescale (`fast imaging') approaches}\label{s_fast}
As mentioned in Section~\ref{s_terminology}, almost all radio transient surveys to date have been designed to detect either `fast' (timescales $<$ seconds) transient behaviour, via single-dish/beamformed observations (with typical time resolution of 10s to 100s of microseconds), or `slow' (timescales $>$ seconds) transient behaviour via imaging (with maximum time resolution of $\sim 10+$ seconds). These two regimes have substantially different observational, technical and computational approaches. 

However, in recent years, the discovery of (a) very slow pulsars \citep[e.g.][]{tan_lofar_2018, caleb_discovery_2022}; (b) long-period radio transients  \citep[e.g.][]{hurley-walker_radio_2022,caleb_emission-state-switching_2024}, and (c) the need to localise fast radio bursts \citep[e.g.][]{chatterjee_direct_2017, bannister_single_2019} has motivated the exploration of the $100\,{\rm ms} - 10\,{\rm s}$ timescales  that fall between these two regimes. 

This missing region of parameter space has been approached from opposite directions. The first has been to take the fast transients approach, but incorporate imaging on (relatively) coarse timescales. These systems build on the success of single dish pulsar projects, by taking advantage of the localisation capabilities of interferometers. For example, the {\it realfast} system \citep{law_realfast_2018}, running on the VLA (and inspired by V-FASTR on the VLBA; \citealt{wayth_v-fastr_2011}), uses an alternative approach to traditional imaging, made possible by the low density of sources in any given field on timescales of $\sim 10\, {\rm ms}$. Unlike the image-subtraction below, this subtracts the mean visibilities on $\sim$\,s timescales to remove constant sources, creates `dirty' images for a range of integration time and dispersion measure, and then uses a matched filter to identify dispersed point sources for subsequent analysis. 
This approach resulted in the first localisation of a fast radio burst \citep{chatterjee_direct_2017} as well as identification of unusual bursts from a Galactic $\gamma$-ray source \citep{anna-thomas_unidentified_2024}.

More recently, \citet{wang_craft_2025} developed the Commensal Realtime ASKAP Fast
Transient COherent (CRACO) system, which is another example of a system that comes from a pulsar-inspired background. CRACO images on $110\,{\rm ms}$ timescales, with plans to increase this to higher resolutions. It works on-the-fly, and can only store a relatively small buffer of images for sources of interest. An early result from CRACO is the detection of a long period transient with a 44.3\,minute period \citep{wang_detection_2025}.

The second approach is to perform standard radio imaging, on much shorter timescales than the observation length. This has been impractical for most earlier radio surveys due to the limited $uv$-coverage of the instruments. Prior to the past year, there were only a small number of projects that did imaging transient searches on very short ($\sim10$~s) timescales \citep[e.g.][]{obenberger_monitoring_2015,kaplan_murchison_2015,tingay_serendipitous_2018}. However, the low angular resolution and relatively poor sensitivity of the instruments used in those studies (LWA and MWA) at the time these searches were conducted meant the ability to detect many transient phenomena was limited. 

The current set of interferometers, with a large number of dishes or antennas, and hence much improved instantaneous $uv$-coverage have made imaging (or re-imaging) on short timescales possible. 
One example of this approach is the fast imaging of MeerKAT data on 8\,s timescales by \citet{fijma_new_2024}. 
Another example is the VASTER fast imaging pipeline \citep{lee_searching_2026}  
which images ASKAP data on 10\,s and 10 min timescales. Both of these projects ran the fast imaging in offline, batch-mode.

A significant step towards online/realtime fast imaging is TRON (Transient Radio Observations for Newbies), which can image MeerKAT data on 8~second timescales \citep{smirnov_tron1_2025}. So far it has been used to search archival imaging data, resulting in the detection of millisecond pulsars \citep{smirnov_tron1_2025} and a stellar flare \citep{smirnov_tron2_2025}.
At the time of writing, work is underway to get TRON running in realtime on MeerKAT (I. Heywood, {\it private communication}) and VASTER is running in realtime on ASKAP (Y. Wang, {\it private communication}),  although results from these modes have not yet been published.

These early explorations have already produced some interesting results. The next step is to make these approaches more computationally efficient, and integrate them into the standard processing pipelines for current and future telescopes. 

\subsection{Multi-wavelength identification}
Radio data alone is rarely sufficient to allow the class of a given transient candidate to be identified. There are a few exceptions, most notably pulsars and fast radio bursts, although if they are discovered in imaging data, beamformed or single-dish observations are needed to confirm them (see \citealt{kaplan_serendipitous_2019} for a pulsar example and \citealt{bhandari_limits_2020} for an FRB example).

Typically, multi-wavelength data is required to confirm the source type of a radio transient. For example the gamma-ray counterpart of a GRB afterglow, the optical counterpart of a radio supernova, the optical/IR counterpart of a star, or the quasar responsible for observed scintillation. Even within the radio regime, multiple radio frequencies may be needed to understand the nature of a new transient (such as its spectral behaviour and evolution).  This multi-wavelength data may exist in archival observations, or may require dedicated follow-up observations. Related to this is dedicated imaging at higher angular resolution (which might require different instruments, different frequencies, or both). Such data can be vital for establishing precise positions for counterpart identification, such as in crowded regions of the Galactic plane \citep{caleb_emission-state-switching_2024} or within a host galaxy \citep{dong_flat-spectrum_2023}.

It is beyond the scope of this paper to discuss the many ways of identifying different classes of objects, and all their associated challenges and subtleties. However, we have summarised some of the common checks that are conducted in Table~\ref{t_mw}. This is intended as a high-level guide to common practice, rather than a detailed implementable method. 

The relatively small number of radio transients detected in surveys to date means that this exploration has been done manually. However, as the number of detections increases, automatic methods are becoming more important, and we discuss these in Section~\ref{s_ml}.

\begin{table*}[t!]
    \caption{A high-level summary of the types of multi-wavelength data and checks used to identify radio transient candidates.}
    \label{t_mw}
    {\tablefont \begin{tabular}{>{\raggedright\arraybackslash}p{1.2cm}
    >{\raggedright\arraybackslash}p{3.2cm}
    >{\raggedright\arraybackslash}p{3.5cm}
    >{\raggedright\arraybackslash}p{3cm}
    >{\raggedright\arraybackslash}p{5cm}} \toprule 
     Waveband & Check & What it tells us & Relevant classes & Notes \\
\hline
    Radio	& Detection in archival reference surveys	& Long-term variability, spectral index	& All	& Check for detections in NVSS, SUMSS, FIRST, RACS, GLEAM, TGSS, VLASS\ldots \\	
    Radio & Detection in archival targeted observations &  Long-term variability, spectral index, astrometry, shape& All &	Check for detections in VLA, ATCA, GMRT archival observations \\
    Radio & Linear and circular polarisation & Polarisation properties	& Stars, pulsars, LPTs & Linear polarisation may need Faraday rotation check\\
    Radio & Dynamic spectrum & Short-timescale variability, spectral structure & Stars, pulsars, LPTs & Need good signal-to-noise ratio \\
    Radio & Lightcurve periodicity & Whether source has periodic behaviour & Stars, pulsars, LPTs & 
    \\
    Radio & Cross-match with pulsar and FRB catalogues & Whether source is a known pulsar or FRB & Pulsars, FRBs & Sources can be poorly localised, need a careful choice of match radius$^{\rm a}$ \\
    Optical/IR	& Detection in archival reference surveys	& Possible counterpart or host galaxy	& All & \\
    Optical/IR	& Detection in telescope observing archives	& Possible counterpart or host galaxy	& All	& Check deepest observations available (e.g. ESO archives) \\
    Optical/IR & Cross-match with stellar catalogue & Possible stellar counterpart &	Stars & Cross-match with Gaia. Care is needed due to high false-positive rate$^{\rm b}$ and large proper motions \\ 
    Optical/IR & Cross-match with quasar catalogue & Possible AGN counterpart or host galaxy & AGN & Check SDSS, milliquas \\
    Optical/IR	& WISE colour-colour plot & 	Broad classification of source type	& Galaxies, AGN, stars & Need to check WISE counterparts are real associations \\
    Optical/IR & Get optical lightcurves & Whether sources is optically variable & All & Check \textit{TESS}, ZTF, etc. \\
    X-ray & Check X-ray source catalogues and archives$^{\rm c}$ & Possible counterpart & All &May need to include variability search within observations \\
    Gamma-ray &  \textit{Fermi} source catalogue & Possible counterpart & GRBs, pulsars, AGN & \\

    General & Cross-match with known objects & Whether source is already catalogued & All & Check for match in SIMBAD, NED, Vizier and other catalogues \\ 
    General & 	Cross-match with known transient events & Possible association with known event & GRBs, SNe, TDEs, XRBs, Novae &	Check GCNs, ATels, Transient Name Server$^{\rm d}$ \\

    General	& All solar system planet ephemerides & Whether source is a planet (or the Moon) &  		Solar system planets & Need accurate position and time of observation \\  

    \botrule
    \end{tabular}}
    \begin{tabnote}
        {$^{{\rm a}}$ The Pulsar Scraper includes both published, and announced-but-unpublished pulsars, and can be accessed at \url{https://pulsar.cgca-hub.org/}.} \\
        {$^{{\rm b}}$ See \citet{driessen_sydney_2024} for a full analysis of cross-matching with stellar catalogues, including how to select a suitable match radius.}\\
{$^{\rm c}$ NASA's High Energy Astrophysics Science Archive Research Center can be accessed at \url{https://heasarc.gsfc.nasa.gov/docs/archive.html}.}\\
        {$^{{\rm d}}$ The Transient Name Server can be accessed at \url{https://www.wis-tns.org/}}.
    \end{tabnote}
\end{table*}

\subsection{VLBI observations}\label{s_vlbi}
For the most part, observations using Very Long Baseline Interferometry (VLBI) function as a follow up technique for radio transients identified elsewhere.  VLBI can provide a critical role in determining the geometry and expansion rate of relativistic explosions (e.g. \citealt{taylor_angular_2004,bloom_possible_2011,mooley_mildly_2018,giarratana_expansion_2024}), separating the components of gravitational lenses (e.g. \citealt{spingola_radio_2016}), identifying jet ejection and kinematics in X-ray binary microquasars \citep[e.g.][]{dhawan_au-scale_2000,dhawan_kinematics_2007,stirling_relativistic_2001,miller-jones_rapidly_2019,wood_xrb_2021}, or eliminating star-formation regions from searches for FRB persistent radio sources (e.g. \citealt{bruni_discovery_2024}).  Similarly, for Galactic sources it can be used to determine parallax distances and help constrain energetics (e.g. \citealt{miller-jones_cygnus_2021}).

However, we do note that VLBA-like systems can be used for transient discovery, such as the VLBA Fast Radio Transients Experiment (V-FASTR; \citealt{wayth_v-fastr_2011}).  In this case it is searching for events like FRBs, evaluating incoherent sums between antennas on ms timescales. A detailed discussion of this approach is outside the scope of this review.

\subsection{Machine learning and classification}\label{s_ml}
The aims of transient searches are to: (a) identify, and study the nature of, individual objects; (b) understand populations of variable sources; or (c) use the transient objects as probes of our Galaxy or extragalactic space. To achieve any of these aims, classifying the source is critical. Typically this involves measuring radio properties of the source either from radio images or its lightcurve, and searching for multi-wavelength associations. Doing this work manually is laborious, and consists of making decisions based on an extensive list of parameters and rules --- some of which are strictly deterministic and others of which require judgement. These factors make transient source classification an ideal task for supervised machine learning. 

Supervised machine learning is the use of algorithms that improve automatically through experience, to classify data \citep{mitchell1997machine}. The basic methodology involves extracting features (measuring properties) from the dataset; then training a model that maps these features to a set of known classes. The model is evaluated using metrics such as precision, recall and overall accuracy. When suitable performance is achieved, the model can then be applied to new datasets.  

Machine learning has been widely used in astronomy: see \citet{ball_data_2010}, \citet{fluke_surveying_2020} and \citet{huertas_ml_2023} for overviews, and see \citet{buchner_how_2024} for a useful how-to guide for beginners. A variety of algorithms have been applied to time-domain astronomy, tackling a wide range of classification tasks. For example, the classification of variable stars using random forests \citep{richards_machine-learned_2011}, photometric supernova classification \citep{lochner_photometric_2016} and separating real from bogus events in the Zwicky Transient Facility \citep{duev_real-bogus_2019}. 

Most of the work in machine learning classification of radio transients has focused on the fast transients domain, where there are many candidates, significant computational challenges, and a need for real-time (or rapid) detection \citep[e.g.][]{wagstaff_machine_2016,agarwal_fetch_2020} because the large data volumes prohibit full storage. In the image domain, efforts have been somewhat limited by a lack of training data, and the lack of surveys producing large numbers of transient and highly variable sources.  We discuss the key work in the following subsections. Most of the efforts to date have been supervised approaches (see Section~\ref{s_supervised}), which can only recover class types that have already been identified. However, there has been some exploration of unsupervised approaches, which we touch on briefly in Section~\ref{s_anomoly}.

\subsubsection{Radio lightcurve classification}\label{s_supervised}

One of the primary data products of a multi-epoch survey is radio lightcurves. Radio data alone does not typically provide enough information for a firm classification for many sources types: this usually requires multi-wavelength information \citep[e.g.][]{mooley_caltech-nrao_2016, rowlinson_search_2022}. However, the lightcurves of different source classes do have different time-dependent features, as shown in Figure~\ref{f_lightcurves}. For example: radio supernovae have a rapid rise followed by a power-law decay; flaring stars show short, discrete bursty behaviour that is sometimes periodic; and scintillation appears as stochastic fluctuations with characteristic amplitudes and timescales. Some sources classes (e.g. syncrotron transients) also show frequency dependent time evolution, and others such as pulsars and long period transients show variability in their polarisation fraction (these dimensions are not captured in Figure~\ref{f_lightcurves}).

\begin{figure*}[ht!]
\centering
\includegraphics[width=0.85\textwidth]{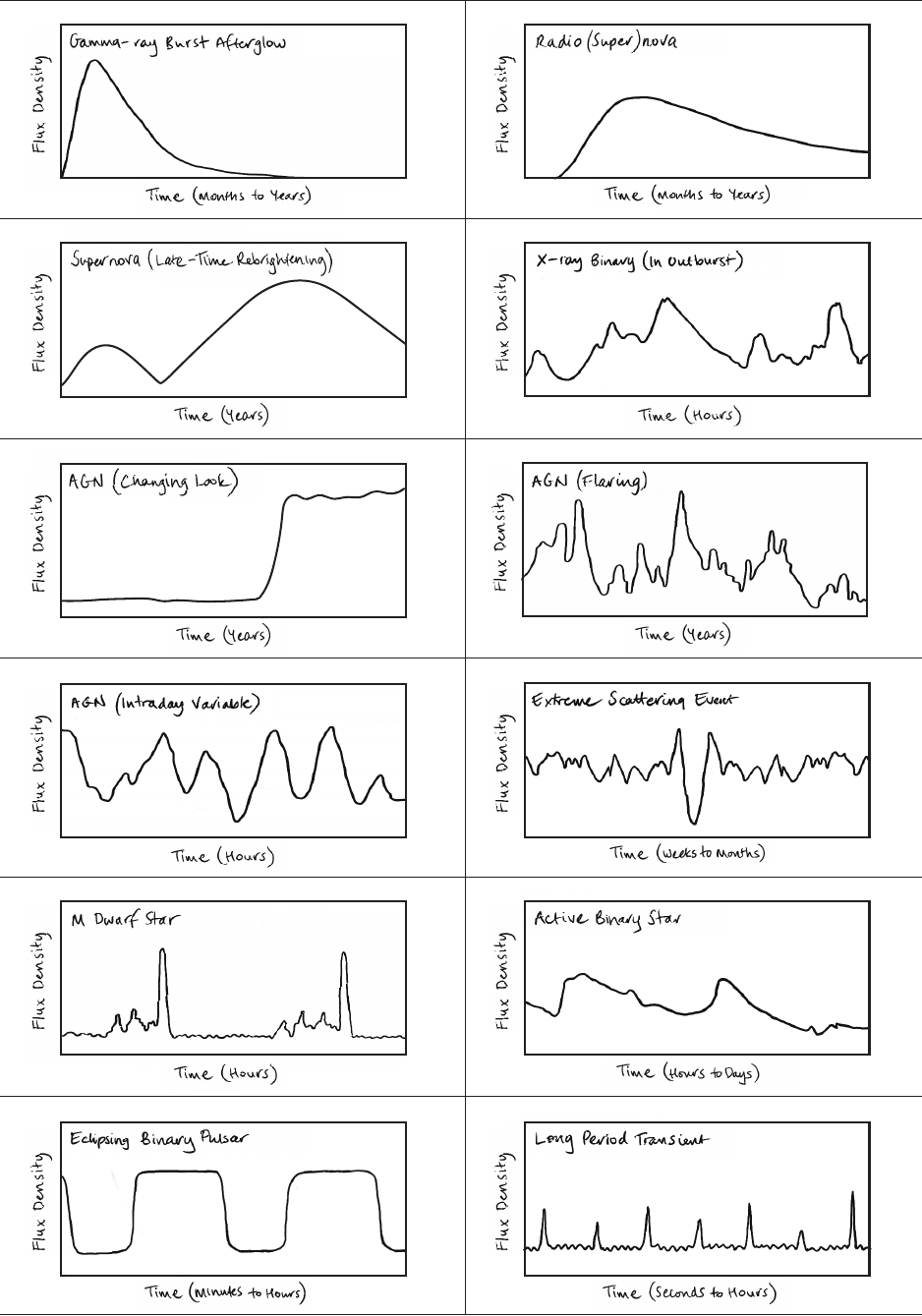}
\caption{Sketches of lightcurves for a range of radio transient source types. The lightcurves capture qualitatively different variability with rough timescales indicated. We have not attempted to capture spectral energy evolution, polarisation fraction, or the typical flux density scale. Not all source types are included; for example other synchrotron transients such as TDEs will show qualitatively the same evolution as GRB afterglows. Note that many source classes exhibit variability on multiple timescales. Most of the sketches are inspired by real data, as listed: AGN intraday variability \citep{bignall_rapid_2003}; AGN flaring \citep{hovatta_long-term_2008}; Changing look AGN \citep{meyer_late-time_2025}; supernova re-brightening \citep{anderson_peculiar_2017}; active binary star \citep{osten_multiwavelength_2004}; X-ray binary \citep{fender_comprehensive_2023}; eclipsing binary pulsar \citep{zic_discovery_2024}; M dwarf star \citep{zic_askap_2019}; extreme scattering event \citep{bannister_real-time_2016}; long period transient (and long period pulsars) \citep{wang_discovery_2025}.} 
    \label{f_lightcurves}
\end{figure*}

As a result of the lack of suitable training data (i.e. observed, well-sampled radio lightcurves from a range of source types), the first efforts in automatic classification of radio transients used simulated datasets \citep[e.g.][]{rebbapragada_classification_2012,murphy_vast_2013}. These were useful as a proof-of-concept, but the simulated lightcurves did not account for the full complexity (or the poor sampling) of real survey data. 

Following this, there were various efforts that focused on specific aspects of classification. \citet{pietka_use_2017} focused on the timescale of variability as a key feature in determining an initial classification for radio transient lightcurves. They automatically measured the rise rate of $\sim 800$ synchrotron flares from a range of source types to define a probability distribution of timescales for each source type. While the predictive power of a single property is naturally limited, the intention was that this would be a useful component of a more comprehensive classification process.

Along similar lines, \citet{stewart_optical_2018}
investigated radio versus optical flux density as a means of making a preliminary classification of radio transients. They showed that different source types do fall (broadly) into different regions of this two-parameter space. For example, quasars tend to have quite a high radio to optical flux density ratio, whereas stars typically have a much lower one.   \citet{wang_discovery_2022} took this further and used radio to near-infrared (to be more robust against extinction) flux ratio versus circular polarisation fraction, showing that the locations of pulsars and stellar sources (the two main classes of sources with significant polarisation) were very different.  And \citet{wang_detection_2025} compared X-ray and radio fluxes (drawing on the radio/X-ray correlation plots for accreting sources; e.g. \citealt{gallo_universal_2003,russell_reproducible_2016,ridder_radio_2023}), showing that LPTs occupied an extreme region of that space distinct from other potentially related populations like low-mass X-ray binaries, stars, and cataclysmic variables, but did resemble magnetars.
As with the timescale of variability, these  metrics alone are not enough to definitively classify objects, but would be useful parts of a classification pipeline.

The most comprehensive attempt to classify radio transient lightcurves to date is that by \citet{sooknunan_classification_2021}, based on an approach presented by \citet{lochner_photometric_2016} for optical lightcurves. They addressed the limitations of the currently available data in several ways: (1) they interpolated a set of real lightcurves (from \citealt{pietka_variability_2015} and \citealt{dobie_turnover_2018}) so they have equivalent sampling; (2) they augmented this set of lightcurves to produce a larger, more balanced training set; and (3) they incorporated some external contextual information, such as whether the object is in the Galactic plane, and the object's optical flux density. The classifier achieved $\sim 75\%$ accuracy for a 5-class problem (AGN, GRBs, XRBs, Novae and SNe), a promising result, but with significant room for improvement. 

Automatic classification of radio transients is on the cusp of becoming a reality for large surveys. The machine learning techniques are well-established in optical time domain surveys, and some of the additional challenges presented by radio surveys have been investigated in isolation. With the much larger datasets resulting from current surveys, it should be possible to train pipelines that work effectively, at least to provide an initial classification that can be further investigated manually.

\subsubsection{Anomaly and outlier detection}\label{s_anomoly}
One of the aims of radio transient surveys is to detect rare classes of objects, or unusual examples of existing classes. Anomaly detection is a type of unsupervised machine learning that involves searching for data points that do not conform to the normal properties of a given dataset (often called outliers), making it well suited to this task. The methods for identifying anomalies range from well-established statistical techniques, through to more sophisticated algorithms. For a comprehensive review of anomaly detection techniques and applications in science, engineering and business, see \citet{chandola_anomaly_2009}.
For a review of anomaly detection methods applied to time series data, see \citet{blazquez-garcia_review_2021}. 

In time-domain astronomy, anomaly detection approaches have been demonstrated on a range of tasks. For example, searching for unusual objects in Kepler light curves \citep{giles_systematic_2019,martinez-galarza_method_2021}. Both of these projects were motivated by the serendipitous discovery of Boyajian’s star KIC 8462852 \citep{boyajian_planet_2016}, and the need to prepare for, and capitalise on, the increased discovery potential for unusual objects that will come with the Vera Rubin Observatory 10-year Legacy Survey of Space and Time \citep[LSST;][]{ivezic_lsst_2019}. 

\citet{lochner_astronomaly_2021} present a general framework for anomaly detection aimed at astronomy applications. It incorporates active learning, with the ability for the expert user to rank and categorise the results. Their approach is demonstrated on a set of 50\,000 simulated lightcurves that represent typical (99\%) and outlier (1\%) sources. \citet{andersson_finding_2025} is the first group to apply this approach to observed radio transient lightcurves. Using data from ThunderKAT, they were able to recover interesting outlier sources (as defined by citizen science volunteers --- see Section~\ref{s_citizen}) with some success, giving results that are complementary to other approaches. This area of research shows promise for future, larger, radio transient surveys.

\subsubsection{Citizen science projects}\label{s_citizen}

Citizen science projects \citep[one of the first, and most influential of which, was SETI@HOME;][]{werthimer_berkeley_2001} are designed to leverage the enthusiasm of members of the public to help complete large scale scientific data collection and classification, either using only their personal computer facilities (as in the case of SETI@HOME) or using the citizens themselves to help analyse data. In astronomy, a pioneering project in this space was Galaxy Zoo \citep{lintott_galaxy_2008} which invited the general public to visually inspect and classify the morphology of nearly a million galaxies from the Sloan Digital Sky survey \citep{lintott_galaxy_2011}. 
The success of the initial Galaxy Zoo project led it to broaden its remit to the Zooniverse. This platform, and other similar approaches, have been applied to a wide range of classification tasks including gravitational wave glitch identification \citep{zevin_gravity_2017},  exoplanet discovery \citep{christiansen_k2-138_2018}, and radio galaxy classification \citep{lukic_radio_2018}.

The first (and, to date, only) citizen science project for classifying image domain radio transients is the `Bursts From Space' project on MeerKAT \citep{andersson_bursts_2023}. They had $\sim 1000$ volunteers who provided $\sim 89\,000$ classifications of sources using images and lightcurves. Sources were classified into five classes: Stable, Extended Blobs, Transients/Variables, Artefacts and Unsure. From this, 142 variable sources were identified, most of which the authors conclude are likely to be AGN. The amount of data available from current and future radio surveys should allow similar projects with more detailed classification schema.

\section{Characterising the time variable sky }\label{s_characterising}

There are two major aspects to characterising the dynamic radio sky. The first is how to measure and compare the variability of individual objects. In Section~\ref{s_metrics} we describe some of the main metrics used to do this, and discuss the advantages and disadvantages of each. The second is how to characterise the population of highly variable radio sources, against the background radio source population. In Section~\ref{s_rates} we discuss this in the context of what we know from current radio transient surveys.

\subsection{The variability of individual objects}\label{s_metrics}

To understand the time variable sky we need a way of identifying and characterising the variability of individual radio sources. There are a range of metrics for identifying variable sources that have been used in the literature. In some cases, the aim is to identify the most highly variable objects for further monitoring, in other cases the aim is to characterise the time behaviour of the entire radio source population. 

In this section we focus on metrics for identifying variability. Classification of sources into different discrete types (e.g. supernovae vs. stars) is covered in Section~\ref{s_ml}.

\subsubsection{Basic metrics}

For surveys with a small number of epochs, the simplest way of measuring the variability of a source is to take the maximum difference between the observed flux density values at different times. For example, \citet{carilli_variability_2003} used:
\begin{equation}
    \Delta S = |S_1 - S_2|
\end{equation}
the absolute change in flux density, $S$, between two pairs of two epochs (in their case, 19 days and 17 months apart). To determine if this difference was significant, they compared the value of $\Delta S$ to the $5\sigma$ image rms noise. Sources that exhibited the most extreme variability of $100\%$, would be detected in one epoch and not the other. These have typically been referred to as transients in the literature (but see Section~\ref{s_terminology} for a discussion of terminology).

This approach was also used by \citet{levinson_orphan_2002} in their search for orphan afterglows between the FIRST and NVSS surveys. When the two epochs have significantly different sensitivity limits, additional care is required. \citet{levinson_orphan_2002} chose a flux density cutoff designed to exclude (most) persistent sources fluctuating around the sensitivity limit. 

A related approach is the fractional variability $V$ used by \citet{gregory_radio_1986}:
\begin{equation}
    V = \frac{S_{\rm max}-S_{\rm min}}{S_{\rm  max}+S_{\rm min}},
\end{equation}
where $S_{\rm min}$ and $S_{\rm max}$ are the minimum and maximum flux density, respectively. This metric is only appropriate for data with a small number of measurements. However, the modulation index below can be considered a more sophisticated version of this metric.

Rather than consider a difference between values, \citet{de_vries_optical_2004} looked at the ratio between two flux densities $S_1$ and $S_2$, which they called the variability ratio, $VR$:
\begin{equation}
    VR = S_1/S_2
\end{equation}
This was modified to account for the varying noise levels in the different epochs, which could be functions of the flux densities themselves (most appropriate for bright sources).    

All of these metrics are inherently limited, and for surveys with many observations of each source, a more sophisticated approach is preferred.

\subsubsection{$\chi^2$ probability}

Another way of measuring overall variability is to
calculate the $\chi^2$ value for a model where the source remained constant over the observing period: 
\begin{equation}
    \chi^2_{\rm lc} = \sum_{i=1}^{n} \frac{(S_i - \bar{S})^2}{\sigma^2_i}
    \label{e_chi2}
\end{equation}
where $S_i$ is the $i$th flux density measurement with variance $\sigma^2_i$ within a source light curve, and $n$ is the total number of flux density measurements in the light curve. $\bar{S}$ is the weighted mean flux density defined as:
\begin{equation}
    \bar{S} = \sum_{i=1}^n \frac{S_i}{\sigma_i^2} / \sum_{i=1}^n \frac{1}{\sigma_i^2}
\end{equation}
This has been used for a number of variability monitoring programs, for example \citet{kesteven_variability_1976,gaensler_long-term_2000,bell_survey_2014}. \citet{kesteven_variability_1976} considered a source `variable' if the probability $p(\chi_{\rm lc}^2)$ of exceeding the observed $\chi_{\rm lc}^2$ by chance is $\le 0.1\%$ and `possibly variable' if $0.1\% \le p \le 1\%$.  Care must be taken to consider the number of sources measured (effectively a `trials factor'), since with 1000 sources a single-source probability of 0.1\% becomes insignificant \citep[e.g.][]{bannister_22-yr_2011}.

The probability $p$ is independent of the number of data points, so this method can be used to compare lightcurves with different numbers of observations. 
However, as identified by the authors, a  limitation of this metric is that it is independent of the order of the flux density measurements, and hence is not sensitive to slowly increasing or decreasing variability with time, much less more complex behaviour (cf.\ \citealt{leung_matched-filter_2023}).

\subsubsection{Modulation index}
Together with $\chi^2$, a popular statistic is the modulation index, which is the standard deviation of the flux density divided by the (weighted) mean \citep[e.g.][]{narayan_physics_1992,rickett_radio_1990}.  This can also be written as
\begin{equation}
    V = \frac{1}{\bar{S}} \sqrt{\frac{N}{N-1}(\overline{S^2}-\bar{S}^2)},
    \label{e_V}
\end{equation}
with
\begin{equation}
    \overline{S^2} = \sum_{i=1}^n \frac{S_i^2}{\sigma_i^2} / \sum_{i=1}^n \frac{1}{\sigma_i^2},
\end{equation}
and where the use of $N-1$ in the denominator means we are using the unbiased estimator of the sample variance. 
This has been used extensively in radio transients analysis \citep[for example, see][]{gaensler_long-term_2000,bannister_22-yr_2011,swinbank_lofar_2015,rowlinson_identifying_2019}.
The modulation index, which is also called the `coefficient of variation,' is often written as $V$, but also as $m$ (especially in the literature around scintillation). It is an extension of the fractional variability that makes use of more than two measurements. Different authors  also use weighted or unweighted means.   

Yet another variant on this is also seen: a `debiased' variability index \citep{akritas_linear_1996,barvainis_radio_2005,sadler_properties_2006}:
\begin{equation}
    V_d = \frac{1}{\bar{S}} \sqrt{\frac{\sum (S_i - \bar{S})^2 - \sum \sigma_i^2}{N}},
\end{equation}
which also has extensive use in the X-ray literature \citep[e.g.,][]{green_molonglo_1999,vaughan_characterizing_2003,markowitz_expanded_2004}.
This is similar to Eqn.~\ref{e_V} in presenting fractional variability.  It typically has not used weighted sums, but attempts to account for the range in individual measurement error through the subtraction of the per-observation variance inside the radical.  A downside to this is that the factor inside the radical can easily be negative for non-variable sources (rather than deal with this issue directly, other authors use a `normalised excess variance' which is $V_d^2$; e.g. \citealt{nandra_asca_1997,vaughan_characterizing_2003}). \citet{barvainis_radio_2005} and \citet{sadler_properties_2006} handle this in slightly different ways.  They generally allow $V_d$ to be negative when the factor inside the radical is negative, and then present the results either as a negative value or use the range of small negative and positive values to define a minimum threshold for variability, and assign this value to all numbers below it.  
However, we note that $\sum (S_i - \bar{S})^2$ is distributed as a scaled  $\chi^2$ distribution for $N-1$ degrees-of-freedom.  For well-detected sources but with minimal excess variability, the fraction of values that have the factor inside the radical negative is $>50$\%.  
Also, as discussed extensively in the literature the exact value of this metric will depend on how the cadence of the observations compares to a typical variability timescale (when known), so in some cases more explicit parameterisations or functional forms could be used (Section~\ref{s_methods}).

\subsubsection{$\eta$-V parameters}
It has become standard to consider both the reduced $\chi^2$:
\begin{equation}
    \eta = \frac{1}{N-1} \chi^2_{\rm lc}
\end{equation}
(where there are $N-1$ degrees of freedom) and the modulation index $V$ together in finding variables which would have high values of $\eta$, $V$, or both \citep[e.g.][]{swinbank_lofar_2015,rowlinson_identifying_2019}.  

Both of these statistics are based on the second moments of the flux density (i.e. variance or standard deviation), and both are independent of data order.  So they should convey some similar information.  However, while correlated, the outliers in $\eta$ or $V$ are not always the same.  As discussed by \citet{rowlinson_identifying_2019}, sources with high $\eta$ usually are brighter, have some large flares or similar excursions from the mean, and have low uncertainties relative to the mean.  Such lightcurves are statistically distinguishable from a constant value, although this relies on accurate calibration and error analysis at high signal-to-noise, which is not always reliable.  In contrast, sources with high $V$ typically have lower average flux densities. 

This is illustrated in Figure~\ref{f_metrics}, which consists entirely of simulated data with Gaussian noise and a range of signal-to-noise ratios (SNRs)\footnote{Note that SNR is used as acronym both for supernova remnant and signal-to-noise ratio in this review, and in the wider literature. The context usually makes it clear which meaning is intended.} but no transients or variables beyond those expected from the noise realisations.  As expected, at high $V$ the data are dominated by low SNR sources that just happened to have slightly higher noise realisations.  The trend with $\eta$ is harder to see, but at a constant $V$ there is a trend toward higher $\eta$ at higher SNR.  We also show the theoretical predictions for the marginal distributions.  In the case of $\eta$ it is a scaled $\chi^2$ distribution (as expected), which indeed resembles a log-normal distribution as discussed by \citet{rowlinson_identifying_2019}.  For $V$ the distribution is largely just the distribution of $1/{\rm SNR}$, which will depend on the underlying flux density distribution: since in this simulation $S$ has a power-law distribution and ${\rm SNR}\propto S$ (constant noise), $V$ also has a power-law distribution up to the maximum value but with an inverted index, $N(>V)\propto V^{1.5}$.   This can change in other samples, though. Finally, we also see a cutoff at high $V$ that depends on $\eta$ as $V_{\rm max}=\sqrt{\eta}/{\rm SNR}_{\rm min}$: such cutoffs are observed in real data (e.g. \citealt{rowlinson_identifying_2019,murphy_askap_2021}).  

This demonstration serves several purposes.  Firstly, the use of both $\eta$ and $V$, while common, is not necessarily optimal since they are related to each other as a function of SNR.  A better second metric in addition to $\eta$ or $V$ might be something optimised for detecting short-timescale variability such as the result of an outlier search or a matched filter, or some other metric that incorporates information about the order of the data.
Secondly, the background distribution of $\eta$ should follow a $\chi^2$ distribution, while that of $V$ depends on the range of SNRs considered, and both can be used to determine thresholds for investigation.  

\begin{figure}
    \centering
    \includegraphics[width=1.0\linewidth]{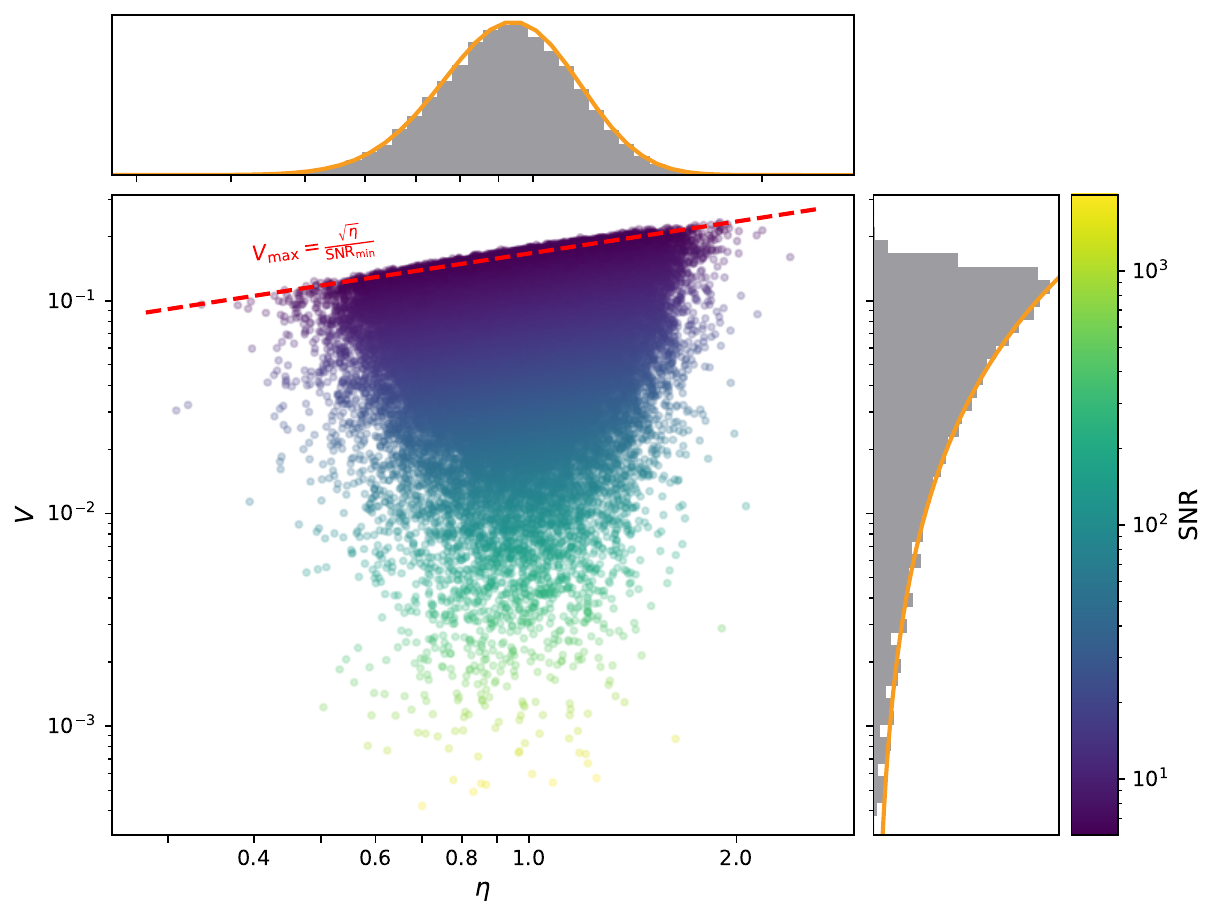}
    \caption{Standard $\eta$-$V$ plot for identifying variable sources, following \citet{swinbank_lofar_2015}.  The data here were simulated and consist entirely of Gaussian noise with constant amplitude.  The distribution of source brightnesses are a power-law with cumulative distribution $N(>S)\propto S^{-1.5}$, appropriate for standard candles in a Euclidean universe \citep[e.g.,][]{hoyle_counting_1961,longair_luminosity_1965}, with SNR ranging from 6 to 2000.   The points are coloured by SNR (with a colour-scale to the right).  We also show marginal distributions for $\eta$ (top) and $V$ (right), along with theoretical predictions (solid orange lines).  Finally, we show the line $V_{\rm max}=\sqrt{\eta}/{\rm SNR}_{\rm min}$ which bounds the top of the distribution (red dashed line). }
    \label{f_metrics}
\end{figure}

\subsubsection{Other variability metrics}
\label{s_methods}
The metrics discussed in the previous section are related: they are both second-order statistics that depend on the degree of variability but ignore any shape/ordering of the lightcurve.  Hence a number of other metrics or analyses have been used in different cases.

Perhaps the simplest is the temporal gradient $\nabla_F \equiv dS/dt$ used by \citet{bell_murchison_2019}, who also divided the gradient by its uncertainty, $\nabla_S \equiv \nabla_F/\sigma_F$ to identify statistically significant changes.  They also found that higher-order polynomial changes were not needed to describe the data.

More general parameterisations use the autocorrelation function:
\begin{equation}
    ACF(\tau) = \langle S(t+\tau)S(t)\rangle,
\end{equation}
which is sometimes replaced by the autocovariance function (also ACF):
\begin{equation}
    ACF(\tau) = \langle [S(t+\tau)-\overline{S}][S(t)-\overline{S}]\rangle,
\end{equation}
where the means are subtracted before calculation.  Both of these determine the $ACF$ as a function of lag $\tau$.  The autocorrelation function can  be used to study the timescale of variations \citep{rickett_interstellar_1977}, especially for diffractive scintillation in pulsars.  A similar concept is the structure function:
\begin{equation}
    D(\tau) = \langle [F(t+\tau) - F(t)]^2 \rangle
\end{equation}
for all measurements in some bin with lag $\tau$,
where $F$ can be the raw flux density measurements \citep{hughes_university_1992,kumamoto_flux_2021}, but also measurements with the mean subtracted and then normalised by the standard deviation \citep{gaensler_long-term_2000}, or just normalised by the mean  \citep{rickett_intensity_1990,kaspi_long-term_1992,lovell_micro-arcsecond_2008}.  This is commonly used to study variability  in AGN \citep{hughes_university_1992,lovell_micro-arcsecond_2008, gaensler_long-term_2000} and pulsars \citep{rickett_intensity_1990,kaspi_long-term_1992,kumamoto_flux_2021} which can be compared against the predictions for refractive scintillation (e.g. \citealt{hancock_refractive_2019}).  Using expectations for the behaviour of the structure function with lag, it can be used to give a better definition of the modulation index and its associated timescale \citep[e.g.][]{kaspi_long-term_1992}.  
In both cases effort must be taken to do this correctly for irregularly-sampled data \citep{edelson_discrete_1988,kreutzer_s-acf_2023}.  

These functions are also intrinsically related to the Fourier power spectrum and its extension for irregularly-sampled data, the Lomb-Scargle power spectrum \citep{scargle_studies_1982,scargle_studies_1989}.  Such power spectra can be estimated directly when searching for explicitly periodic emission, such as from eclipsing pulsar systems \citep[e.g.][]{zic_discovery_2024,petrou_investigating_2025}. 
There are also many refinements to these broad techniques such as Gaussian processes \citep{aigrain_gaussian_2023}, but in the cases of radio lightcurves with only a few tens of data-points for most examples, the differences between the  techniques are small.  

Furthermore, Fourier techniques can be seen as matched filters with sine/cosine kernel functions with unknown period.  This can be explored further with more specific kernels that are appropriate to specific transient types.  For example, \citet{leung_matched-filter_2023} designed a matched filter around the expected shape of a GRB afterglow, especially orphan afterglows, parameterised as a power-law lightcurve with an optional break.  Compared to selection via a standard techniques selecting sources compared to a constant model (effectively the $\chi^2_{\rm lc}$ in Equation~\ref{e_chi2}),  none of the five selected sources from that paper would have passed a more basic $3\sigma$ threshold due to their slow evolution.  Such a strategy can of course be adopted for an arbitrary light-curve shape, such as a generic burst.  All of these techniques, though, require richer data sets with at least tens (if not more) samples, which are only becoming common with the latest generation of surveys.

\subsubsection{Burst detection}

In time-domain surveys with small numbers of temporal samples (observing epochs or otherwise), there is in general no need for explicit burst detection beyond the variability analysis presented above.  But as the number of samples grows, especially with short integrations from longer images (Section~\ref{s_differencing}), the need for explicit burst detection grows.  

Historically, there has not been a large focus on burst detection from radio images, largely due to the small number of samples available.  This is in contrast to the significant literature in finding and classifying bursts in other domains such as FRBs \citep[e.g.,][]{rajwade_needle_2024} or XRBs and GRBs \citep[e.g.,][]{barthelmy_burst_2005,meegan_fermi_2009}.  Those cases are made further difficult by the need to examine large data-sets in real time --- leading to efficient hardware implementations, machine learning classifiers, and distributed architectures --- something which radio imaging surveys are only now starting to confront.

Beyond  initial approaches using the metrics above or variants, there can be techniques that segment the data into `background' and `burst'.  At the simplest level, iterative sigma clipping \citep[e.g.,][]{leroy_robust_1987}
is often used for outlier excision \citep{vallisneri_taming_2017}, but can be used for outlier identification as well.  More complex approaches can be taken to the same problem \citep[e.g.,][]{hogg_data_2010} with robust statistical models that attempt to  assign probabilities to each data point regarding whether it is part of the background or not.  However, these approaches may be limited in that they consider each observation independently.  If the burst duration is less than the sampling time that is fine, but if bursts can last longer then techniques which take into account the temporal relation between data points may be preferred.

The basic techniques used in those cases are typically variants of matched filtering (also see Section~\ref{s_techniques}), which is also commonly used for FRB detection (after dedispersion).  Typically a Gaussian or boxcar filter (if there is \textit{a priori} knowledge that the burst shape should be different that could of course be used instead, or a more complex basis like wavelets if there are enough samples)  of a given width is convolved with the data to identify bursts, and then this is repeated for a range of widths.  And as with the outlier detection metrics above, a more formal Bayesian approach can be taken here too.  For instance, \citet{scargle_studies_2013} offer an efficient algorithm that can identify segments of arbitrary length which differ from the background, and which can be applied to data arriving in real time and to data which have already been collected.  This algorithm achieves similar  sensitivity to the theoretical limit.

\subsection{Populations and rates}\label{s_rates}
In many previous papers, where transients were identified from 2-epoch searches in small numbers, often years after the fact (limiting followup possibilities), the results were presented as a limit or constraint on the rate (areal density) of transient sources \citep[e.g.,][]{bower_search_2011,croft_allen_2010,croft_allen_2011,stewart_lofar_2016,rowlinson_limits_2016,bell_automated_2011,bell_search_2015,bell_survey_2014,bhandari_pilot_2018,mooley_sensitive_2013}; see \citet{mooley_caltech-nrao_2016} for a nice compilation that has been recently updated\footnote{See \url{http://www.tauceti.caltech.edu/kunal/radio-transient-surveys/index.html}.}.  This served to help develop expectations for improved surveys, as well as understand the discovery rate relative to the expected rates of various source populations \citep[e.g.,][]{metzger_extragalactic_2015}.  This analysis particularly applied to synchrotron transients, which have very similar lightcurves.

However, the changes outlined in Section~\ref{s_trends} have changed this practice. Firstly, using multi-epoch surveys makes the distinction between transients and variables less clear.  Although some sources are clearly explosive in origin and hence intrinsically `transient', many of the sources identified by earlier surveys are more clearly examples of variable sources when richer data sets are used \citep[e.g.,][]{murphy_askap_2021}.  This also means that lightcurve shape can influence the inferred transient rate explicitly, as described in detail by \citet{carbone_new_2016,carbone_calculating_2017}.  Those works include more powerful but also more specific techniques to constrain the rates of individual source populations. These approaches rely on assumptions about the underlying source properties and distributions, as well as detailed knowledge of the observing strategy (similar to \citealt{feindt_simsurvey_2019}). For a general review such as this, it is difficult to do a comprehensive analysis for a large set of previous surveys and sources.
Secondly, faster identification  has allowed much more reliable followup and classification \citep[e.g.,][]{law_discovery_2018,anderson_caltech-nrao_2020,dong_transient_2021,dong_flat-spectrum_2023}.  Thirdly, searches for shorter-duration events and polarised sources have allowed discovery of new classes of sources besides synchrotron transients \citep[e.g.,][]{hurley-walker_radio_2022,hurley-walker_long-period_2023,caleb_emission-state-switching_2024,lee_interpulse_2025,wang_discovery_2021,wang_detection_2025,dong_chime_2025}.  
Finally, the latest generation of surveys (VLASS, VAST, LoTSS, MWA GPM) are still ongoing, with individual sources being published and final tallies of discoveries still to come.  All of these make the continuation of overall rate projections less valuable.  Rates are extremely valuable still for studying individual classes of objects (e.g., \citealt{beniamini_evidence_2023}) but we do not focus on the `transient rate'.   

These recent surveys have allowed us to start to characterise the overall populations of radio transients and variables for the first time. Figure~\ref{f_pop} shows the number of each source type found in a selection of recent surveys. Most surveys found AGN/galaxies as the dominant class (except for \citet{dobie_radio_2023} who deliberately excluded them), followed by radio stars and pulsars.

\begin{figure}
    \centering
\includegraphics[width=1.0\linewidth]{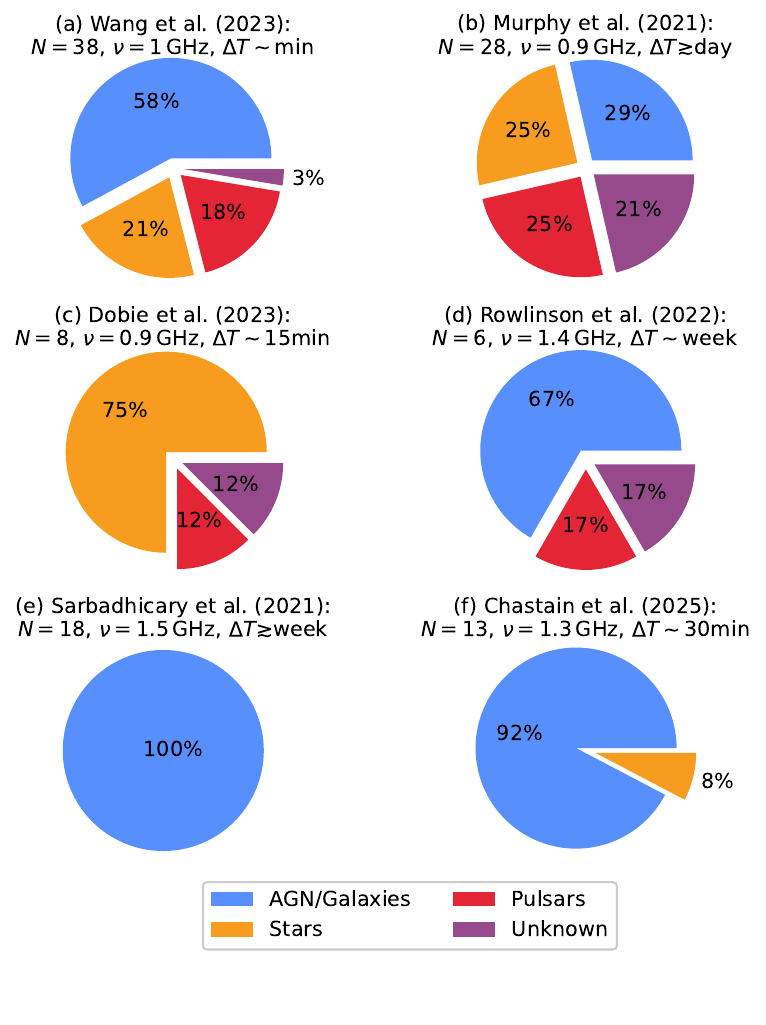}
    \caption{Source class distribution for the transient and variable sources found in a selection of recent image-domain surveys: 
    \citet{wang_radio_2023}, who searched for minute-scale variability at frequencies near 1\,GHz with ASKAP; \citet{murphy_askap_2021}, who searched for longer-scale variability at 888\,MHz with ASKAP; \citet{dobie_radio_2023}, who searched for fast variability at 943\,MHz with ASKAP; \citet{rowlinson_search_2022}, who searched for variability on timescales of weeks at 1.4\,GHz with MeerKAT;      
    \citet{sarbadhicary_chiles_2021}, who searched for longer-scale variability in deep images with the VLA at 1--2\,GHz; and \citet{chastain_commensal_2025}, who searched for short-term variability with MeerKAT at 1.3\,GHz.
    Most surveys show AGN \& galaxies as the dominant class, except for \citet{dobie_radio_2023} who deliberately excluded them.  Since each survey had very different fields-of-view and total numbers of epochs we do not present absolute numbers, but just relative populations, and even those will change for different surveys (e.g. stars and pulsars will be over-represented on shorter timescales).
    }
    \label{f_pop}
\end{figure}

We have looked at the potential discovery volume from different past/ongoing surveys (Figure~\ref{f_fom}) but more to show the overall progression with time, rather than to allow direct comparison.  The proliferation of survey strategies probing different timescales with different techniques has made such comparisons increasingly hard (e.g., \citealt{rowlinson_limits_2016}).  In the following section, we look toward future facilities (for which the  details of potential surveys are still unclear) to see broadly where they might excel, but we still do not give detailed projections for yields.  This is best left to in-depth analyses that examine specific source classes.

\section{Previewing future transient surveys }\label{s_future}

The substantial amount of effort that has been put into conducting image domain radio transient surveys over the past 15 years has resulted in significant steps forward in our understanding of individual radio transient source types, of the overall populations, and of the advantages and limitations of various approaches to transient identification and classification. 

It is critical that we draw on this experience to optimise the scientific benefit of the surveys that will be possible on upcoming radio telescopes. In this section we outline the specifications of three major future instruments, summarise some of the other planned facilities, and reflect on lessons learned from current transient surveys. 

We summarise the basic properties of the telescopes in Table~\ref{t_future}.  Note that this does not contain details of prospective surveys.  Some of the facilities will be primarily survey-driven, while others will be used more for PI-led science including transient followup.  So focusing on the `survey speed' may not be the most relevant metric.  Nonetheless, we do present a survey speed calculated for each facility as:
\begin{equation}
    {\rm SS} \equiv \Omega_{\rm inst} \frac{A_{\rm eff}^2}{T_{\rm sys}^2}.
\label{e_ss}
\end{equation}
Unlike in Section~\ref{s_fom}, $\Omega_{\rm inst}$ here is the instantaneous telescope field-of-view, since we are discussing telescopes and not surveys.
In principle this should actually be an integral of the sensitivity over solid angle \citep[e.g.,][]{hotan_australian_2021}, but such information is not always available.  In some uses survey speed can also include a bandwidth factor (so that it more directly reflects the time needed to achieve a given flux density limit over a given area), but we do not do that here.

This survey speed is intended to be proportional to the expected speed it would take to survey the sky down to a constant limiting flux density (and higher is better), although it does ignore some elements like changing bandwidth between telescopes or other sources of noise like confusion (compared to Eqn.~\ref{e_merit2} which uses the delivered noise).
Note that we can relate the survey speed to the system equivalent flux density:
\begin{equation}
    {\rm SEFD} \equiv \frac{2 k_B T_{\rm sys}}{A_{\rm eff}}
\label{e_sefd}
\end{equation}
via
\begin{equation}
    {\rm SS} = \Omega_{\rm inst} \frac{4k_B^2}{{\rm SEFD}^2}
\end{equation}
and the continuum sensitivity
\begin{equation}
    \sigma = \frac{{\rm SEFD}}{\sqrt{N_p \Delta t \Delta \nu}}
\end{equation}
with $N_p=2$ the number of polarisations, $\Delta t$ the integration time, and $\Delta \nu$ the bandwidth (and this ignores any loss of sensitivity from $uv$ tapering or weighting).  So we also have
\begin{equation}
    {\rm SS} = \Omega_{\rm inst} \frac{4 k_B^2}{N_p \Delta t \Delta \nu \sigma^2}
\end{equation}

In addition, comparisons between telescopes operating at very different frequencies are complicated by the changing sensitivities and bandwidths, together with expected source characteristics.  Instead of an explicit spectral correction like in  Eqn.~\ref{e_merit2} we separate the high ($\sim 1.4\,$GHz) and low ($\lesssim 200\,$MHz) facilities (especially as the contribution from  Galactic synchrotron emission in the latter can be highly significant and highly variable with position).

In Figure~\ref{f_future} we plot the capabilities of the future facilities from Table~\ref{t_future} (which we discuss more below) along with current-generation facilities.  We include both the survey speed, relevant for wide-field discovery, as well as the sensitivity (inverse of the SEFD), which is relevant for pointed followup.  It is clear that the future facilities will have big improvements in both metrics, although the survey speed improvements will be most significant, and overall the boost at lower frequencies with the SKA-Low will be more significant than at higher frequencies.

\begin{figure*}
    \centering
    \includegraphics[width=1.0\textwidth]{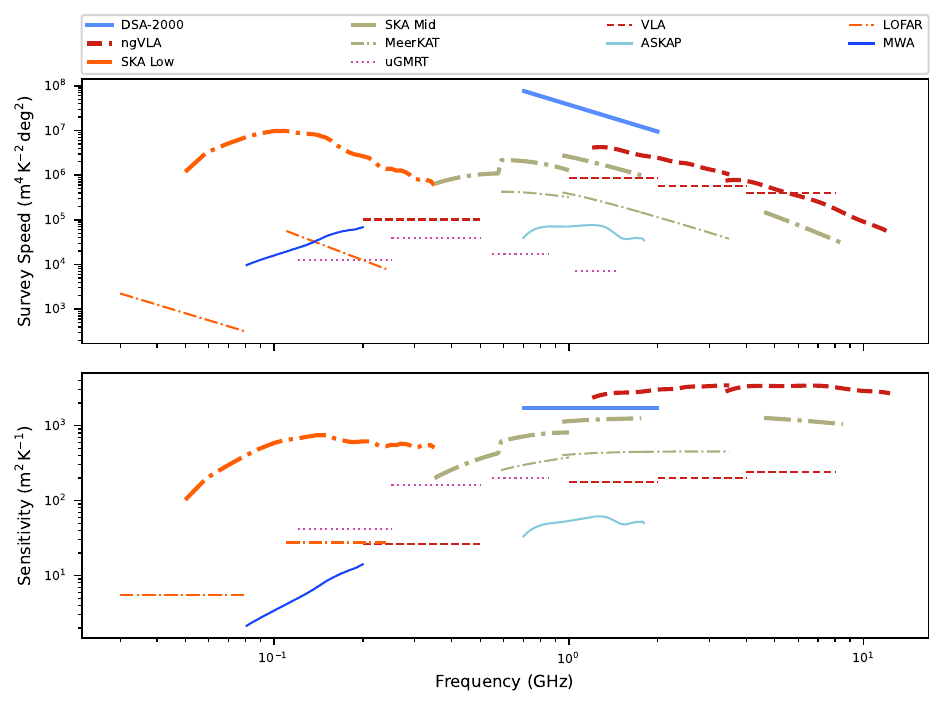}
    \caption{Sensitivity of planned future radio facilities.  We show the survey speed (Eqn.~\ref{e_ss}) in the top panel and the sensitivity ($A_{\rm eff}/T_{\rm sys}$, from Eqn.~\ref{e_sefd}) in the bottom panel as a function of frequency.  Current facilities are shown with thinner lines, while future facilities are with thicker lines.   The data are a mix of detailed calculations and simplistic projection (e.g., ignoring effective area changes within a frequency band).  Data sources for future facilities are: DSA-2000 from G.~Hallinan (pers.\ comm.), ngVLA from \url{https://github.com/dlakaplan/}, SKA1-low from \url{http://skalowsensitivitybackup-env.eba-daehsrjt.ap-southeast-2.elasticbeanstalk.com}, and SKA1-mid from \url{https://gitlab.com/ska-telescope/ost/ska-ost-senscalc}.  Data sources for current facilities are: MeerKAT from \url{https://gitlab.com/ska-telescope/ost/ska-ost-senscalc}, upgraded GMRT (uGMRT) from \citet{gupta_upgraded_2017}, VLA from \url{https://science.nrao.edu/facilities/vla/docs/manuals/oss/performance}, ASKAP from \citet{hotan_australian_2021} and E.~Lenc (pers. comm.),  LOFAR from \citet{van_haarlem_lofar_2013} for the Dutch array but with double the number of LBA antennas per station, and MWA from \citet{ung_determination_2019} via \url{https://github.com/ysimonov/MWA-Sensitivity}.  Facilities at $<300\,$MHz include contributions from the sky temperature, although we use a pointing location away from the Galactic plane.}
    \label{f_future}
\end{figure*}

\subsection{The SKA Observatory}
The SKA Observatory (SKAO) consists of two telescopes (SKA-Low and SKA-Mid), currently under construction\footnote{The specifications given in this Section are based on the official SKAO documents available in this Zenodo collection: \url{https://zenodo.org/communities/skaokeydocuments/}, in particular the `SKAO Staged Delivery, Array Assemblies And Layouts' document: \citet{seethapuram_skao_2025} and the `Design Baseline Description': \citet{dewdney_ska1_2022}.}. SKA-Low will consist of 512 low frequency aperture array stations, spread over an area 80\,km in diameter, located in Western Australia. Each station will have 256 wire antennas in a 38-m diameter area. It will observe in the frequency range 50--350\,MHz. SKA-Mid will consist of 197 mid frequency dishes (64 from the existing MeerKAT array) spread over an area \red{160\,km} in diameter. It will be located in the Karoo region of South Africa. The telescope will operate in 6 frequency bands (3 initially) ranging from \red{350\,MHz} to 15.4\,GHz. Full details and technical specifications of the telescopes are given in the Design Baseline Description and the SKAO Staged Delivery, Array Assemblies And Layouts documents and summarised in Table~\ref{t_future}.

The SKAO telescopes are being constructed in stages, as outlined in the SKAO construction timeline\footnote{\url{https://www.skao.int/en/science-users/timeline-science}}. A key milestone will be the completion of the AA$^*$ arrays, which will consist of 307 SKA-Low stations and 144 SKA-Mid dishes. These arrays are expected to be completed in 2029 and 2031, respectively, with the specifications summarised in Table~\ref{t_future}.

\begin{table*}[]
    \caption{Specifications for future radio telescopes} 
        \label{t_future}
{\tablefont
    \begin{tabular}{c|c c c | c c c}
    \toprule
    Telescope & LOFAR 2.0 LBA$^{\rm a}$& SKA Low & LOFAR2.0 HBA$^{\rm a}$ & SKA Mid$^{\rm b}$ & DSA-2000 & ngVLA$^{\rm c}$ \\\hline 
    Collecting area (m$^2$) &70\,192& 419\,000&41\,226& 33\,000 &49\,000 & 54\,000 \\
Frequency range (GHz) & 0.01--0.08&0.05--0.35 & 0.12--0.24 &\red{0.35}--15.4 & 0.7--2.0 & 1.2--116 \\
Bandwidth (MHz) &48& 300 &48& 720 & 1300 & 2300 \\
Angular resolution (arcsec) & 0.5& 3.3--23& 0.2 &0.7 & 3 & 0.08 \\
Field-of-view (\degsq) & 12 & 2.3--113& 12 & 0.7 & 7.0 & 0.55 \\
1\,h continuum sensitivity (\ujybeam) & 3000& 14--26&85 &0.9 & 1$^{\rm d}$ & 0.27$^{\rm d}$ \\
Survey speed$^{\rm e}$ ($10^{6}\,$\,\degsq\,m$^4$\,K$^{-2}$) &$\sim 10^{-3}$& 0.6--10&0.01--0.1& 2.6 & 13 & 2.1 
\botrule
    \end{tabular}}
        \begin{tabnote}
    {$^{\rm a}$ Multiple configurations trading bandwidth between HBA and LBA arrays and multiple beams are possible.  Most specifications are for the full EU array.}\\
        {$^{\rm b}$ Based on \url{https://www.astron.nl/telescopes/square-kilometre-array/}, \citet{dewdney_ska1_2022}, and \url{https://sensitivity-calculator.skao.int/mid}.  Most specifications are for Band 2, containing 1.4\,GHz.  Sensitivity and resolution can be changed with different weighting schemes.}\\
        {$^{\rm c}$ Based on \url{https://ngect.nrao.edu}.  Most specifications are for the `main' (core plus spiral out to 1000\,km) array and for Band 1 (centred at 2.4\,GHz).}\\
        {$^{{\rm d}}$ At 1.4\,GHz.} \\
        {$^{\rm e}$ Defined as $\Omega A_{\rm eff}^2/T_{\rm sys}^2$.}\\    
        \end{tabnote}
\end{table*}

The science case for the SKAO telescopes has been outlined in two paper collections. Firstly, in {\it `Science with the Square Kilometre Array'} \citep{carilli_science_2004}, and then, a decade later, in {\it `Advancing Astrophysics with the Square Kilometre Array'} \citep{braun_advancing_2015}. A new science book, stemming from the 2025 SKAO General Science Meeting, {\it `A new era in astrophysics: Preparing for early science with the SKAO'} is currently in preparation.

The exact operating model of the SKAO telescopes (including how time is allocated, and potential survey design) is yet to be determined. However, the current plan is that PI-led science would start in 2030 on AA$^*$ and larger observing programs led by the Key Science Projects would start several years later.

\subsection{DSA-2000}
\label{s_dsa2000}
The DSA-2000 telescope is a  planned instrument that, as originally scoped, will consist of $2000 \times 5$-m dishes spanning an area of 19 km $\times$ 15 km in Nevada, USA \citep{hallinan_dsa-2000_2019}. However, the array specification has changed to $1650 \times 6.15$-m (along with a new name, still to be determined), which will have a similar survey speed but a better point-source sensitivity (G.~Hallinan, pers. comm.).
The telescope will operate at radio frequencies of 700\,MHz to 2\,GHz . The baseline reference design specifications are summarised in Table~\ref{t_future}. The aim is to survey the sky repeatedly with a logarithmic cadence, with more of a focus on the Galactic plane.

The science case for DSA-2000 has been outlined in the Community Science Book\footnote{\url{https://www.dropbox.com/scl/fi/579zwhw5r8pt5o45ftsn0/DSA-2000_Community_Science_Book.pdf}}. There are four main science areas, two of which are directly relevant to this review: `{\it the dynamic radio sky}' and `{\it multi-messenger astronomy}'. Unlike the SKAO and ngVLA, there are no receivers to switch, so the DSA-2000 is planning on surveying the sky for its main science goals (with some time devoted to  followup of multi-messenger transients).  This, combined with the small element diameter and consequent large field-of-view, gives DSA-2000 a considerable advantage in survey speed compared to the other facilities at the same frequency.

\subsection{The next-generation VLA}
The next-generation Very Large Array \citep[ngVLA;][]{murphy_science_2018} is a  planned instrument that will consist of a Main array of $214\times 18\,$m off-axis parabolic antennas distributed across
the USA and Mexico, out to baselines of 1000\,km, plus another $30\times 18\,$m dishes on continental baselines (out to 9000\,km) in the Long Baseline Array (LBA) and $19\times 6\,$m dishes in the Short Baseline Array for diffuse sources (which is of less interest for transient science). The array will operate in a frequency range from 1 to 116~GHz, with maximum bandwidth 20~GHz. 

In contrast to the SKAO and DSA-2000, the ngVLA is designed to be primarily a PI-driven instrument. It has 5 key science drivers, the most relevant to this review being: `{\it Understanding the Formation and Evolution of Stellar and Supermassive Black Holes in the Era of Multi-Messenger Astronomy}'.  For the most part it is not planned that there will be large-scale surveys with the ngVLA, but it will certainly participate in dedicated time-domain followup studies, and the capabilities of the LBA for astrometry and precision measurements enable new science compared to what SKA and DSA-2000 offer on their own (e.g. Section~\ref{s_vlbi}).

\subsection{LOFAR 2.0}
The LOw-Frequency ARray \citep[LOFAR;][]{van_haarlem_lofar_2013} is a network of 24 Core stations located near Dwingeloo in the Netherlands, 14 Remote stations distributed across the Netherlands and 14 larger International stations distributed over Europe. Each station consists of low band antennas (LBAs) that operate at 10--90~MHz and high band antennas (HBAs) that operate at 110--240 MHz.  

LOFAR 2.0\footnote{See \url{https://www.lofar.eu/lofar2-0-documentation/}.} is a major upgrade to the LOFAR telescope. Two new stations will be added.  New low-noise amplifiers will be added to all antennas \citep{wang_real-time_2025}, and processing improvements will allow simultaneous use of both LBAs and HBAs, or using two beams  for each HBA station.  There are also improvements in the clock distribution and central processing facility used for calibration and imaging, as well as a 24/7 transient buffer for imaging and localisation.  The net result will be a considerably more powerful and flexible system capable of optimising its operation for many different transient cases (timescales, sensitivities, etc), as well as general improvements in image fidelity and processing  (enabling faster followup with better localisation).

The science case is outlined in the LOFAR 2.0 White Paper\footnote{\url{https://www.lofar.eu/wp-content/uploads/2023/04/LOFAR2_0_White_Paper_v2023.1.pdf}}. Transient science (including supernova, gamma-ray bursts, stellar activity and exploring the unknown), forms a major part of the science case.

\subsection{All-sky radio transient monitors}
\label{s_allsky}
To maximise the chance of detecting very rare, very bright events (e.g. fast radio bursts, long period transients, extreme stellar bursts, or coherent signals from neutron star mergers) it is desirable to maximise the time on sky.  These can be achieved with existing and proposed facilities that can continuously monitor large fractions of the sky \citep[e.g.][]{prasad_aartfaac_2016,bochenek_stare2_2020,sokolowski_southern-hemisphere_2021,connor_galactic_2021,lin_burstt_2022,luo_fast_2024,kosogorov_allsky_2025},  which are being developed with increasing sensitivity and other capabilities. 
This strategy achieved notable success in detecting a FRB-like burst from a Galactic magnetar \citep{bochenek_fast_2020}, which served to motivate future experiments.  

All-sky monitors have typically been designed to  detect `fast' transients, using multiple receivers or digital beamforming techniques rather than imaging.  This leads to good performance for short-duration signals, especially those that have detectable dispersion across the band (helping with interference rejection), but means they have reduced sensitivity to longer-duration signals due to fluctuations in the receiver electronics.  Nonetheless, there is overlap in the type of sources these systems can detect and those discussed in this review, such as the discovery of a LPT with period of 14\,min by the CHIME FRB system \citep[][although note that the same source was independently detected through LOFAR imaging by \citealt{bloot_strongly_2025}]{dong_chimefast_2025}. This discovery was helped by the relatively short pulse period and narrow duty cycle, such that individual pulses were short enough to be detected by the FRB system. Future explorations of the overlap between all-sky beamformed searches and more targeted imaging searches will help map out the underlying distribution of periods and fluences from sources like these. 

\section{Summary and outlook} 
\label{s_conclusions}
Time domain studies have been important since the dawn of radio astronomy, and give us insight into physical phenomena that span orders of magnitude in luminosity (from stars in our local neighbourhood through to distant gamma-ray bursts); in timescale (from nanosecond pulses through to afterglows that evolve over years); and in physical properties (from highly dense, magnetised compact objects to scintillation caused by interstellar plasma).  These regimes are often inaccessible via other observations and have led to numerous breakthroughs.

The field of image-domain transients has evolved from targeted studies of individual objects, to serendipitous coverage by generic surveys, through to custom designed large scale surveys. In the current decade (the 2020s), the plans mapped out many years ago have started to come to fruition, with a suite of telescopes reaching the sensitivity, resolution, and survey speed required to detect a wide range of radio transients at megahertz and gigahertz radio frequencies. 

Some of the scientific highlights of the past decade (from both targeted and untargeted observations) include (with tentative results in italics):
\begin{itemize}
\item Detection of an extreme scattering event in real time \citep{bannister_real-time_2016}; 
\item Identification of persistent radio sources associated with Fast Radio Bursts \citep{chatterjee_direct_2017};
\item Radio detection and monitoring of the afterglow from GW170817 \citep{hallinan_radio_2017}; 
\item {\it Detection of a possible gamma-ray burst orphan afterglow \citep{law_discovery_2018,mooley_late-time_2022}};
\item Discovery of fast blue optical transients \citep[FBOTs;][]{ho_at2018cow_2019};
\item {\it Tentative signs of radio emission from star-planet interactions \citep{vedantham_coherent_2020}}.
\item Detection of delayed radio flares from a tidal disruption event \citep{horesh_delayed_2021}; 
\item Precision measurement of the distance to and mass of the black hole X-ray binary Cygnus X-1 \citep{miller-jones_cygnus_2021};
\item Detection of merger-triggered core collapse supernova \citep{dong_transient_2021};
\item Discovery of long period transients \citep{hurley-walker_radio_2022,caleb_discovery_2022}; 
\item First detection of radio emission from Type Ia supernova \citep{kool_radio-detected_2023};
\item Substantial expansion of the known radio star population \citep{driessen_sydney_2024}; 
\item Identification of Type II radio burst from an M dwarf, suggestive of a coronal mass ejection \citep{callingham_radio_2025}.
\end{itemize}
Many of these discoveries are only the first in their class (e.g. long period transients and fast blue optical transients), and have already been supplemented by further discoveries.  Future surveys will continue to detect more examples, allowing us to understand the full underlying populations. 

The diversity of these results show that we are long past the point of exploring the radio transient sky in a way that is uninformed by the properties of the underlying populations. While early surveys emphasised searching in ways that were independent of the time domain properties of the potential populations, we can now characterise the populations on very different timescales (contrast long period transients, with periods of seconds to hours; with gamma-ray bursts that evolve over months).   Rather than counting the number of `transients', we can instead study multiple classes.

Much of the planning over the past few decades has been framed in the context of preparation for SKAO-era radio transient surveys. With this in mind, we end with some reflections on lessons learned from the pathfinder and precursor surveys that we hope feed into the plans for upcoming surveys with the telescopes discussed in Section~\ref{s_future}. Some of the things we think will be important are:

\begin{itemize}
\item {\bf Real-time imaging and transient detection:} No major radio transient imaging survey so far has had the capability to detect transients in real time (unlike FRB surveys). This will be critical in opening up shorter timescales (by enabling rapid follow-up or more complex analysis), which are currently proving very productive in finding new classes of radio transients, and in enabling rapid follow-up of new detections. 

\item {\bf Commensal transient surveys:}
Serendipity is an important part of transient astronomy. To discover rare objects, it is important to maximise the time on sky. Hence continuing to have full commensal access to other large surveys, with dedicated rapid processing, for the purpose of transient detection is critical, and helps maximise the scientific output of our telescopes. Of course data processing in the era of massive datasets is not `free', but the benefits of commensal transient surveys are likely to make this relatively minimal extra cost worthwhile.

\item {\bf Custom-designed transient surveys:}
It is important to state that commensal observing is not enough to reveal and study populations of radio transients. We have now begun to characterise the time domain, spectral, and polarisation properties of a range of source classes. To maximise the chance of detection of the most interesting sources, custom-designed surveys that consider the time domain properties of specific populations are required. 

\item {\bf Radio follow-up capability:}
Large-scale widefield surveys are key to discovering rare events. However, understanding them often requires detailed radio monitoring across a range of frequencies. For example, to establish the spectral evolution of synchrotron sources, or establish periodic behaviour in stellar objects. Hence it is critical in the SKAO era that we maintain access to facilities that can be used for this follow-up, including VLBI. 

\item {\bf Multiwavelength coverage:}
Multiwavelength (optical, near-infrared, X-ray, gamma-ray) survey data is vital for assessing the nature of new discoveries.  In general this does not have to be contemporaneous, although this is beneficial for some scientific goals. For example, stellar flares show significant structure within an event, so modelling requires multi-wavelength data, but triggering is impractical. For synchrotron transients, optical/IR data is required to identify a host galaxy, and archival high-energy data can be vital to determining whether it is a novel event like an orphan afterglow.

\item {\bf Automatic classification of radio transients:}
Now we are detecting significant numbers of radio transients in untargeted surveys, it is important to have automatic methods for removing poor quality data, and for classifying transients and identifying the most interesting outliers. There are well-established approaches in optical transient surveys that are in relative infancy in radio. These should be developed and refined.
\end{itemize}
In addition, the capabilities identified in Section~\ref{s_trends} are important considerations for future surveys. 

Enormous progress has been made in the field of radio transient surveys in the past decades. The surveys (and targeted monitoring) conducted to date have been fruitful, and in some cases revealed new populations of objects. However, there are still many unanswered questions that we will be able to address when the telescopes currently under development come online. With such a substantial investment of time and resources, it is important we incorporate the lessons learned from our existing experience as we head into this new era of radio transients.

\section*{Acknowledgements}
Thank you to the anonymous reviewer whose comments were very helpful in improving this manuscript, and to the editor, Minh Huynh, for advice and assistance throughout the publication process. 

Huge thanks to Ron Ekers, Elaine Sadler, Dougal Dobie and Iris de Ruiter for feedback on parts of our draft manuscript. 
In preparing this review we drew on the very useful compilation of radio transient surveys and rates by Kunal Mooley. Thanks to Kunal (and Deepika Yadav) for updating their online table in time for this review. Thank you to Laura Driessen for permission to adapt the colour-magnitude diagram in Figure 5. Romy Pearse created the lightcurve sketches in Figure 8. Thank you also to Ashna Gulati for help preparing Figures 4 and 8. Finally, thanks to Liroy Lourenço for his help with references and the formatting of this manuscript. 
Of course any remaining errors in the text are our own.

We had useful interactions and discussions with many people in the preparation of this review. Particular thanks to Gemma Anderson, Shari Breen, Manisha Caleb, Fernando Camilo, Lauren Carter, Shami Chatterjee, Barnali Das, Paul Demorest, Iris de Ruiter, Dougal Dobie, Laura Driessen, Ashna Gulati, Gregg Hallinan, Natasha Hurley-Walker, Scott Hyman, Francois Kapp, Emil Lenc, Walter Max-Moerbeck, Eric Murphy, Joshua Pritchard, Kovi Rose, Antonia Rowlinson, Sarrvesh Sridhar, Randall Wayth, Michael Wheatland and Andrew Zic. 

This research has made use of the Astrophysics Data System, funded by NASA under Cooperative Agreement 80NSSC21M00561.
In conducting this review, generative AI tools (ChatGPT-4) were used to help search for relevant papers in the literature. No information or text directly from generative AI was used in the preparation of this manuscript. 

This research was supported
by the Australian Research Council Centre of Excellence for Gravitational Wave Discovery (OzGrav), project number CE230100016, and by National Science Foundation grant AST-2511757.

\bibliography{merged_sorted_noabstracts}

\end{document}